\documentclass[preprint,12pt]{elsarticle}

\usepackage{amssymb}
\usepackage{amsmath}
\usepackage{graphicx} % Required for inserting images

\usepackage{subcaption}  
\usepackage{bm}  % Bold math
\usepackage{amsmath}

\usepackage{natbib} % to use \cite{}, \citet{}
\usepackage{mathrsfs} % mathscr command
\usepackage{hyperref}
\usepackage{xcolor} % to use \textcolor{}{}
\usepackage[ruled,vlined,english, algosection]{algorithm2e} % for the pseudo-code

\newcommand{\enstr}{\omega^2}

\NewDocumentCommand{\evalat}{sO{\big}mm}{%
  \IfBooleanTF{#1}
   {\mleft. #3 \mright|_{#4}}
   {#3#2|_{#4}}%
}
\newcommand{\ensavg}[1]{\langle #1 \rangle}
\journal{Journal of Computational Physics}

\begin{document}

\begin{frontmatter}

%% Title, authors and addresses

%% use the tnoteref command within \title for footnotes;
%% use the tnotetext command for theassociated footnote;
%% use the fnref command within \author or \affiliation for footnotes;
%% use the fntext command for theassociated footnote;
%% use the corref command within \author for corresponding author footnotes;
%% use the cortext command for theassociated footnote;
%% use the ead command for the email address,
%% and the form \ead[url] for the home page:
%% \title{Title\tnoteref{label1}}
%% \tnotetext[label1]{}
%% \author{Name\corref{cor1}\fnref{label2}}
%% \ead{email address}
%% \ead[url]{home page}
%% \fntext[label2]{}
%% \cortext[cor1]{}
%% \affiliation{organization={},
%%             addressline={},
%%             city={},
%%             postcode={},
%%             state={},
%%             country={}}
%% \fntext[label3]{}

\title{Physics-based localization methodology for Data Assimilation by Ensemble Kalman Filter}

%% use optional labels to link authors explicitly to addresses:
%% \author[label1,label2]{}
%% \affiliation[label1]{organization={},
%%             addressline={},
%%             city={},
%%             postcode={},
%%             state={},
%%             country={}}
%%
%% \affiliation[label2]{organization={},
%%             addressline={},
%%             city={},
%%             postcode={},
%%             state={},
%%             country={}}

\author{Sarp ER} %% Author name
\author{Marcello MELDI} %% Author name

%% Author affiliation
\affiliation{organization={Univ. Lille, CNRS, ONERA, Arts et Metiers ParisTech, Centrale Lille, UMR 9014- LMFL- Laboratoire de Mecanique des fluides de Lille -Kampe de
Feriet, F-59000 Lille, France}%Department and Organization
}

%% Abstract
\begin{abstract}
A physics-based methodology for the determination of the localization function for the Ensemble Kalman Filter (EnKF) is proposed. The spatial features of such function evolve dynamically over time according to the relevant instantaneous flow features of the ensemble members with the objective, to reduce the computational cost of the Data Assimilation (DA) procedure when applied with solvers for Computational Fluid Dynamics (CFD). The validation of the methodology has been carried out by the analysis of two test cases exhibiting different features. This permits to investigate different physical features, tailored for each test case, which affect the localization function. The flow over a two-dimensional square cylinder at $Re=150$ is the first case investigated. It has been shown that the proposed localization procedure leads to a more cost-effective DA process by reducing the size of the assimilated regions while keeping the same level of accuracy. The capabilities of the methodology are further demonstrated by the investigation of the turbulent flow around a three-dimensional circular cylinder for $Re=3900$. Again, the methodology exhibits an excellent trade off in terms of accuracy versus computational requirements.
\end{abstract}

%%Graphical abstract
\begin{graphicalabstract}
\includegraphics[width=1.0\textwidth]{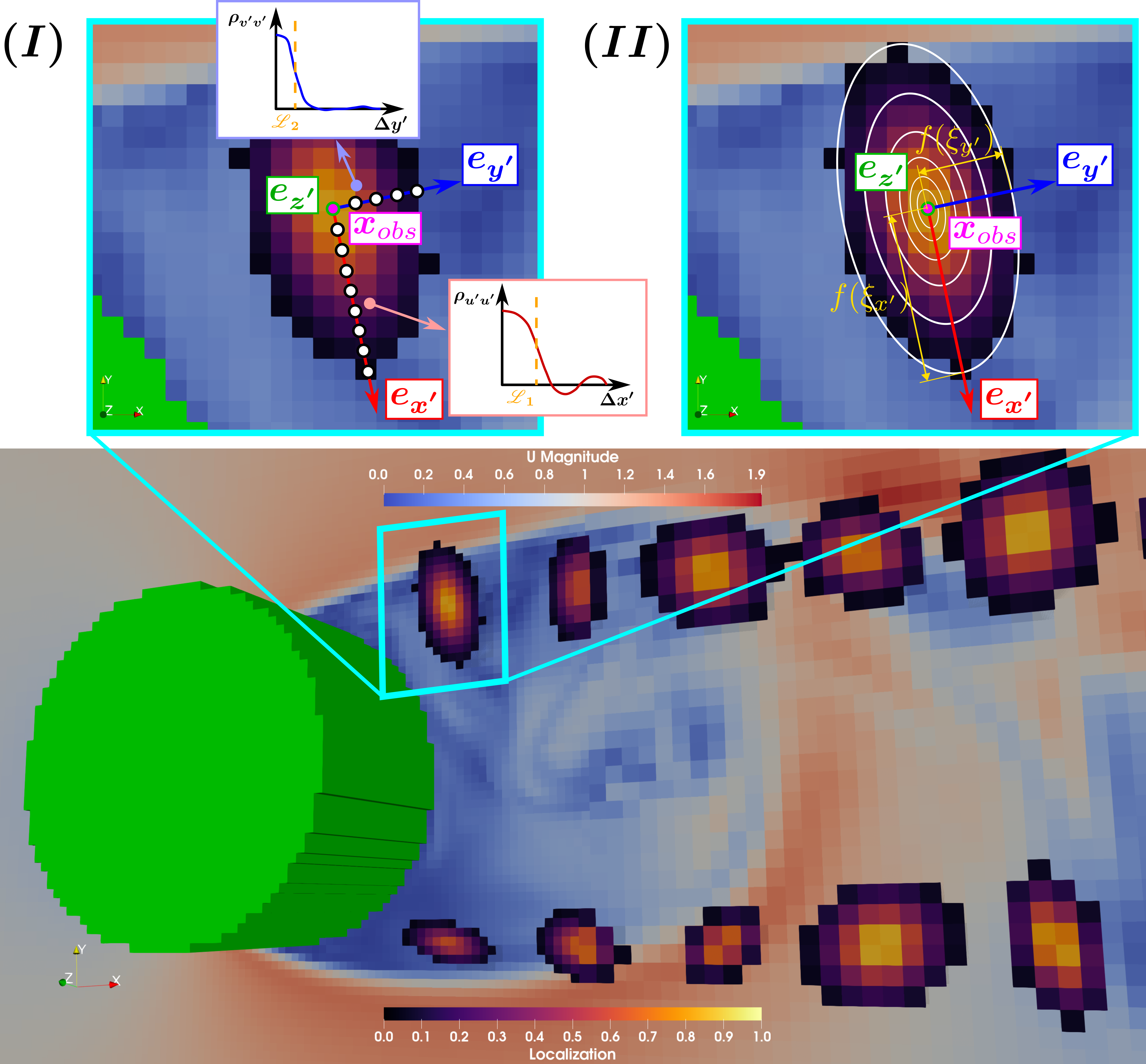}
\end{graphicalabstract}

%%Research highlights
\begin{highlights}
\item A physics based online-localization procedure is developed for the localization of the regions of data assimilation using EnKF.
\item The proposed methodology is applied for the synchronization of the virtual model with a reference simulation in two different cases, namely 2D unsteady flow around square cylinder and 3D turbulent flow around circular cylinder.
\item A significant reduction of the computational costs associated with DA procedure is observed due to the optimal choice of the DA regions at each analysis stage, without any loss in the synchronization performance as a result of the reduction of DA regions.
\item The relationship between the optimized parameters and the flow physics have been demonstrated.
\end{highlights}

%% Keywords
\begin{keyword}

DA \sep EnKF \sep localization \sep CFD \sep synchronization
%% keywords here, in the form: keyword \sep keyword

%% PACS codes here, in the form: \PACS code \sep code

%% MSC codes here, in the form: \MSC code \sep code
%% or \MSC[2008] code \sep code (2000 is the default)

\end{keyword}

\end{frontmatter}

%% Add \usepackage{lineno} before \begin{document} and uncomment 
%% following line to enable line numbers
%% \linenumbers

%% main text
%%

%% Use \section commands to start a section
\section{Introduction}
\label{sec:Introduction}
%% Labels are used to cross-reference an item using \ref command.

Applications of Data Assimilation (DA) \cite{Daley1991_cambridge,Asch2016,Evensen2022_Springer} to fluid mechanics have seen a rapid raise in the last decade \cite{Suzuki2012_jfm,Rochoux2014_nhess,Mons2016_jcp,Meldi2017_jcp,Zhang2020_jcp,Chandramouli2020_jcp,Mons2021_prf,DeMarinis2021_fcm,Zhao2022_be,Villanueva2024_sync_channel,Valero2025,Pillai2025_pof}. Thanks to the massive advancement in the availability of computational resources and in the development of efficient numerical architectures, one of the open challenges in the field is represented by the possibility to create digital twins \cite{Semeraro2021_ci,Rasheed2020_IEEE} of complex flow applications using numerical simulations. In these applications, the numerical replica of the real flow would receive streaming information of the latter obtained at localized sensors and, thanks to its capacity to resolve large physical domain, would be able to produce control techniques to anticipate or delay unwanted physical occurrences. Because of the relevance of such applications in numerous fields such as urban and transport engineering, energy harvesting and pharmaceutical applications, research development for these technologies has seen a rapid increase in recent times. DA tools are perfectly tailored for this kind of tasks, as they are able to handle multiple streaming sources of information such as predictions obtained by the physical and the digital twin. Works available in the literature on this topic mainly deal with the usage of reduced order models (ROM) based on data-driven techniques \cite{Brunton2020_arfm} to perform the numerical tasks of a digital twin \cite{Donato2024}. While the computational cost required by ROM is currently suitable for real-time investigation and coupling with physical phenomena, their performance over long time window may diverge due to intrinsic structural limitations and to quality of the data used for the training. The usage of high fidelity tools such as scale-resolving numerical simulation is still prohibitive for the application to digital twins for two main reasons. The first one is associated with the computational resources required to perform Direct Numerical Simulation (DNS) or Large Eddy Simulation (LES) \cite{Pope2000_cambridge}, which is prohibitively large at the present time. New paradigms in computational sciences, such as quantum computing, may however change the way calculations are performed in the foreseeable future. The second issue is associated with the computational costs of Data Assimilation to handle physical systems described by a large number of degrees of freedom (DOF). While DA has been successfully used to perform flow synchronization \cite{Villanueva2024_sync_channel} and parametric optimization of the numerical model \cite{Villanueva2024_Urban,Villanueva2024_sync_channel,Valero2025}, which are two cornerstones elements for efficient applications with digital twins, computational costs can be prohibitive if the DA algorithm is applied to the full physical domain. Some techniques of localization can be used to select specific regions of application of the DA procedures, potentially reducing the computational burdens required. Among these techniques, \cite{Villanueva2025_arxiv_JFM} have published a localization algorithm referred to as \textit{hyperlocalization}. This strategy, which was developed for the Ensemble Kalman Filter (EnKF) \cite{Evensen2022_Springer}, a well known sequential DA algorithm, is able to reduce the computational requirement of more than 100 times with equivalent accuracy than classical DA approaches. The technique consists in applying the DA procedure only in selected region where important information is obtained by sensors. However, the size of this region is somehow arbitrary and the shape of the region has been selected to be spherical for practical reasons.

In the present work, the hyperlocalization strategies previously introduced are augmented via infusion of the physical features of the flow. More precisely, the DA regions are shaped according to physical criteria related to the flow features observed in the region. The initial volume of such DA region, which is set to be spherical, is modulated to an ellipsoid whose axes and size are determined according to the physical criteria imposed. The new technique, which will be referred to as physics-based online localization (PBOL), is applied to investigate two unsteady test cases for external flows. The main difference between the two applications is the Reynolds number of investigation, which produces a laminar flow in the first case and a turbulent regime in the second application. Different physical criteria are applied to determine the size of the DA region for the two test cases, exploiting the features of the flow. 

The present article is structured as follows. In section \ref{sec:numerical_details}, the details of the numerical method for the solution of the governing equations are given, along with the ensemble Kalman filter algorithm used for the data assimilation stage and the various methods used for the localization of DA in the literature. In section \ref{sec:onlineLoc_methodology} the novel approach for the localization, namely physics-based online-localization (PBOL), is introduced with a description of the hyperlocalization methodology. In section \ref{sec:Results}, the results of the two test cases are presented. Firstly the 2D flow around the square cylinder, for the validation and the primary demonstration of the proposed PBOL methodology. Secondly the turbulent flow around 3D circular cylinder, where the application of the PBOL methodology is carried out for a fully turbulent test case. Finally, in section \ref{sec:conclusion}, the conclusions and final remarks are given.

\section{Numerical methods and algorithms} \label{sec:numerical_details}

This section is devoted to the presentation of the numerical tools used in this work. This includes the CFD solvers to simulate the flow dynamics as well as the state of the art DA algorithm, for which the physics-based localization approach is incorporated.

\subsection{Fluid state solver} \label{sec:numerical_details_CFD}

The dynamic behavior of an incompressible, Newtonian fluid flow is described by the continuity equation and the Navier-Stokes equations for mass and momentum conservation,

\begin{equation}
    {\nabla} \cdot \bm{u} = 0,
    \label{eq:continuity}
\end{equation}

\begin{equation}
    \frac{\partial \bm{u}}{\partial t} + \bm{u} {\nabla} \bm{u} = - \frac{1}{\rho} \nabla p + \nu \nabla^2 \bm{u} + f_p,
    \label{eq:NS_equation}
\end{equation}
where $\bm{u}$, $p$ and $\nu$ are the velocity field, pressure field and the kinematic viscosity of the fluid and $f_p$ is a volume forcing term. In this work, $f_p$ is used to account for the presence of immersed solid bodies via the immersed boundary method (IBM). A classical penalization approach \cite{Angot1999} has been used for the IBM where the forcing term is defined as
\[
    f_p(\bm{x}) = 
\begin{cases}
    -\nu \bm{D} (\bm{u - \bm{u}_{b}}),& \text{if } \bm{x} \in \Omega_{b}\\
    0, & \text{otherwise}
\end{cases}
\]
where $\Omega_b$ stands for the volume of the solid body, $\bm{u}_b$ is the velocity of this immersed body i.e. $\bm{u}_b = \bm{0}$ if it is stationary, and $\bm{D}$ is a tensor containing penalty coefficients, which can be determined for the best compromise between accuracy and the stability of the numerical solver \cite{Angot1999}.

Simulations are performed using the open-source, finite-volume platform OpenFOAM \cite{UserManual_OpenFOAM} and the implementation of this IBM tool within the solver used in the present study is detailed in the work by \citet{Valero2025}. More precisely, the IBM is integrated in the PimpleFOAM solver, a segregated solver where the velocity-pressure coupling is obtained using the PIMPLE algorithm \cite{UserManual_OpenFOAM}.
The calculations are performed using second order centered discretization schemes for all the spatial derivatives, while the time integration is carried out using the backward scheme, which is second order accurate in time. Because of the relatively low Reynolds number $Re$ of the test cases analyzed and thanks to the grid refinement selected, no turbulence modeling or subgrid-scale closure \cite{Pope2000_cambridge} is included in the numerical solvers. 

%\subsection{Data assimilation: EnKF algorithm} \label{sec:numerical_details_EnKF}

\subsection{Data assimilation: ensemble Kalman filter} \label{sec:EnKF}

Data Assimilation (DA) techniques are devoted to obtain an \textit{augmented state estimation} taking into account measurements and predictions coming from different sources of information. For studies in fluid mechanics, classical applications combine results from a numerical \textit{model} such as CFD with available \textit{observation}. The latter is usually supposed to provide high-fidelity measurements which are however sparse or limited in the physical domain, such as samples from sensors in experiments. DA methods can broadly be categorized in variational and statistical approaches. Variational methods, such as 3D-Var and 4D-Var originate from optimization theory and rely on minimizing a cost function that quantifies the discrepancies between model predictions and observations \cite{Daley1991_cambridge}. On the other hand statistical approaches, including the Kalman filter (KF), particle filter, and ensemble Kalman filter (EnKF), which are usually based on Bayesian inference \cite{Asch2016}. These methods not only integrate observation but also account for the uncertainties associated with both the model and the observation.

\begin{figure}[!htb]
    \centering
    \includegraphics[width=0.75\textwidth,trim={0 0 0 1.2cm},clip]{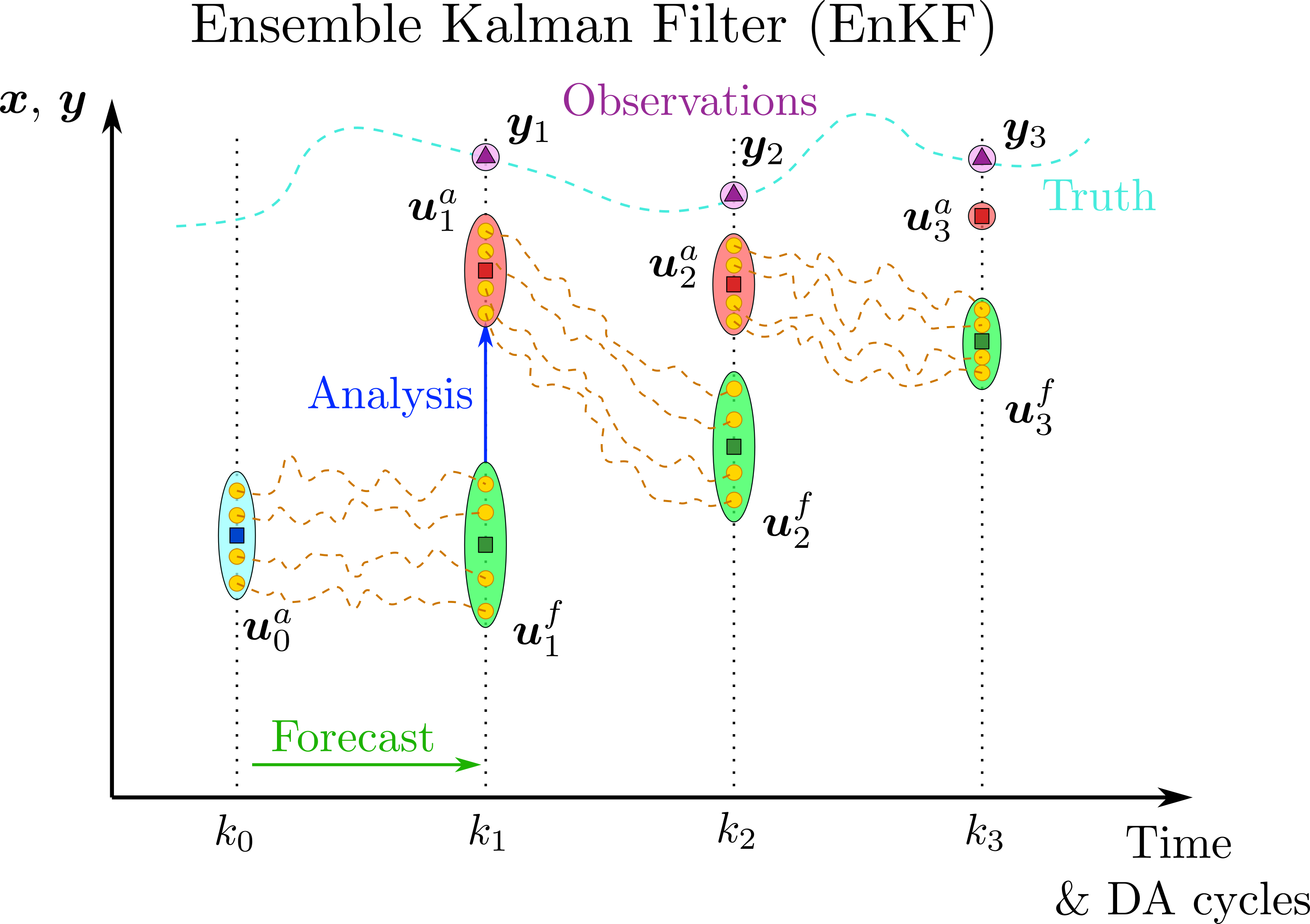}
    \caption{Scheme providing a qualitative representation of sequential DA using the EnKF.}
    \label{fig:schematic_EnKF}
\end{figure}

The EnKF is an ensemble version of the KF method which is particularly efficient for applications dealing with high-dimensional, non-linear systems. 
The EnKF methodology relies on two steps, namely the \textit{forecast} and \textit{analysis} steps. The forecast step involves the evolution of the system states of the ensemble members by the model equation while the analysis step produced the augmented state via the assimilation of the observation and the model prediction. One relevant feature of this approach is that uncertainties in the state predicted by the model and in the observation can be rigorously taken into account. The complete sequential DA process, which can integrate several forecast-analysis steps, is shown in figure \ref{fig:schematic_EnKF}.

Let us consider $N_e$ system state vectors $\bm{u}_i$ with $i \in [1, N_e]$, where $N_e$ is the number of ensemble members. These states are physical solutions of the flow investigated and they are obtained via numerical simulation using the \textit{model}. The state vectors advanced in time from the time step $k-1$ to $k$ are: 
\begin{equation}
    \bm{u}^f_{i,k} = \mathcal{M}_{i,k:k-1} \, \bm{u}^f_{i,k-1},
    \label{eq:DA_forecast_stage}
\end{equation}
where the superscript $f$ denotes the forecast using the model and $\mathcal{M}$ stands for the model equations i.e. for the present study, $\mathcal{M}$ is the numerical CFD algorithm resolving the governing equations described in the section \ref{sec:numerical_details_CFD}.

If observation is available at time $k$, an analysis phase can be performed to obtain a DA state estimation. This state is then used to update the ensemble members' system states for the following forecast steps. The update is affected by the discrepancy between the forecast and the observation, which drives the structure and magnitude of the so called Kalman gain $K_k$. The state update performed during the analysis (superscript $a$) reads as:
\begin{equation}
    \bm{u}^a_{i,k} = \bm{u}^f_{i,k} + \bm{K}_{k} \left( \bm{y}_k - \mathcal{H}(\bm{u}^f_{i,k-1}) \right).
    \label{eq:DA_analysis_stage}
\end{equation}
Here $\bm{y}_k$ is the observation vector at the time step $k$ and $\mathcal{H}$ is a mathematical operator used to project the model state to the observation space. The Kalman gain governs the update of the system states and it accounts for the level of confidence/uncertainty for each observation and for the forecast system state vector. It is computed as
\begin{equation}
    \bm{K}_{k} = \bm{X}_{k}^f(\bm{S}_{k})^T \left[\bm{S}_{k}(\bm{S}_{k})^T + \bm{R}_{k}\right]^{-1},
    \label{eqn:KalmanGain_general}
\end{equation}
with
\begin{equation}
\begin{aligned}
    \bm{X}_{k}^f &= \frac{\bm{u}_{i,k}-\langle \bm{u}_{k} \rangle}{\sqrt{N_e-1}},\, \\
    \bm{S}_{k} &= \frac{\bm{s}_{i,k}- \langle \bm{s}_{k} \rangle}{\sqrt{N_e-1}},
    \label{eq:AnomalyMatrices_general}
\end{aligned}
\end{equation}
where $\langle \cdot \rangle$ is the ensemble average operation over the ensemble members. In order to perform a consistent comparison between the ensemble of model realization and observation, data sampled for the latter is perturbed by a bounded Gaussian noise based on the covariance matrix of the measurement error $\bm{R}_{k}$ to obtain an $N_e$ set of values. This operation is performed at each analysis stage \textit{k} to obtain a well-posed mathematical problem. 

One of the main advantages of the EnKF when compared with the KF is that the error covariance matrix is not evolved explicitly in time but it is obtained via a Monte-Carlo approach i.e. running a set of $N_e$ ensemble members with different initial system states (and/or parameters if present) and sampling the distribution of their system states after the application of the model equation, see eq. \ref{eq:DA_forecast_stage}. This lower-rank approximation of the error covariance matrix is then computed from the statistics of the ensemble distribution by using \ref{eq:AnomalyMatrices_general}. While this procedure dramatically reduces the computational costs, it is also responsible for drawbacks which will be discussed in Sec. \ref{sec:localization}.

\subsection{Localization} 
\label{sec:localization}

The approximation of the error covariance matrix performed in the EnKF leads to important advantages when handling non-linear systems as it does not need an adjoint model nor linearization procedures. 
On the other hand, the number of ensemble members $N_e$ that can be reasonable generated via numerical applications is limited when dealing with numerical problems described by a very large number of degrees of freedom, due to heavy computational burden. The representation of the full error covariance matrix of a very large, complex system using a limited number of snapshots (i.e. ensemble members) leads to sampling errors \cite{Hamill2001,Houtekamer2001,Asch2016}. The restricted size of the ensemble pool leads to non-physical, long-distance correlations, which can negatively impact the performance of the DA procedure \cite{Hamill2001,Hunt2007}.

Several methods have been reported in the literature to reduce and control these sources of error \cite{Asch2016}. One well known technique is the physical localization, which is based on the selection of a partition of the physical domain where DA is applied. In the study of \citet{Houtekamer2001}, multiple DA analyses are performed grouping available observation in batches. The state update is performed sequentially summing up the contributions for each batch. Similarly in \citet{Hunt2007}, multiple DA analyses are carried out for each element of the grid used for the model forecast. In this case not every available observation is used, but a limited subset is selected according to criteria based on the distance of each sensor to the grid element. This strategy is equivalent to a physical localization as the DA problem is solved by subsets of observations and/or data points determined by physical proximity. More recently, \citet{Zhang2024} have presented a methodology based on non-overlapping domain decomposition used for the partition of the global EnKF problem into smaller regions, which acts as a physical localization mechanism, eliminating the non-physical correlations arising from low-rank representation of the error covariance matrix. They employed total variation regularization to account for the missing covariance information at the decomposition boundaries, leading to an acceleration in convergence and improvement of the accuracy of the inferred field.

Another family of strategies used to control sampling errors is known as covariance localization. The localization is here achieved by the Schur product of a localization function, with certain features e.g. Gaussian shape, decaying to zero at a specified distance \cite{Hamill2001}, with the error covariance matrix. This procedure acts as a filter suppressing the non-physical correlations appearing at long-distance separations\cite{Hamill2001,Houtekamer2001}. Classical applications set the value for the localization function to the unity at the observation location. The function then tends to zero with the increasing distance from the observation location, based on the expected decrease of the correlation value with distance \cite{Houtekamer2001}. Because of the very large number of degrees of freedom simulated by CFD approaches and the consequent large computational resources required, localization is a key tool to improve DA stability and efficiency in terms of accuracy versus costs.

\subsection{CONES: Coupling OpenFOAM with numerical environments} 
\label{sec:CONES}

A key element governing the DA efficiency when using CFD is related to the possibility of performing multiple tasks in an \textit{online} environment. The EnKF is a sequential DA methodology, meaning that the DA step is carried out whenever a new observation is available. Thus the DA procedure consists of multiple DA cycles with forecast (model advancement using CFD) and analysis steps (calculations using a DA software). If the $N_e$ CFD simulations are stopped at the end of the forecast and restarted at the end of the analysis phase, this can request prohibitive computational costs in terms of writing output files and loading the computational grids and variables. Therefore, the possibility to \textit{pause} the CFD simulations while the DA code is running represent a major advantage in terms of computational efficiency. In the present work, the forecast step in eq. \ref{eq:DA_forecast_stage} is carried out by the OpenFOAM algorithm. The analysis step for the computation of the corrected state matrix $\bm{u}^a_{i,k}$ is carried out by an independent DA algorithm, coded in Python language and follows the procedure presented in algorithm \ref{alg:EnKF}. The online coupling of the OpenFOAM code and the Python EnKF algorithm is performed using the CONES library \cite{Villanueva2024_Urban,Valero2025} based on the message passing interface (MPI) \cite{mpi41} and mpi4py \cite{mpi4py_Dalcin2005} libraries.

\begin{figure}[!htb]
    \centering
    \includegraphics[width=1\textwidth]{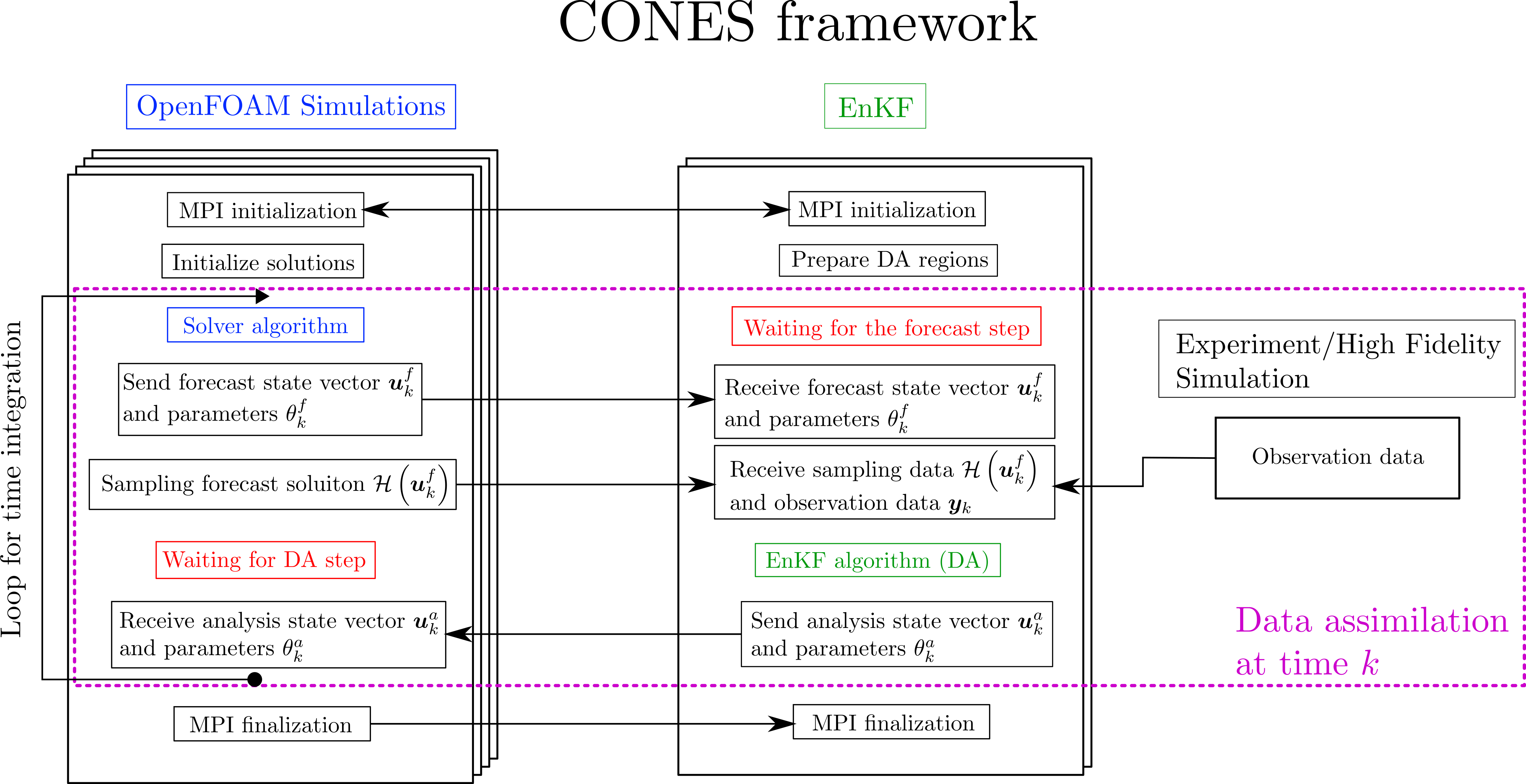}
    \caption{Scheme representing the work environment of the library CONES used for DA purposes.}
    \label{fig:schematic_CONES}
\end{figure}

\begin{algorithm}
    \caption{Algorithm for the Ensemble Kalman Filter (EnKF)}
    \label{alg:EnKF}

    \nl MPI initialize \\
    \nl Read grid info \\
    \nl Prepare objects containing EnKF \& simulation parameters \\
    \nl Read clipping info from \textit{topoSet} and construct DA regions data structure \\
    \nl Read observation data and associate them with DA regions \\
    
    \For{$k = 1$ to $K$}{ 
        \nl Receive forecast state vectors from OpenFOAM simulations $\bm{u}_{i,k}^f$ \\
        %\nl \textit{(optional)} Receive parameters $\Theta_{i,k}$ \\
        \nl Receive model sampling data $\bm{s}_{i,k} = \mathcal{H}(\bm{u}_{i,k}^f)$ \\

        \For{$j = 1$ to $N_r$}{
            \nl Creation of an observation matrix from the observation data by introducing observation errors:\\
            \qquad$\bm{y}_{i,k} = \bm{y}_{k} + \bm{e}_{i, j}$, with $\bm{e}_{i, k} \sim \mathcal{N}(0,\bm{R}_{k})$ \\
            \nl Calculation of the ensemble means:\\
            \qquad$\langle \bm{u}_{j,k}^f \rangle = \frac{1}{N_e}\sum_{i = 1}^{N_e}\bm{u}_{i,j,k}^f$,\,
            $\langle \bm{s}_{j,k} \rangle = \frac{1}{N_e}\sum_{i = 1}^{N_e}\bm{s}_{i,j,k}$,\\
            \nl Calculation of the anomaly matrices:\\
            \qquad$\bm{X}_{k}^f = \frac{\bm{u}_{i,k}-\langle \bm{u}_{k} \rangle}{\sqrt{N_e-1}}$,\,
            $\bm{S}_{k} = \frac{\bm{s}_{i,k}- \langle \bm{s}_{k} \rangle}{\sqrt{N_e-1}}$,\\
            \nl Calculation of the Kalman gain:\\
            \qquad$\bm{K}_{k} = \bm{X}_{k}^f(\bm{S}_{k})^T \left[\bm{S}_{k}(\bm{S}_{k})^T + \bm{R}_{k}\right]^{-1}$ \\
            \nl \textit{(optional)} Apply localization function:\\
            \qquad$\bm{K}_{k}^L = \bm{L} \odot \bm{K}_{k}$ \\
            \nl Update state matrix:\\
            \qquad$\bm{u}_{i,k}^a = \bm{u}_{i,k}^f + \bm{K}_{k}^{L}(\bm{y}_{i,k}- \bm{s}_{i,k})$
    }

    \nl Send analysis state vectors to OpenFOAM simulations $\bm{x}_{i,k}^a$ \\
    }
\end{algorithm}

As it can be seen in algorithm \ref{alg:EnKF}, the DA software starts by the initialization of the MPI environment and reading the inputs such as the OpenFOAM grid information, initial parameters (if present) and DA hyperparameters. After the preparation of the data-structure, the loop over the DA stages starts for the index $k=0$. The DA algorithm waits for the completion of the forecast step carried out by the OpenFOAM and receives the forecast state matrices of the ensemble members $\bm{u}_{i,k}^f$ via CONES when they are ready. The model samples in the observation space $\bm{s}_{i,k} = \mathcal{H}(\bm{u}_{i,k}^f)$ are received from each ensemble member along with the parameters $\Theta_{i,k}$ if they are present. With all the information from the ensemble members are now gathered, the DA algorithm is performed while CFD simulations are paused. Observation matrix, ensemble means, anomaly matrices, Kalman gain and localization matrix are computed. 
Once the augmented state is obtained, the information is sent back to the $N_e$ simulations and physical states are updated. The hold on the CFD simulations is removed and the DA code is paused as the next forecast begins.  

\section{Hyperlocalization and Physics Based Online-localization} 
\label{sec:onlineLoc_methodology}

\subsection{Hyperlocalization algorithm}
\begin{figure}[!htb]
    \centering
    \includegraphics[width=0.6\textwidth]{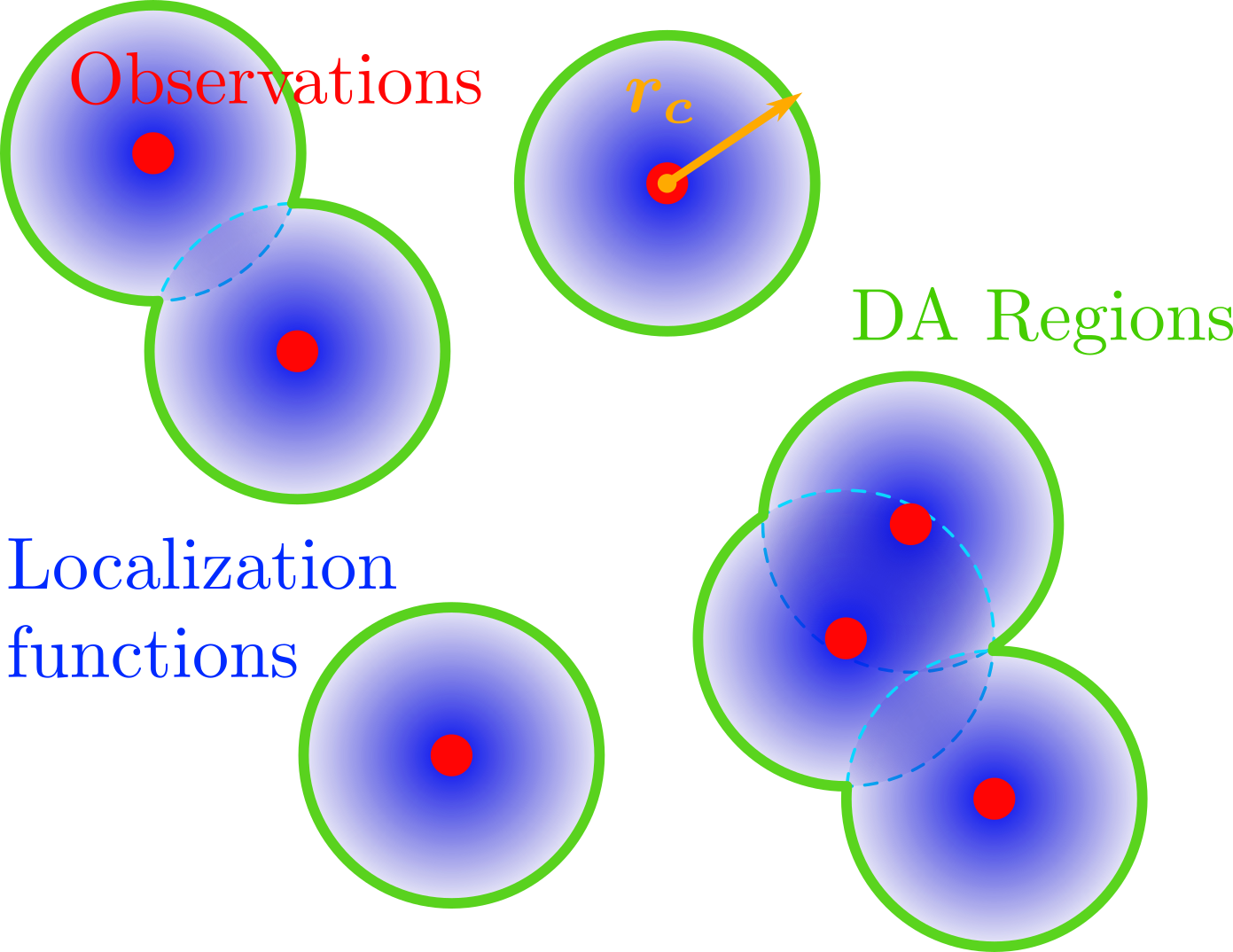}
    \caption{Scheme representing DA regions selected for the hyperlocalization (HL) algorithm.}
    \label{fig:schematic_DA_regions}
\end{figure}

Discussion in Sec. \ref{sec:localization} highlighted how localization is important in terms of numerical stability and computational efficiently. Recently, \citet{Villanueva2025_arxiv_JFM} proposed a localization technique tailored for CFD applications, which is referred to as hyperlocalization (HL). This technique is reminiscent of the works by \citet{Hunt2007} and \citet{Houtekamer2001} but, unlike the latter, it is not inherently sequential but it can be performed in parallel. This last point permits to accelerate the DA analysis procedure, therefore providing improved efficiency in critical applications using CFD. For the HL localization procedure, DA regions around sensors are identified as spherical regions of radius $r_c$ as shown in figure \ref{fig:schematic_DA_regions}. The usage of these regions for DA purposes is justified by the assumption that in fluid flows, especially in turbulent regimes, the correlation function of physical quantities between two points decays rapidly with distance  \cite{Pope2000_cambridge}.

The definition of such zones permits to define $N_R$ DA zones, where the number of degrees of freedom is limited and therefore the size of the matrices in eq. \ref{eqn:KalmanGain_general} and eq. \ref{eq:AnomalyMatrices_general} is dramatically reduced. In addition, the resolution of the $N_R$ analyses can be performed in parallel, therefore providing an additional potential speed-up in computational costs. If sensors are relatively far from another (i.e. the distance between them is larger than $r_c$), then $N_R$ is both the number of DA regions and of sensors. This is the case investigated in the present analyses, which is chosen for sake of simplicity. However, if the probes are close, one single DA region can cluster multiple sensors as shown in fig. \ref{fig:schematic_DA_regions}. 

For the case of HL-based EnKF, the state update provided in eq. \ref{eq:DA_analysis_stage}  is recast in a parallel formulation where $N_R$ updates are performed:
\begin{equation}
\mathbf{u}_{i,k}^a = \mathbf{u}_{i,k}^f + \sum_{j=1}^{N_R}  (\mathbf{K}_{k})_{[j]}\left( (\bm{y}_{i,k})_{[j]} - (\mathcal{H}(\bm{u}^f_{i,k}))_{[j]} \right).\label{eq:regions_DA_analysis_stage}
\end{equation}
Here $[j]$ indicates that calculations of the matrices used to perform the analysis step are calculated using the information obtained from the $j^{th}$ DA region, as shown in the Algorithm \ref{alg:EnKF}. Therefore, the operations performed for each $[j]$ also include projections from and to the complete physical state to the reduced state in each of the $j$ DA regions. The process is repeated $N_R$ times. It must be stressed (see fig. \ref{fig:schematic_DA_regions}) that the DA regions are constructed so that the contributions from each DA region $j$ are independent of each other. Therefore, the procedure can be written as the linear sum of the local DA updates.

It was previously indicated that the HL localization technique works with spherical regions described by a prescribed radius $r_c$. Discontinuities of the flow field at the surface of the region may appear due to the state update. In order to avoid these discontinuities, covariance localization is used to artificially damp the state update term to zero in the proximity of the surface of the DA region. This result is obtained pre-multiplying the Kalman Gain by a localization matrix:
\begin{equation}
    \bm{K}_{j,k}^L = \bm{L}_j \odot \bm{K}_{j,k},
    \label{eq:multiplication_localization}
\end{equation}
where $\odot$ stands for element-wise multiplication. Given that the current DA methodology is developed for applications dealing with continuous physical variables, a reasonable choice is to use a distance-based localization function for covariance localization. For this reason, a normal distribution is used following recommendation in the literature \cite{Asch2016}:
\begin{equation}
    L(\bm{x}_c) = \exp(-\frac{1}{2}\left[\left( \bm{x}_c - \bm{x}_o \right)/\xi \right]^2),
    \label{eq:HL_localization_fct}
\end{equation}
where $L$ is the value of the localization function, $\bm{x}_c$ is the location of the model estimation i.e. the location where the localization function is evaluated, and $\bm{x}_o$ is the location of the probe for observation. The decay rate of the localization function is determined by a length scale $\xi$. This parameter is set so that the state elements of the localized Kalman gain $\bm{K}_{j,k}^L$ tend to zero in the proximity of the surface of the DA region, therefore adequately smoothing the intensity of the state update \cite{Villanueva2024_Urban, Villanueva2024_sync_channel}. The value of $\xi$ is constant in time and it is set at the beginning of the DA procedure.

The HL procedure has been validated against classical EnKF without localization for classical test cases such as a lid-driven cavity and a turbulent plane channel flow, obtaining similar results in terms of accuracy \cite{Villanueva2024_PhDThesis}. However, due to the reduced number of degrees of freedom considered and the possibility to easily parallelize the analysis step, computational resources required are significantly reduced. An interesting point is that this cost reduction becomes proportionally more important for systems described by a larger number of degrees of freedom, therefore showing potential for complex flow applications of industrial interest.

\subsection{Physics-based online localization}

The main limitation of the initial proposal for the HL technique relies on the geometric features of the DA regions. In fact, the shape (usually spherical) and the size must be prescribed at the beginning of the DA run and they cannot be changed during the process. While the choice of the size, driven by the sphere radius $r_c$, can be performed according to apriori estimation of flow statistics, the geometry cannot adapt to instantaneous variations of the features of the velocity and pressure fields. This constraint does not exploit to its fullest the availability of data in a sequential DA procedure such as the EnKF.  

In the present study, a dynamic localization strategy is developed proposing time variation of the geometric features of the DA regions. The method, which will be referred to as physics-based online-localization (PBOL), identifies the shape and size of the DA regions via observation of predicted flow features by the ensemble realizations of the model. These flow features are used to identify regions which respond to physical criteria selected by the users. Therefore, this localization technique has the potential to infuse physical knowledge of the flow to improve the efficiency of the DA method. One appealing characteristic is that, thanks to the sequential features of the EnKF, shape modifications of the regions can be performed at each analysis phase. Therefore, the PBOL technique can provide the means to perform a dynamically changing DA process which adapts to the instantaneous features of the flow. For the present analysis, optimized flow regions will be chosen among ellipsoid shapes corresponding to the constitutive equation:
\begin{equation}
\label{PBOL-ellipsoid}
\frac{{x^\prime}^2}{{\xi_{x^\prime}}^2} + \frac{{y^\prime}^2}{{\xi_{y^\prime}}^2} + \frac{{z^\prime}^2}{{\xi_{z^\prime}}^2} = 1
\end{equation}
where $x^\prime$, $y^\prime$ and $x^\prime$ is a system of reference described by the vectors $\bm{e}_{x'}$, $\bm{e}_{y'}$ and $\bm{e}_{z'}$. The size of the ellipsoid is governed by the size of the semi-axes $\xi_{x^\prime}$, $\xi_{y^\prime}$ and $\xi_{z^\prime}$. Using this form for the DA regions, the online procedure targets optimization for six parameters: 
\begin{itemize}
\item The three angles providing the transformation from the base $[\bm{e}_{x}, \, \bm{e}_{y}, \, \bm{e}_{z}]$ to the base $[\bm{e}_{x^\prime}, \, \bm{e}_{y^\prime}, \, \bm{e}_{z^\prime}]$ 
\item The shape of the elliptical localization function in the three semi-axes $\xi_{x^\prime}$, $\xi_{y^\prime}$ and $\xi_{z^\prime}$
\end{itemize}

\begin{figure}[!htb]
    \centering
    \includegraphics[width=0.48\textwidth]{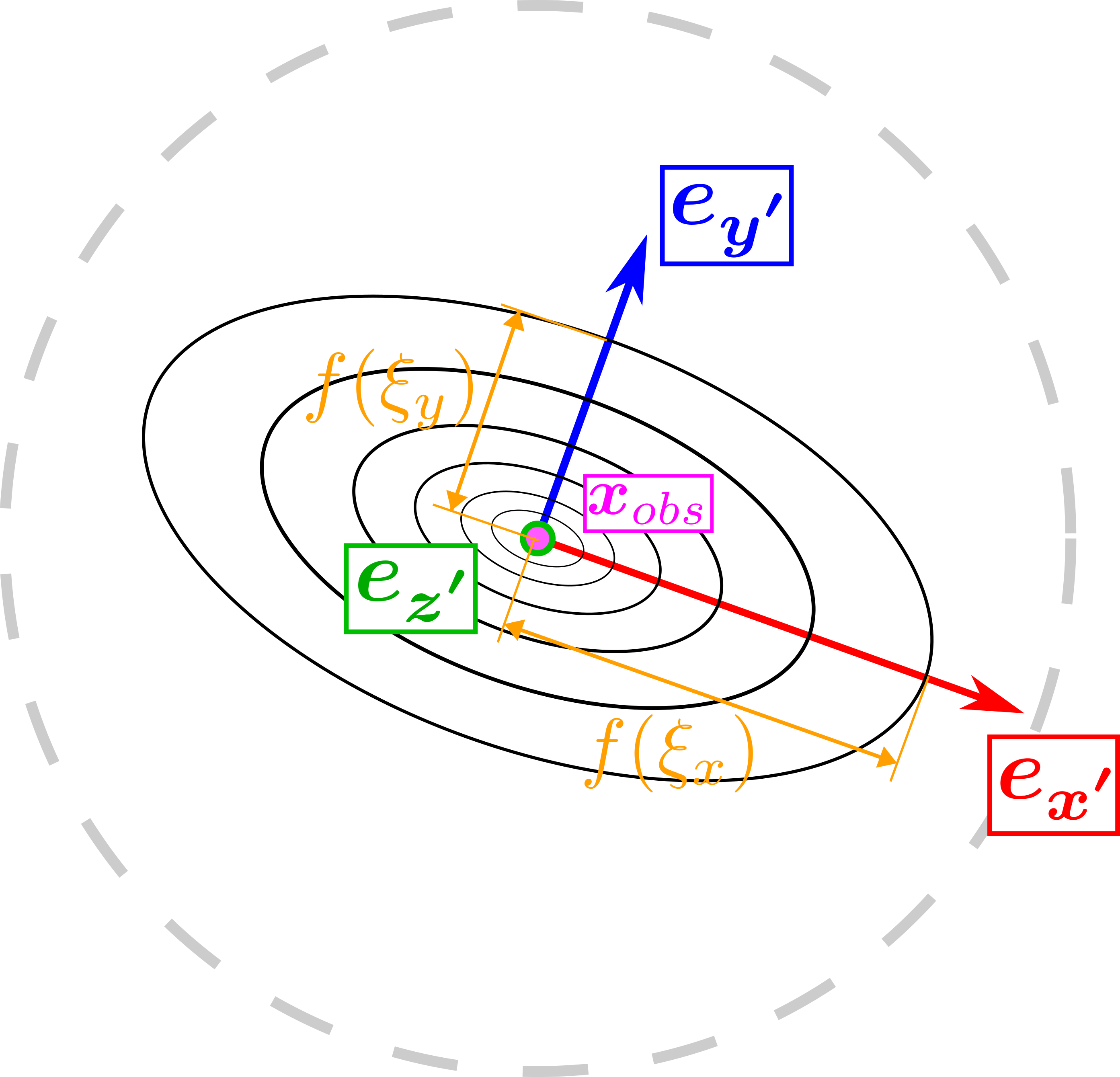}
    \caption{Scheme representing the PBOL technique. The parameters to be optimized are shown.}
    \label{fig:schematic_onlineLoc_function}
\end{figure}

The six parameters are optimized for each of the $N_R$ DA regions considered. This permits to tailor the DA process not only to the instantaneous features of the flow, but also to different locations where the flow may provide specific features. A sketch of representing the PBOL technique and the relevant parameters is shown in the figure \ref{fig:schematic_onlineLoc_function}. For a two-dimensional problem, the number of parameters to be determined is reduced to four, which are the angles providing the rotation to obtain $\bm{e}_{x'}$, $\bm{e}_{y'}$ and the semi-axes $\xi_{x^\prime}$, $\xi_{y^\prime}$.

As it was previously discussed, the optimization of the coefficients is performed using physical information of the flow in the DA regions. This information is used as a constraint to determine the appropriate size and orientation of each ellipsoid. Therefore, the physical features selected can be changed for different test cases of investigation, in order to maximize the quality of the DA process. The PBOL technique is validated via the analysis of two different test cases, and the physical criteria selected for each one of them are detailed in sections \ref{sec:squareCyl2D} and \ref{sec:sync_Cyl3D}.

While the selected criteria are fixed for each test case, the PBOL algorithm is developed with a modular structure which permits an easy implementation of localization strategies using appropriate metric, depending on the physics of the problem.

Similarly to the HL proposition, the PBOL methodology may provide a discontinuity in the state update if covariance localization is not included in the algorithm. In this case, the coefficients driving the exponential functions are determined as a function of the optimized values of $\xi_{x^\prime}$, $\xi_{y^\prime}$ and $\xi_{z^\prime}$. In this way, one can grant that the state update tends to zero in the proximity of the surface of the ellipsoid.
As a result, the localization function is obtained in the shape of a three-dimensional multivariate normal distribution and computed using the formula
\begin{equation}
    L_j = \exp(-\frac{1}{2}\left[\left( \bm{x}_c - \bm{x}_o \right)^T \Sigma_j^{-1} \left( \bm{x}_c - \bm{x}_o \right) \right]),
    \label{eq:OL_localization_fct}
\end{equation}
where the $\Sigma_j$ is the covariance matrix which contains the information related to the orientation of the axis of the ellipsoid with respect to the Cartesian axes. Its magnitude is computed as
\begin{equation}
    \Sigma_j = \mathcal{R}_j D_j \mathcal{R}_j^T.
    \label{eq:Sigma_OL_eqn}
\end{equation}
Here, the $\mathcal{R}_j$ is the rotation matrix containing the local orthonormal axes corresponding to the principal axes of the ellipsoid $\mathcal{R}_j = [\bm{e}_{x'}, \bm{e}_{y'}, \bm{e}_{z'}]$.  $D_j$ is a diagonal matrix defined by $D=diag(\xi_{x^\prime}, \xi_{y^\prime}, \xi_{z^\prime})$. Additional scaling coefficients can be included to tune the covariance localization effects moving away from the center of the ellipsoid.

The PBOL technique differs from the HL localization for three main aspects:
\begin{enumerate}
\item The usage of an ellipsoid instead of a sphere allows for more complex DA correction accounting for a level of anisotropy in the correlation between variables, while keeping competitive computational requirements. 
\item Thanks to the modularity of the technique and the possibility to perform parallel DA analyses for each region, the PBOL procedure leads to different localization features due to variations in the local flow field. 
\item The DA region independently evolve in time as the underlying local flow field changes. This aspect is particularly interesting for statistically non-stationary (i.e. accelerating flows) or unsteady flows where the optimization procedure can adapt the features of the DA region to instantaneous requirements and constraints.  
\end{enumerate}

\section{Results} \label{sec:Results}

The performance of the physics-based online-localization procedure described in section \ref{sec:onlineLoc_methodology}, is now validated via the analysis of two test cases. The investigation will consider the accuracy of the results obtained by the DA algorithm, as well as the speed of the rate of convergence of the method and the computational resources required. The two configurations investigated deal with external flows, and the main difference consists in the flow regime governed by the Reynolds number $Re$. More precisely, the first test case investigated is a low $Re$ flow (unsteady laminar regime) while the second one is for moderate $Re$, where a fully turbulent wake is observed. The investigation of a laminar test case and a turbulent flow also permits to test different metrics to determine the shape of the localization region, which are tailored to efficiently take into account the physical features of the flow.  

\subsection{2D flow around a square cylinder, $Re=150$} \label{sec:squareCyl2D}

\begin{figure}[!htb]
    \centering
    \includegraphics[width=0.9\textwidth]{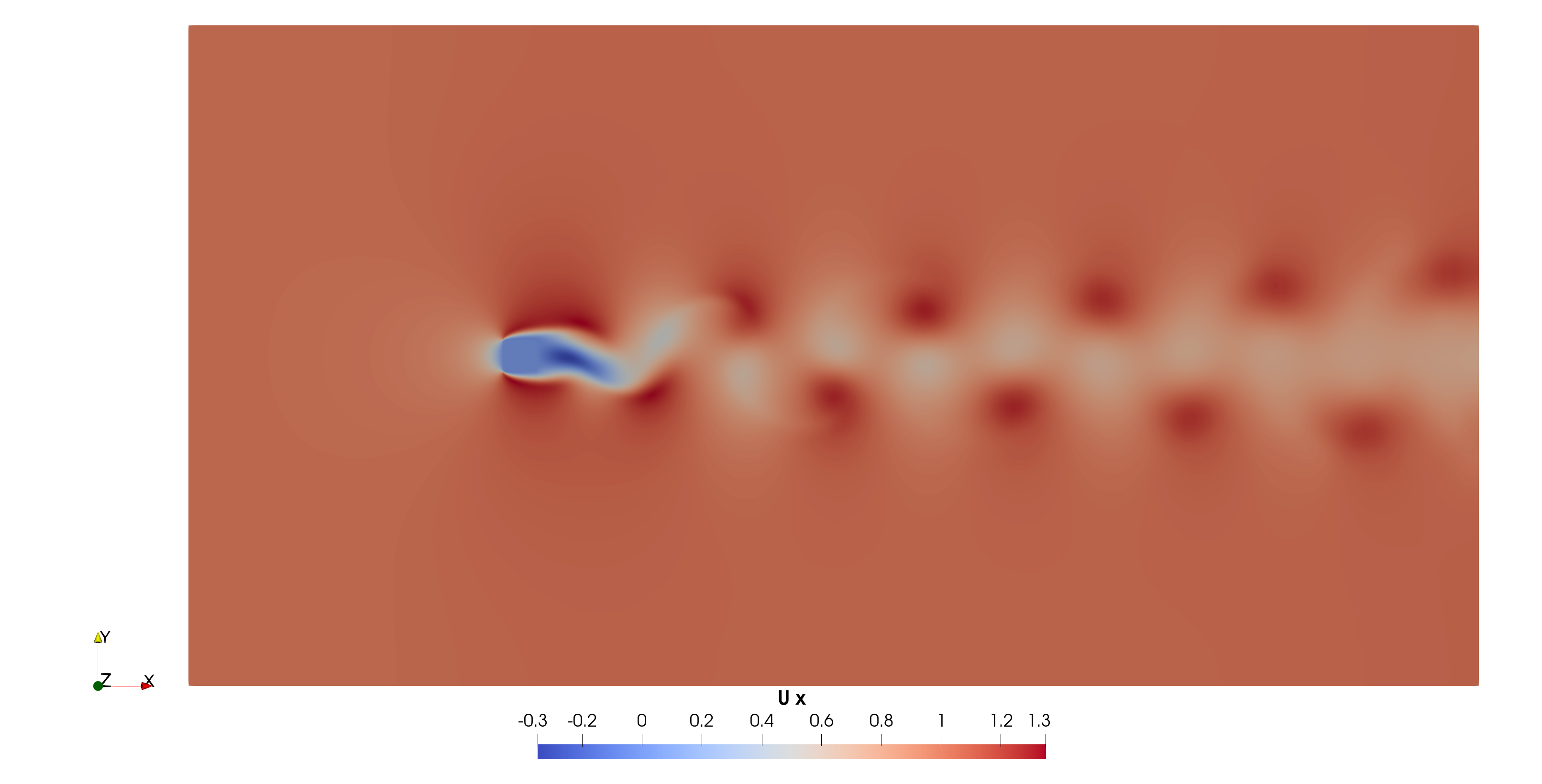}
    \caption{Instantaneous iso-contours of the streamwise velocity $u_x$ and the vortex street downstream the wake region.}
    \label{fig:sqCyl_u_vel_contours}
\end{figure}

The first test case investigated is the flow around a cylinder at $Re=150$. The Reynolds number $Re=\frac{U D}{\nu} = 150$ is here based on the uniform velocity at the inlet $U$, the edge length of the square cylinder $D$ and the kinematic viscosity of the fluid $\nu$. This laminar flow configuration, which exhibits two-dimensional unsteady features, has been extensively studied in the literature \cite{Sohankar1998, Sohankar1999, Saha2003}. The Karman-vortex street observed downstream the cylinder is responsible for fluctuations of the drag coefficient $C_D$ and the lift coefficient $C_L$, which exhibit a sinusoidal behavior whose frequency is governed by the Strouhal number $St$. A visual representation of the flow is provided in \ref{fig:sqCyl_u_vel_contours}. 

The test case consists of a square cylinder with its center placed in the proximity of the origin of the coordinate system at $x=0.5D, \, y=0$. The extent of the two-dimensional domain, which is $[-10, \, 31] \times [-10.5, \, 10.5]$ in terms of $D$, is selected taking into account previous studies reported in the literature \cite{Sohankar1998, Sohankar1999, Breuer2000}. In this problem, $x$ is the streamwise direction, while $y$ is the cross-stream direction. A constant velocity inlet condition $\mathbf{u} = (U, 0)$ is imposed for $x/D = -10$ while a mass conserving advective outlet condition is used at $x/D=31$. This distance has been chosen so that the outflow boundary condition has no effects on the flow field near the cylinder and at least five vortex pairings are observed \cite{Sohankar1998}. Periodic boundary conditions are imposed in the cross-stream direction.

\begin{figure}[!htb]
    \centering
    \includegraphics[width=0.9\textwidth]{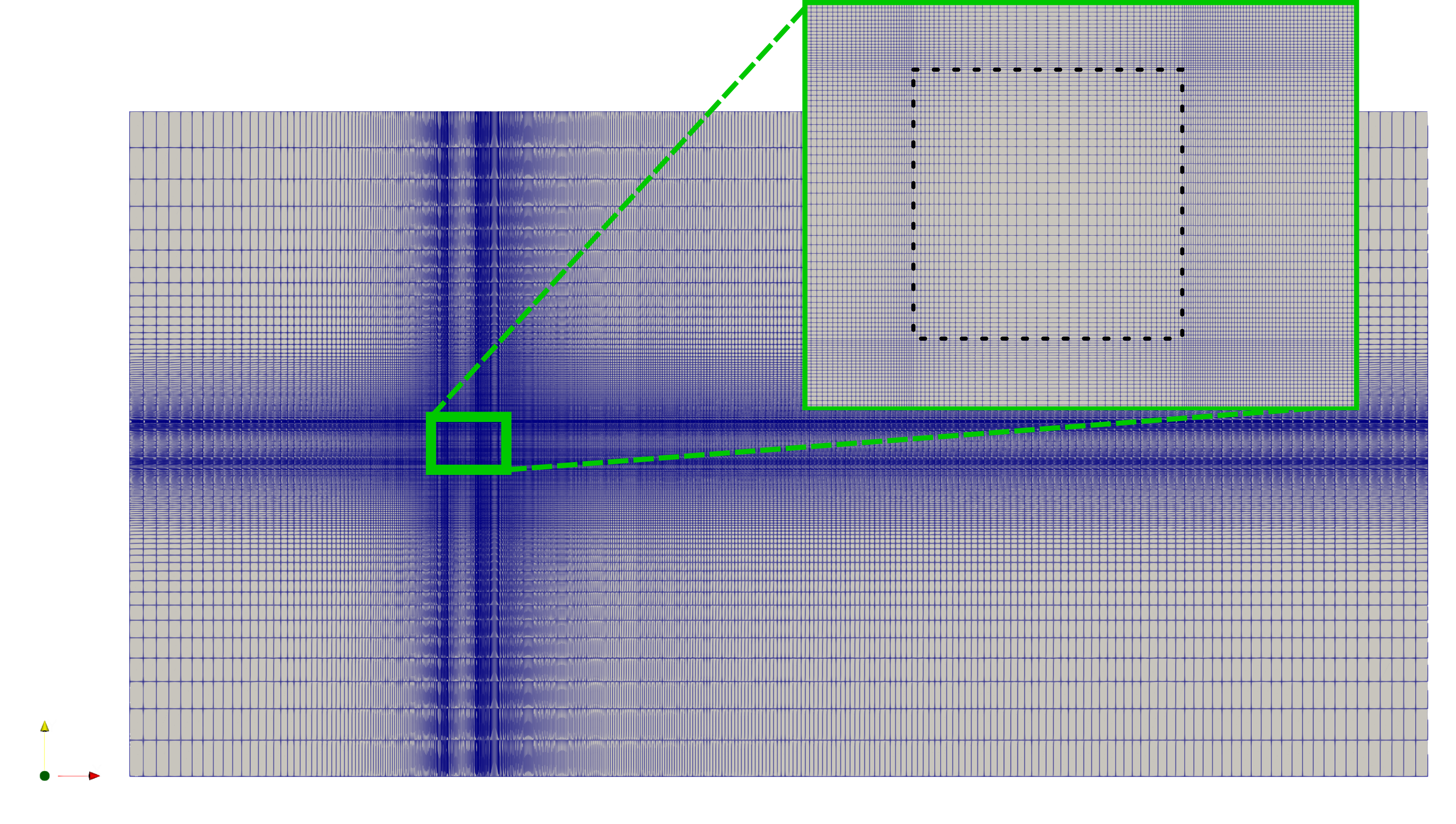}
    \caption{Grid used to discretize the physical domain investigated for the 2D flow around a square cylinder. A zoom corresponding to the region of the immersed body is included.}
    \label{fig:sqCyl2D_IBM_grid}
\end{figure}

The grid features are now described. As discussed in section \ref{sec:numerical_details_CFD} an IBM tool is used to account for the presence of the immersed body, therefore a continuous cartesian grid is used for the analysis. The elements are most refined in the proximity of the surface of the immersed body, as shown in figure \ref{fig:sqCyl2D_IBM_grid}. The thickness of the first element in the proximity of the wall is $\delta_{y=0} = 0.008D$, following recommendation in the literature \cite{Sohankar1998,Sohankar1999}. Moving away from the surface of the cylinder, the grid elements become progressively coarser. The total number of computational cells is $92.3 \times 10^{3}$. The numerical schemes used for numerical simulation are a second-order accurate scheme for the spatial discretization and the time advancement is carried out by the second-order accurate backward scheme. 

\begin{figure}[!htb]
    \centering
    \begin{subfigure}[b]{0.48\textwidth}
    \includegraphics[width=\textwidth]{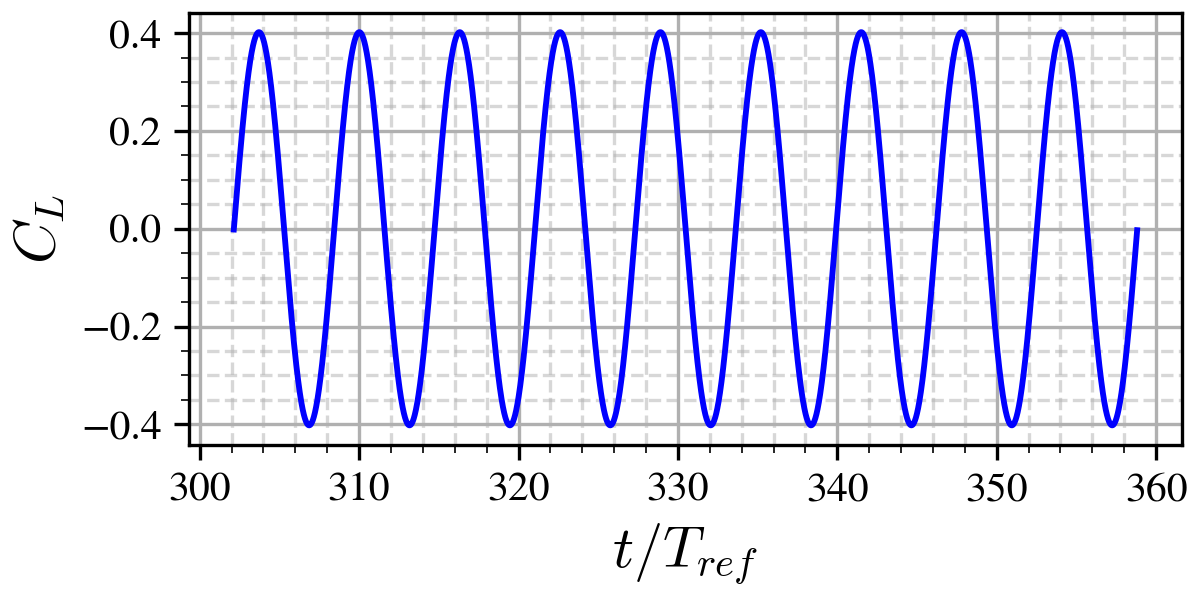}
    \caption{Lift coefficient $C_L$.}
    \label{fig:sqCyl2D_CL}
    \end{subfigure} %\\
    \begin{subfigure}[b]{0.48\textwidth}
    \includegraphics[width=\textwidth]{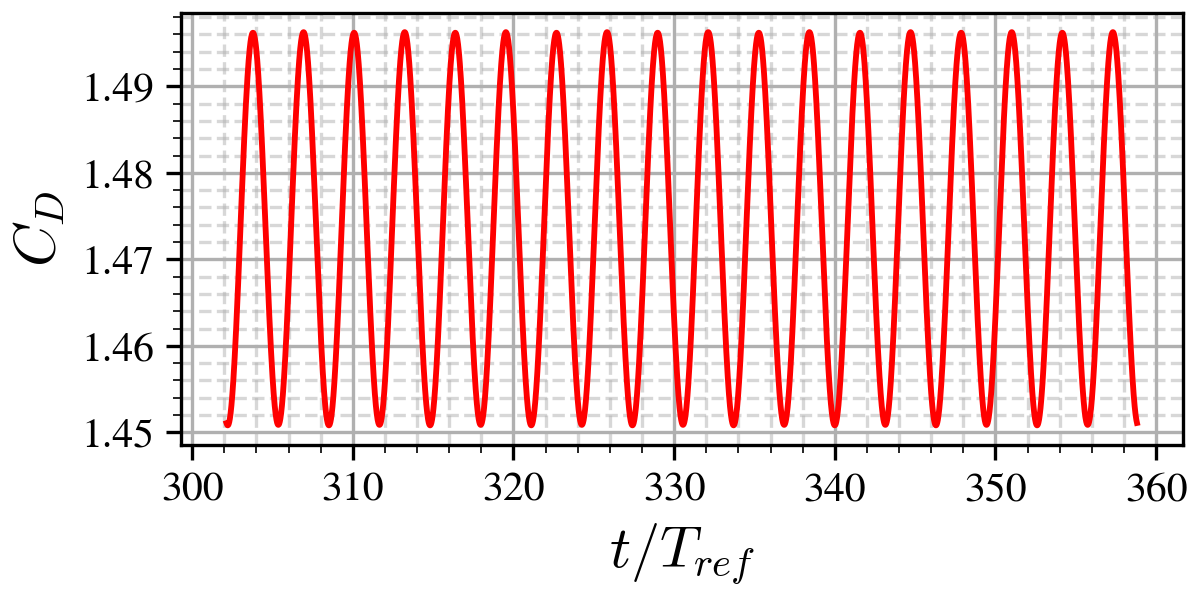}
    \caption{Drag coefficient $C_D$.}
        \label{fig:sqCyl2D_CD}
    \end{subfigure}
    \caption{Lift and drag coefficients $C_L$ and $C_D$ evolving in time.}
    \label{fig:sqCyl2D_forceCoeffs}
\end{figure}

A preliminary simulation is performed to validate the code and to obtain reference data to be used to assess the DA procedure. This preliminary simulation is run for $500 T_{ref}$, where the $T_{ref} = D/U$ is the characteristic advection time. The time step used for time advancement is constant and equal to $dt = 0.004 T_{ref}$. The contours of the instantaneous streamwise velocity component are shown in figure \ref{fig:sqCyl_u_vel_contours} where the vortex street forming downstream of the square cylinder can be seen. The periodic vortex shedding from the downstream corners of the square cylinder induces periodic oscillations in the drag and lift coefficients, which are qualitatively shown shown in figure \ref{fig:sqCyl2D_forceCoeffs} in the time range $t/T_{ref} \in [300-360]$, between two zero crossings of $C_L$. The statistics of the force coefficients are computed using data over $400T_{ref}$. This window begins after a simulation time of $100T_{ref}$, in order to exclude the effects of initial conditions. The mean value of the drag coefficient is $\overline{C_D}=1.473$ while, as expected, the lift coefficient $\overline{C_L} \approx 0$. The standard deviation of $C_L$ and $C_D$ are found to be $\sigma_{C_L} = 0.284$ and $\sigma_{C_D} = 0.016$ respectively. The Strouhal number determined from the lift coefficient is $St=0.159$. The statistics obtained in the present study are in agreement with previous results available in the literature \cite{Sohankar1998, Sohankar1999, Saha2003}.

The DA procedure used for this investigation is now described. The analysis will focus on the usage of state estimation for synchronization purposes, similar to the previous studies in the literature \cite{Villanueva2024_sync_channel}. This procedure, which is essential in emerging technologies such as digital twins \cite{Semeraro2021_ci}, aims to update the physical state predicted by the model so that it follows variations in time of the observation. The synchronization performance becomes important when the system state foreseen by the model deviates from the physical system due to various reasons such as external perturbations or the emergence of extreme events. The DA ingredients used are the following:
\begin{itemize}

\item \textbf{Observation} is sampled from the reference simulation in the form of instantaneous velocity measurements on a limited number of sensors. These sensors are positioned in the locations $(x,y)$ equal to $(2D, 0.5D)$, $(4D, 0.5D)$ and $(6D, 0.5D)$. Therefore, the observation array contains six values (two velocity components $(u_x,u_y)$ measured at three sensors). The sampling has been performed over a time window of $400 T_{ref}$ at each time step, after the flow structures arising from the initial condition are completely advected out of the domain. DA step is performed after windows of $4$ time-steps during the DA test, meaning that a smaller subset of the total observation data have been used for the DA procedure.

\item The \textbf{Model} is the same CFD solver used to perform the preliminary analysis and therefore to sample observation. The same grid and boundary conditions are used, therefore expecting virtually identical results. The $N_e=10$ simulations of the ensemble have been initialized by different flow fields from the preliminary simulation with $1 \, T_{ref}$ time difference between consecutive members. It is here reminded that, considering $St=0.159$, a full shedding cycle lasts around $6.2 \, T_{ref}$, and therefore the initial conditions selected are associated with different instantaneous flow dynamics.
\end{itemize}

The DA problem is set so that the model runs are identical to the preliminary simulation but, because of the periodic behavior of the von Karman street, the usage of different initial conditions for each member creates a discrepancy in the phase of the model prediction. The EnKF is therefore used to update the state of each model simulation using the observation, therefore eliminating the discrepancy in phase due to the initial conditions. 

Three DA runs are performed. First, a run using classical EnKF without localization is done. Second, a run is performed with a localized EnKF using the HL strategy \cite{Villanueva2024_PhDThesis}, with HL regions having a circular shape with the radius of $r_c=0.5$ around each observation probe. Third, a run using the PBOL for the dynamic sizing and orientation of the DA regions for the localization of the EnKF procedure is performed. These runs will be referred as \textit{DA-FD}, \textit{DA-HL} and \textit{DA-PBOL} respectively. 
 
The metric function including physical criteria which is used to drive the PBOL procedure is now detailed. For the present test case, the standard deviation of the enstrophy, $\enstr = |\bm{\omega}|^2$ among the ensemble members is used as the metric variable for the PBOL process, where $\bm{\omega}$ is the vorticity defined as the curl of the velocity, $\bm{\omega} = |\nabla \times \bm{u}|$. It is useful to remind that only one component of the vorticity is present in the current test case as the flow field is two-dimensional. The local fields of $\enstr$ are computed independently in the DA regions for each ensemble member. The PBOL procedure requires the calculation of the parameters $\bm{e}_{x^\prime}$, $\bm{e}_{y^\prime}$, $\xi_{x^\prime}$ and $\xi_{y^\prime}$ to optimize the shape and orientation of the ellipsoid representing the DA region. This procedure is here performed in two steps:
\begin{itemize}
\item \textbf{Determination of $\bm{e}_{x^\prime}$ and $\bm{e}_{y^\prime}$}. Starting from an initially imposed spherical region, the mean and the standard deviation values of $\enstr$ are obtained by the ensemble averaging of the computed fields along all the members, and are denoted as $\ensavg{\enstr}$ and $\sigma_{\enstr}$. The principal axis $\bm{e}_{x'}$ of the localization function, for a given DA region is determined by
\begin{equation}
    \bm{e}_{x'} = \bm{x}_{cm} - \bm{x}_o
\end{equation}
where $\bm{x}_{cm} = [x_{cm}, \, y_{cm}, \, z_{cm}]$ is the location of the the center of mass of the metric field and $\bm{x}_o = [x_{o}, \, y_{o}, \, z_{o}]$ is the location of the sensor sampling observation at the center of the DA region. With the metric function taken as $\phi = \sigma_{\enstr}$ for the present case, the center of mass $\bm{x}_{cm}$, for each DA region, is computed as
\begin{equation}
    \bm{x}_{cm} = \frac{\sum^{N_c}_{i=1} \bm{x}_{c,i} \phi_i V_{c,i}}{\sum^{N_c}_{i=1} \phi_i V_{c,i}}
\end{equation}
Here the summation operations are carried out for the cells constituting the DA region considered. $N_c$ is the total number of cells constituting a given DA region, $\bm{x_{c,i}}$ are the coordinates of the $i^{th}$ grid element, $\phi_i$ is the value of the metric function at the $i^{th}$ cell and $V_{c,i}$ is its volume. The determination of $\bm{e}_{x'}$ permits to identify the orientation of the ellipsoid region, considering that the test case investigated in two-dimensional. The angle $\theta$ is computed between $\bm{e}_{x'}$ and the streamwise unit vector $\bm{i}$ and $\bm{e}_{y'}$ is determined as the orthogonal vector to $\bm{e}_{x'}$.
\item \textbf{Determination of $\xi_{x^\prime}$ and $\xi_{y^\prime}$}. The size of the ellipsoid is obtained studying the behavior of the function
\begin{equation}
    \rho_{\phi}(\xi_{x^\prime},\xi_{y^\prime}) = \frac{\sum^{N_c}_{i=1} L(\xi_{x^\prime},\xi_{y^\prime}) \phi_i V_{c,i}}{\sum^{N_c}_{i=1} \phi_i V_{c,i}}.
\end{equation}
which takes into account the distribution of the metric field over the DA region. Here, $L$ is the covariance localization function in the form of multivariate normal distribution in two-dimensions. The selected values for $\xi_{x^\prime}$ and $\xi_{y^\prime}$ maximize the value of $\rho_{\phi}$, therefore selecting the shape which leads to the most efficient assimilation region. The optimization is performed via a grid search, where $\rho_{\phi}(\xi_{x^\prime},\xi_{y^\prime})$ is computed using five values for $\xi_{x^\prime}$ and five values for $\xi_{y^\prime}$, in the range leading to $[0.2 r_c, \, 1.0r_c]$ in terms of extent of DA region. The set of $(\xi_{x^\prime},\xi_{y^\prime})$ maximizing the value of $\rho_{\phi}(\xi_{x^\prime},\xi_{y^\prime})$ is chosen for a given DA region at an instant. The maximization of $\rho_{\phi}$ by choosing the set of $(\xi_{x^\prime},\xi_{y^\prime})$ ensures that the DA procedure is carried out in the regions with highest variance between the ensemble members, where the contributions by the state estimation DA are significant.   
\end{itemize}

The determination of the parameters $\theta$, $\xi_{x^\prime}$ and $\xi_{y^\prime}$ defines the size and shape of the ellipsoid region used for DA purposes. These parameters also provide information for function used for covariance localization, in order to damp the state estimation term close to the surface of the ellipsoid. This function is expressed as a two-dimensional normal distribution:
\begin{equation}
    L(\bm{x}_c) = \exp(-c_1 (x_c-x_o)^2 + 2 c_2 (x_c-x_o) (y_c-y_o) + c_3 (y_c-y_o)^2).
\end{equation}
where $\bm{x}_c = (x_c, \, y_c)$ is the location of the evaluation of the localization function, and $c_1$, $c_2$, $c_3$ are computed as
\begin{align}
    c_1 &= \cos(\theta)^2/(2 \xi_{x^\prime}^2) + \sin(\theta)^2/(2 \xi_{y^\prime}^2), \\
    c_2 &= -\sin(\theta)\cos(\theta)/(2 \xi_{x^\prime}^2) + \sin(\theta) \cos(\theta)/(2 \xi_{y^\prime}^2), \\
    c_3 &= \sin(\theta)^2/(2 \xi_{x^\prime}^2) + \cos(\theta)^2/(2 \xi_{y^\prime}^2).
\end{align}

\begin{figure}[!htb]
    \centering
    \begin{subfigure}[b]{0.78\textwidth}
    \includegraphics[width=\textwidth]{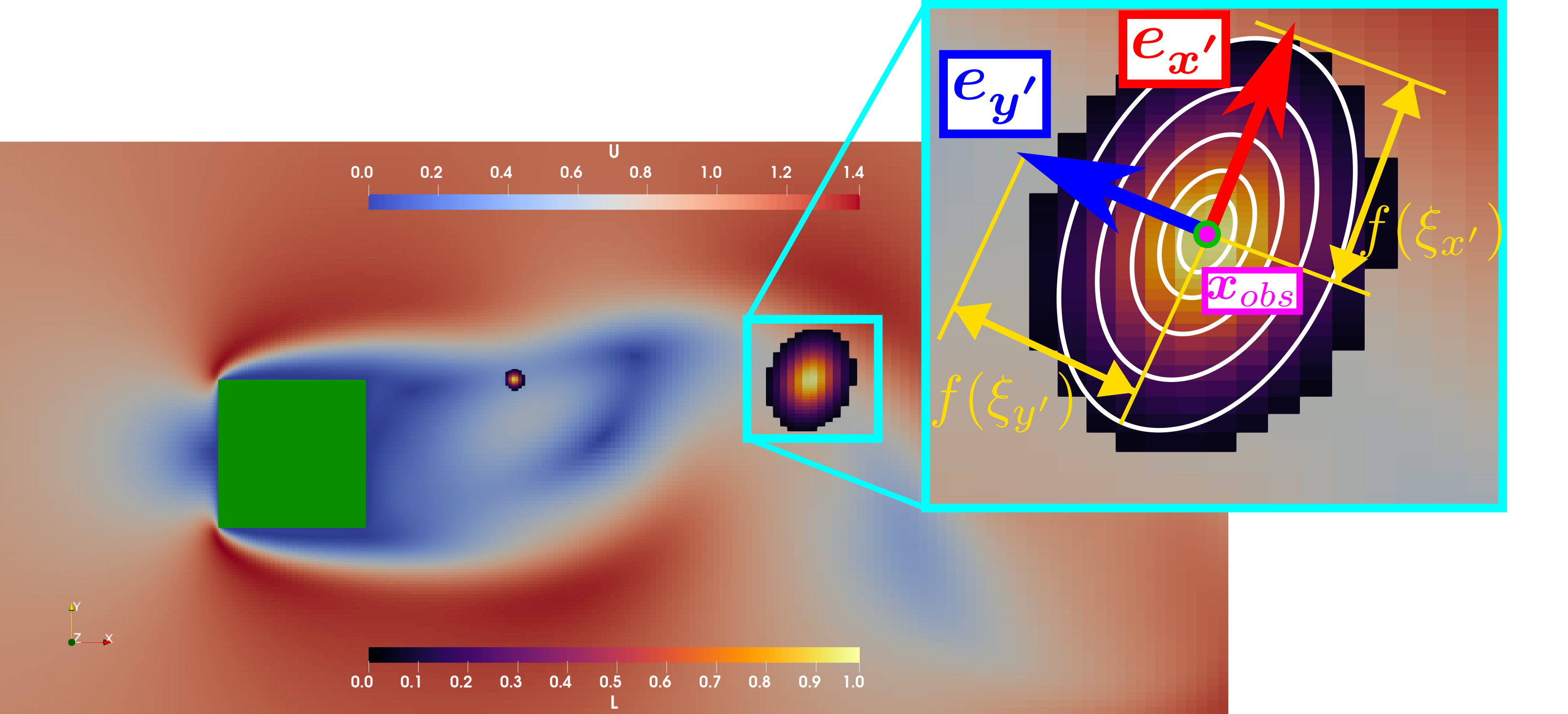}
    %\caption{Example of ellipsoid DA region with covariance localization.}
    \caption{}
    \label{fig:sqCyl2D_onlineLoc_localization_fig}
    \end{subfigure}
    \begin{subfigure}[b]{0.94\textwidth}
    \includegraphics[width=\textwidth]{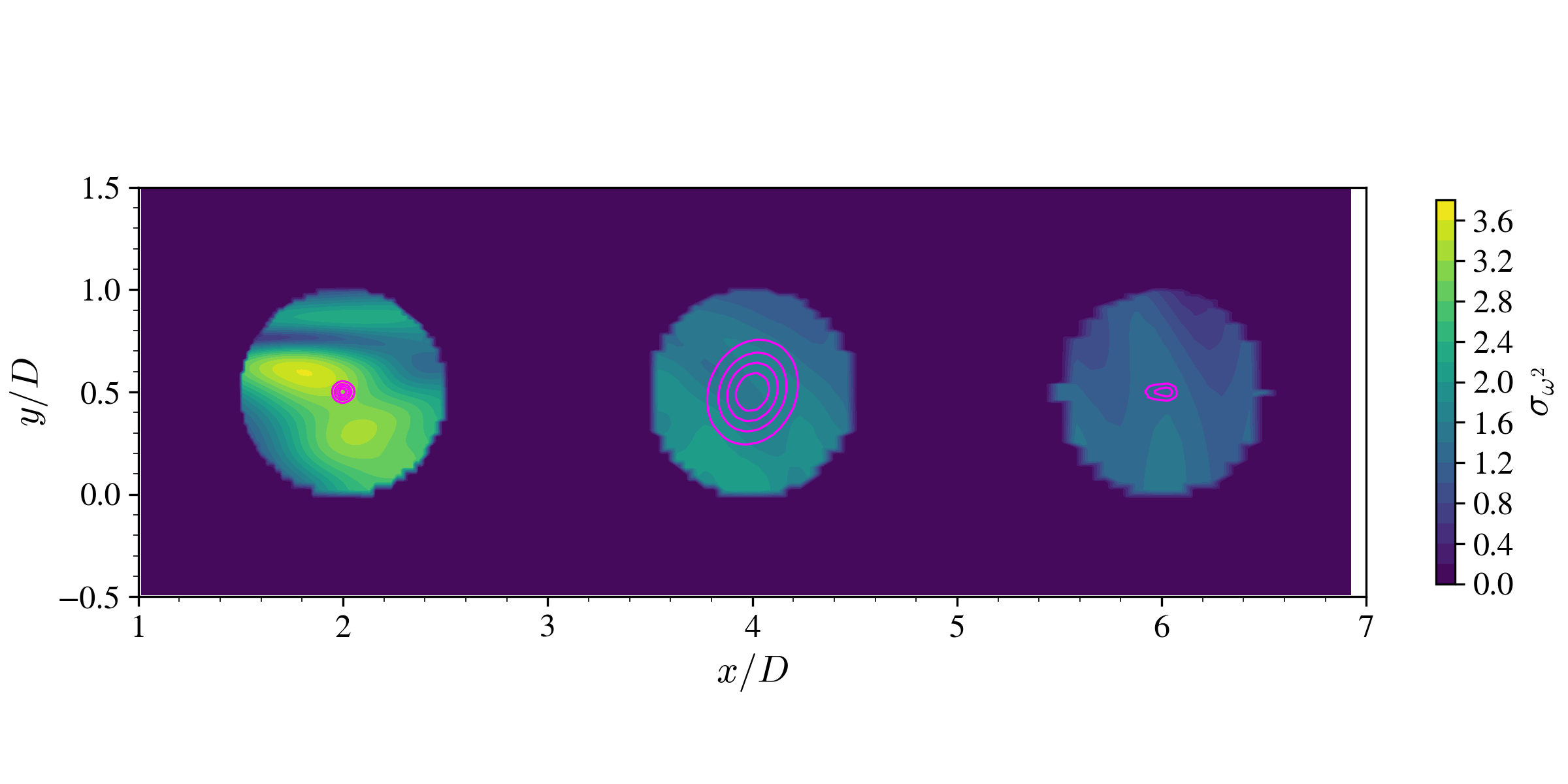}
    %\caption{Iso-contours of the covariance localization functions plotted over the initial circular region. The value of the metric function $\phi = \sigma_{\enstr}$ are also shown as the contour fields inside circular DA regions.}
    \caption{}
    \label{fig:sqCyl2D_onlineLoc_metric_field}
    \end{subfigure}
    \caption{(a) An example of the localization functions in DA regions are shown on top of the contours of the underlying velocity field with the figure inset zooming in one of the regions and (b) the contour field of the local metric fields and the iso-contours of the localization values (in magenta) are plotted near three observation sensors downstream of the square cylinder.}
    \label{fig:sqCyl2D_loc_fig_and_metric_fields}
\end{figure}

The results of application of the PBOL methodology, which is performed for the three DA regions around the selected sensors, is presented in figure \ref{fig:sqCyl2D_loc_fig_and_metric_fields} for an analysis phase. Figure \ref{fig:sqCyl2D_onlineLoc_localization_fig} shows the ellipsoid region with the values of the covariance localization function employed. One can see that the state update is damped to zero at the borders of the ellipsoid. Figure \ref{fig:sqCyl2D_onlineLoc_metric_field} shows the iso-contours of the covariance localization function (in magenta) superposed on the contour field of the standard deviation of $\enstr$, $\sigma_{\enstr}$, which is chosen as the metric function in the current test case.

%=========================================================
% [[[ Square Cyl2D - full dom./HL/PBOL ]]]
%=========================================================
\begin{figure}[!htb]
    \centering
    \begin{subfigure}[b]{0.48\textwidth}
    \includegraphics[width=\textwidth,trim={0cm 0cm 0cm 0cm},clip]{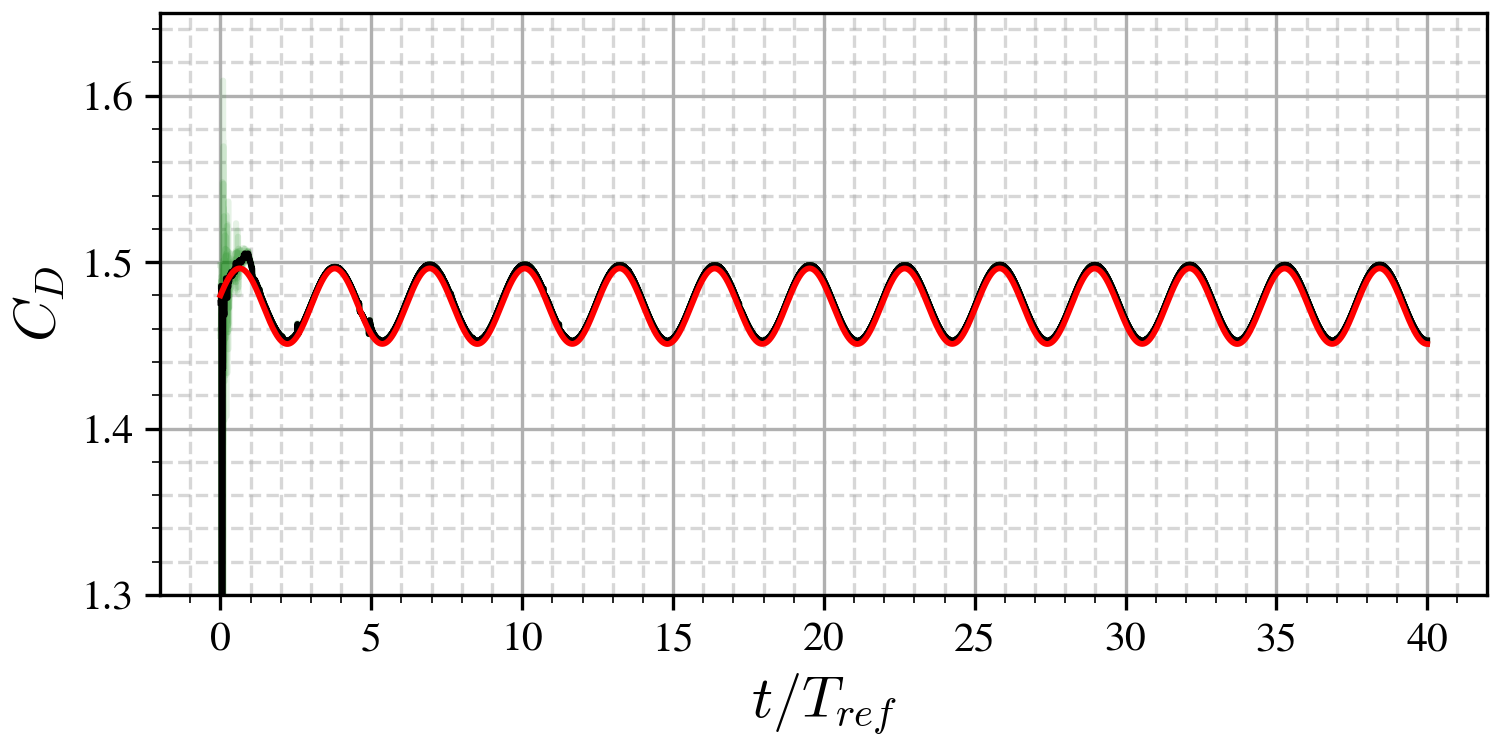}
    \caption{DA-FD, drag coefficient, $C_D$.}
    \end{subfigure} %\\
    \begin{subfigure}[b]{0.48\textwidth}
    \includegraphics[width=\textwidth,trim={0cm 0cm 0cm 0cm},clip]{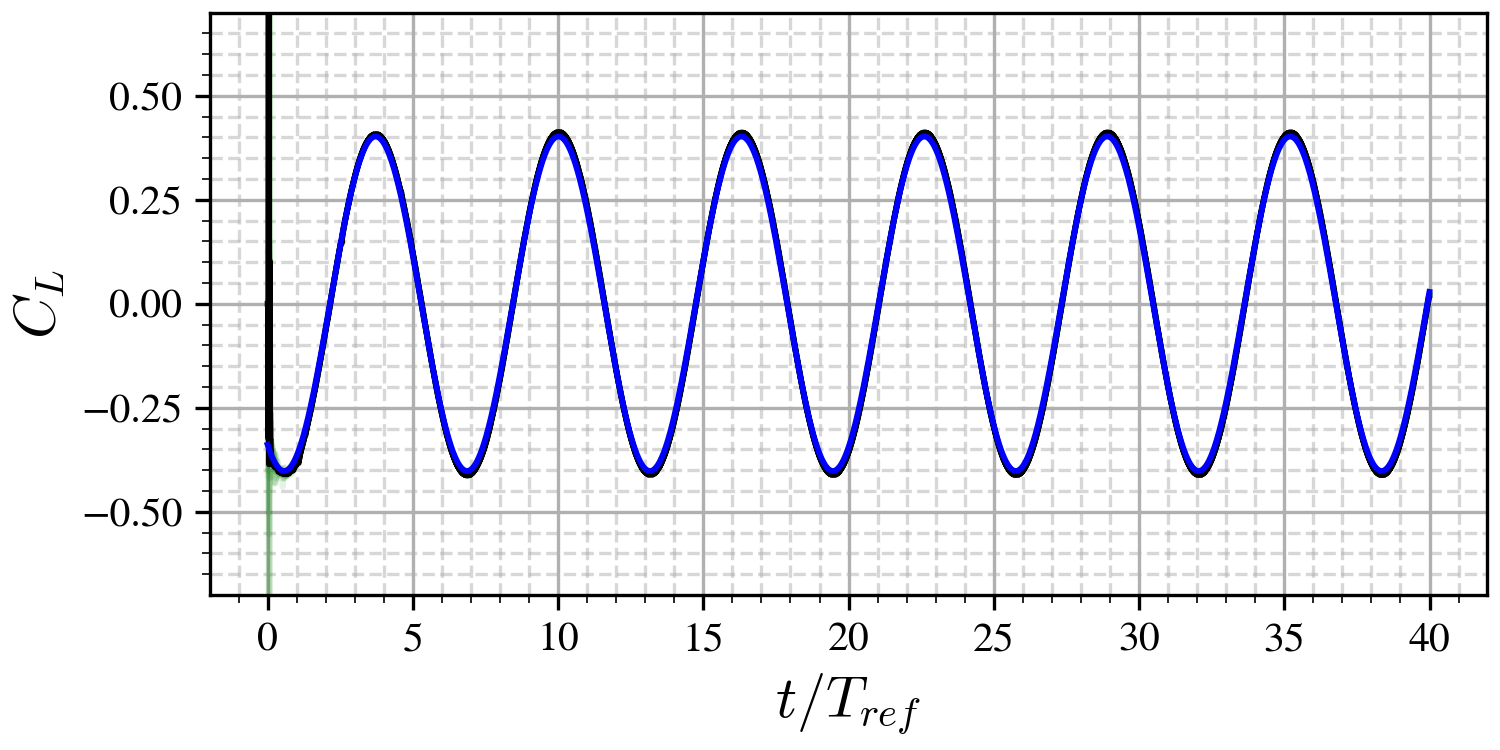}
    \caption{DA-FD, lift coefficient, $C_L$.}
    \end{subfigure}
    \centering
    \begin{subfigure}[b]{0.48\textwidth}
    \includegraphics[width=\textwidth,trim={0cm 0cm 0cm 0cm},clip]{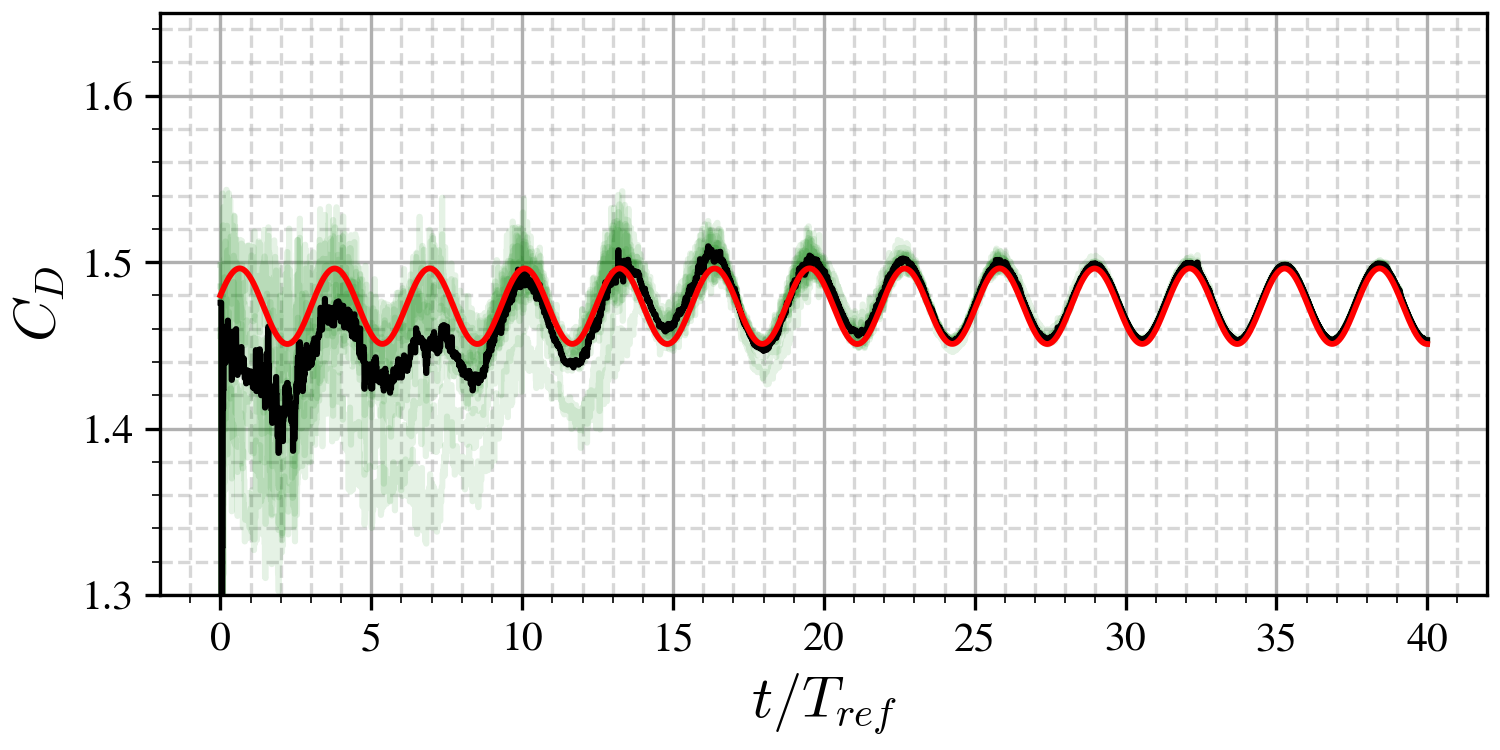}
    \caption{DA-HL, drag coefficient, $C_D$.}
    \end{subfigure} %\\
    \begin{subfigure}[b]{0.48\textwidth}
    \includegraphics[width=\textwidth,trim={0cm 0cm 0cm 0cm},clip]{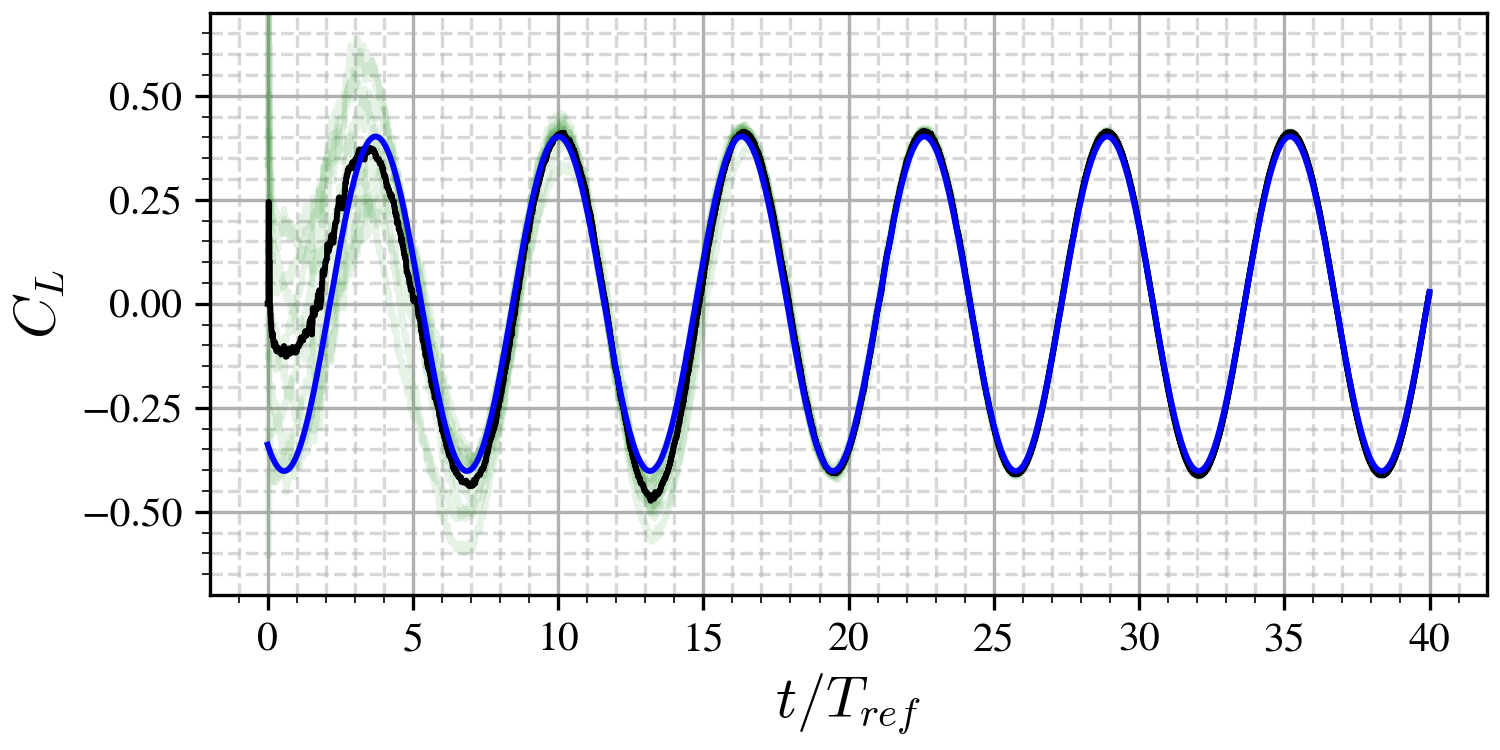}
    \caption{DA-HL, lift coefficient, $C_L$.}
    \end{subfigure}
    \centering
    \begin{subfigure}[b]{0.48\textwidth}
    \includegraphics[width=\textwidth,trim={0cm 0cm 0cm 0cm},clip]{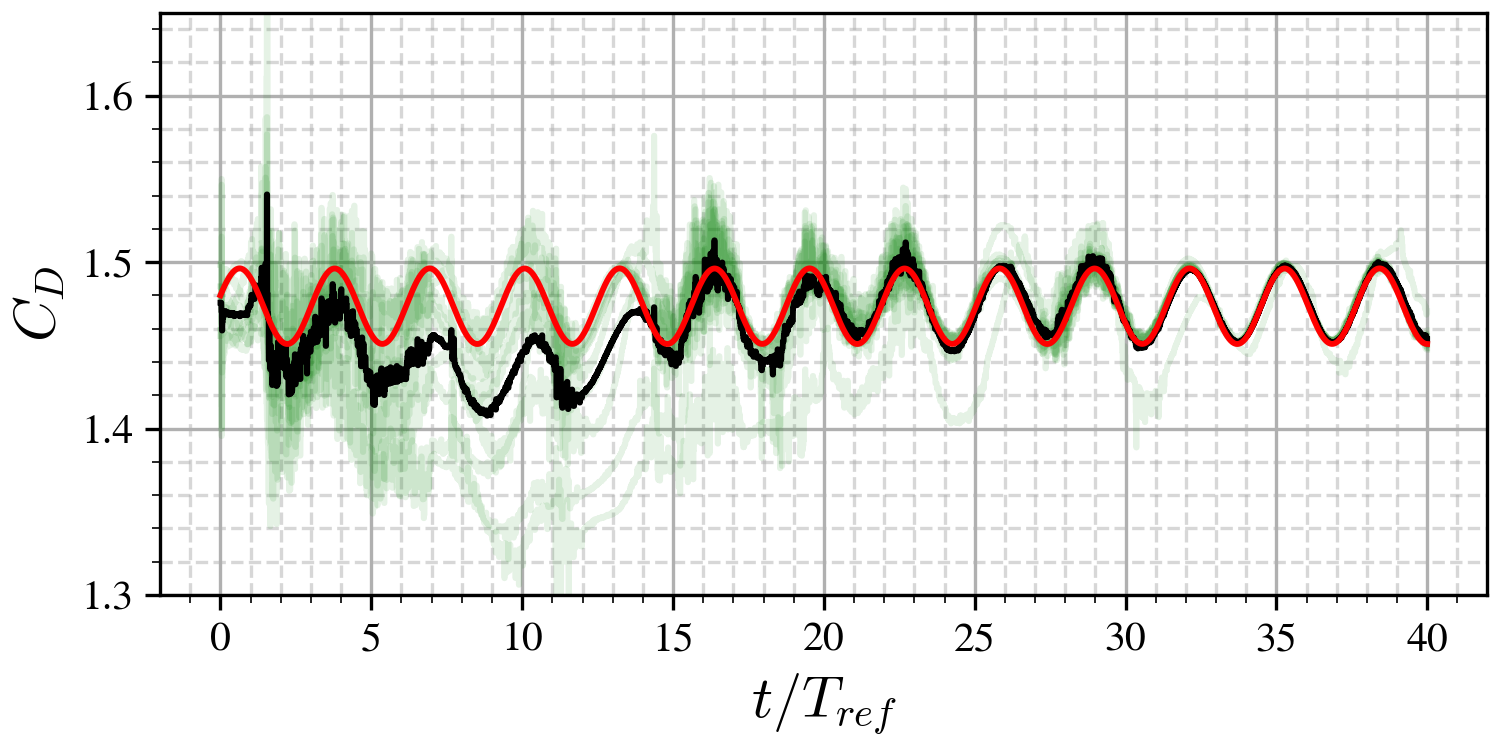}
    \caption{DA-PBOL, drag coefficient, $C_D$.}
    \end{subfigure} %\\
    \begin{subfigure}[b]{0.48\textwidth}
    \includegraphics[width=\textwidth,trim={0cm 0cm 0cm 0cm},clip]{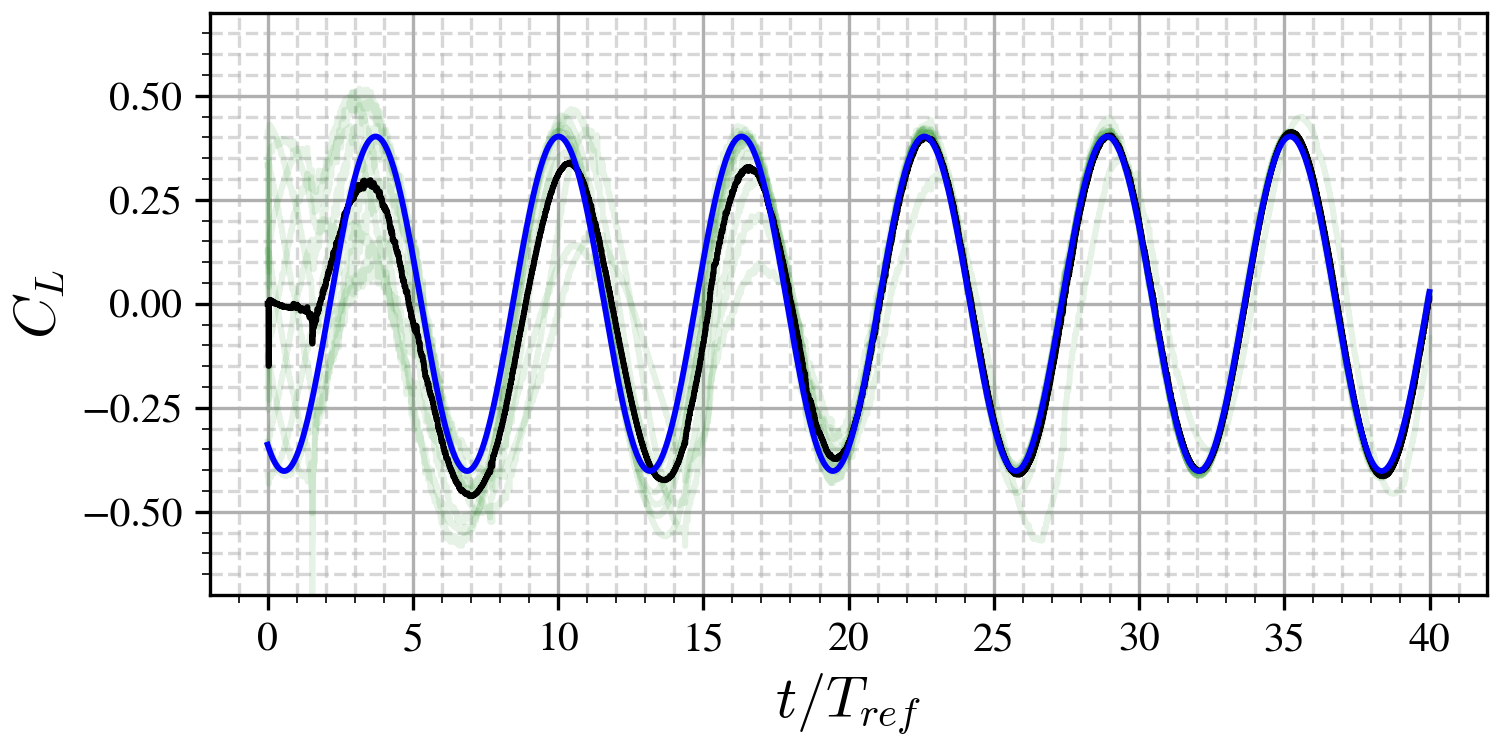}
    \caption{DA-PBOL, lift coefficient, $C_L$.}
    \end{subfigure}
    \caption{DA-based synchronization of the force coefficients $C_D$ and $C_L$ for the runs DA-FD, DA-HL and DA-PBOL.}
    \label{fig:sqCyl2D_CL-CD_compare_full_dom-HL-PBOL}
\end{figure}

\subsubsection{Results}

\begin{figure}[!htb]
    \centering
    \includegraphics[width=0.7\textwidth]{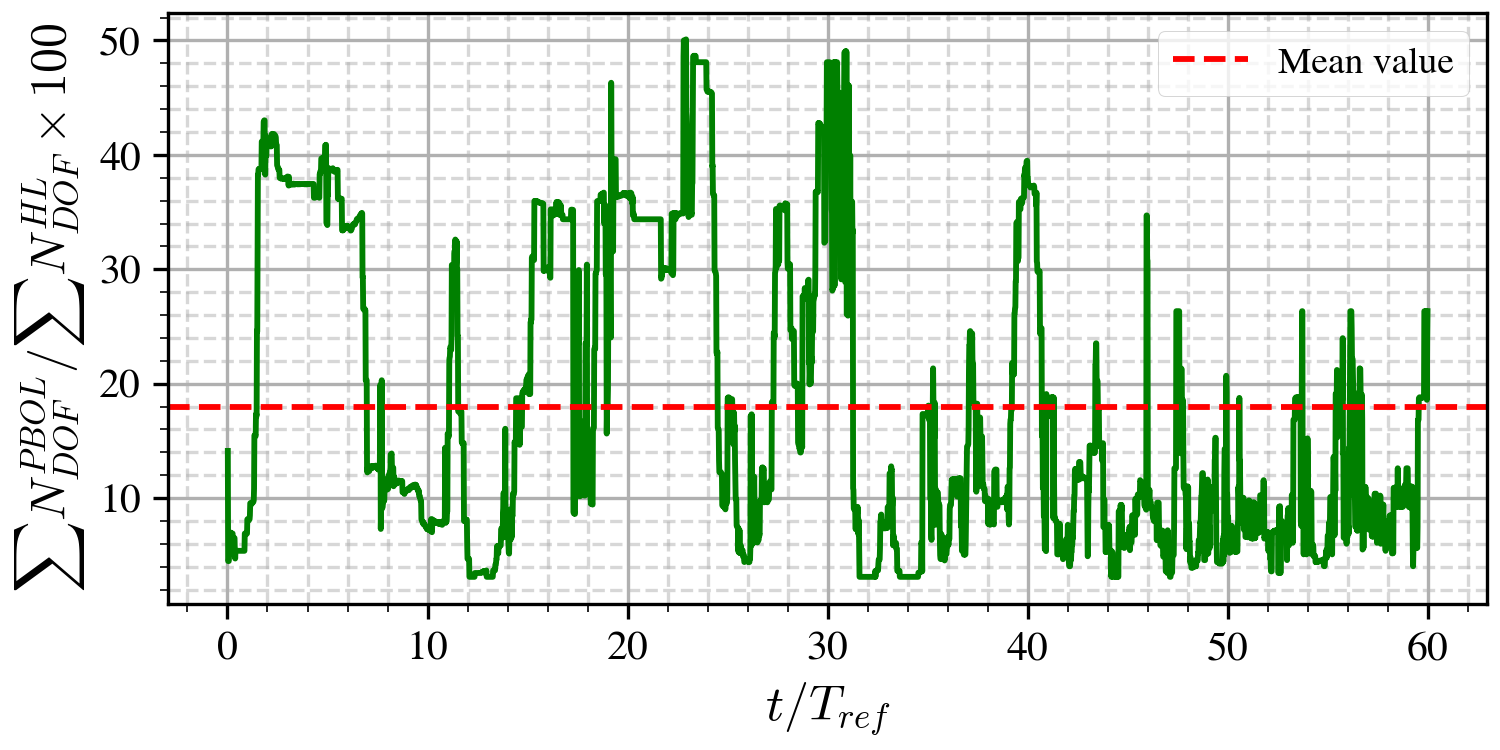}
    \caption{Total number of cells being assimilated in each DA stage with PBOL, compared to the HL procedure.}
    \label{fig:sqCyl2D_nCells_reduction_HL_PBOL}
\end{figure}

Results obtained from the three DA runs are now compared. The synchronization of the force coefficients is presented in the figure \ref{fig:sqCyl2D_CL-CD_compare_full_dom-HL-PBOL} for runs using DA with full domain state estimation, HL, and PBOL. Generally speaking, the three DA runs are able to obtain satisfying synchronization of the flow while using a very limited amount of observation (two velocity components from three sensors). This result is partially due to the laminar nature of the flow and the selected positioning of the sensors, as observation contains relevant information about the physical state. Because of these favorable features of the test case, the EnKF state update is successfully correcting the flow field in every grid element considered in the DA procedure. As a consequence, one can see that the synchronization is nearly instantaneous when no localization is used and the DA analysis is carried out in the complete domain (see figures \ref{fig:sqCyl2D_CL-CD_compare_full_dom-HL-PBOL}(a) and \ref{fig:sqCyl2D_CL-CD_compare_full_dom-HL-PBOL}(b)). On the other hand, the simulation time required to obtain satisfying synchronization increases when localization is applied and therefore the state update is applied to limited regions (see figures \ref{fig:sqCyl2D_CL-CD_compare_full_dom-HL-PBOL}(c,d) and \ref{fig:sqCyl2D_CL-CD_compare_full_dom-HL-PBOL}(e,f)). More precisely, synchronization is obtained after $2.5$ shedding cycles when using the HL localization and after $3$ shedding cycles for the PBOL method. However, even for such a simple test case, the usage of localization reduces the costs associated with the DA steps. The DA run using the HL technique requires in the order of $10$ times less computational effort than the EnKF without localization \citep{Villanueva2024_PhDThesis}.
A comparison between the usage of the HL and PBOL techniques in shown if figure \ref{fig:sqCyl2D_nCells_reduction_HL_PBOL}. Here, the ratio of the total number of cells receiving a state update at each DA stage (i.e. sum of the grid elements affected by DA update for all the DA regions) between the PBOL and the HL methods is shown. Results are shown as a percentage of grid elements used by the PBOL over time versus the total number of assimilated cells in the HL case. Time evolution is observed because the PBOL method is recalculated for each analysis phase. A significant reduction in the number of grid elements receiving a DA state update is obtained for the PBOL procedure when compared with the HL localization, with very similar results in terms of accuracy. Over the time window $t/T_{ref} \in [0,60]$, the number of grid elements used by the PBOL method are, on average, $18\%$ of the cells in the DA regions for the HL case. This highlights how the usage of a physical-based criterion can be helpful in identifying critical regions to perform efficient Data Assimilation.
Figure \ref{fig:sqCyl2D_nCells_reduction_HL_PBOL} also shows that the variable $\sum N^{PBOL}_{DOF}$ measuring the number of effective DA grid elements is larger for the first analysis phases, while it sensibly reduced for $t/T_{ref} > 30$. This result can be explained considering the initial higher variability of the flow (i.e. higher $\sigma_{\enstr}$ values). The selection of this physical constraint permits to account for uncertainty in the physics represented by the model, therefore obtaining larger DA regions to account for several physical features captured by the elements. As the simulation start to synchronize and the flow features become similar, the PBOL procedure is able to identify more clearly the regions of interest and therefore reduce accordingly the number of grid elements used in the DA process.

The optimization in time of the parameters characterizing the PBOL procedure is now investigated. To this purpose, the two variables $A = \sqrt{\xi_{x'} \xi_{y'}}$ and $\varepsilon = \sqrt{1-\frac{(\xi_{x'}, \xi_{y'})_{min}^2}{(\xi_{x'}, \xi_{y'})_{max}^2}}$ are considered. $A$ is an indicator of the total volume of the DA region determined by the PBOL, while $\epsilon$ describes the eccentricity of the DA region measuring the difference in size of the larger and smaller axes of the ellipsoid. $\epsilon \approx 1$ when the ellipsoid is more eccentric and $\epsilon \approx 0$ when the shape is quasi-circular.

\begin{figure}[!htb]
    \centering
    \begin{subfigure}[b]{0.8\textwidth}
    \includegraphics[width=\textwidth,trim={0cm 0cm 0cm 0cm},clip]{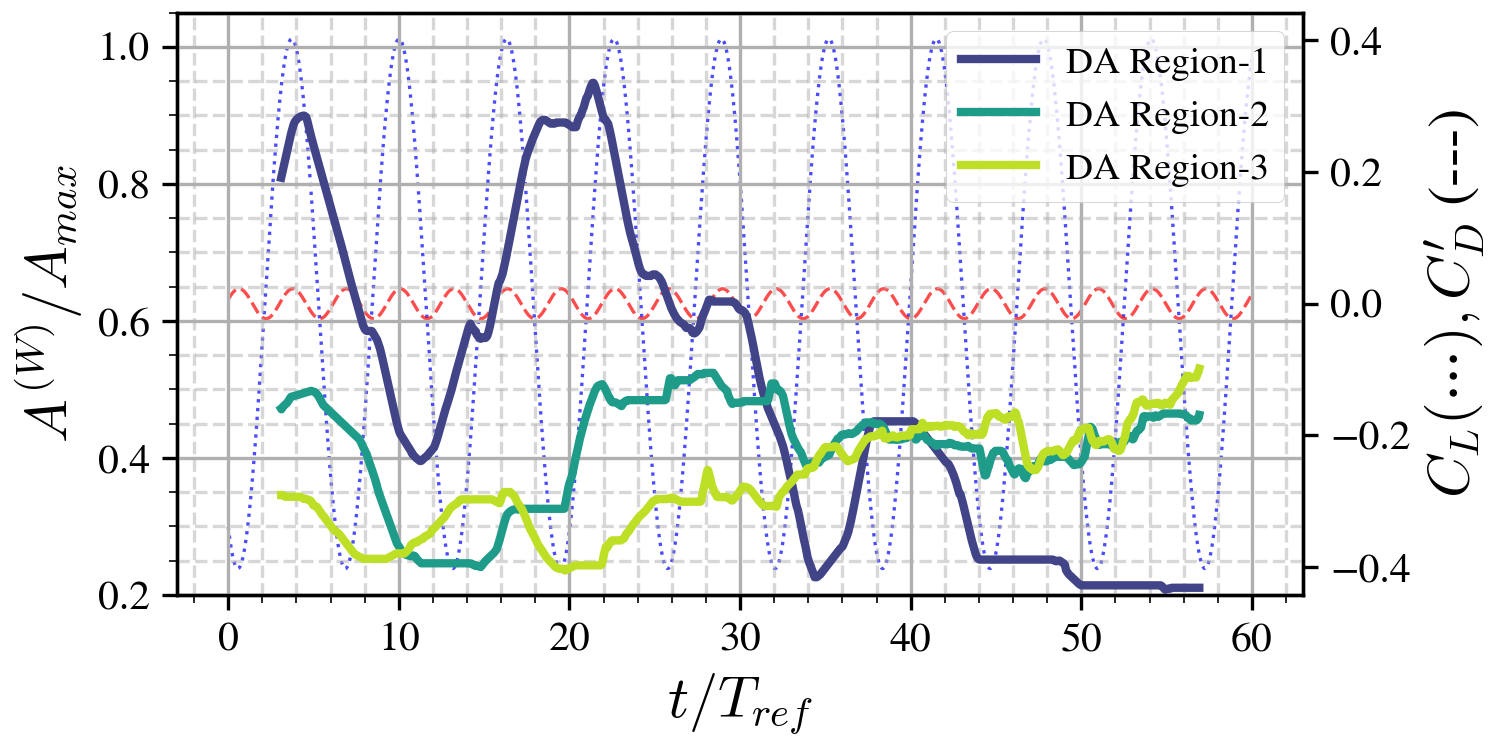}
    \caption{Size of the DA regions.}
    \label{fig:sqCyl2D_CL-PBOL-DA-volume}
    \end{subfigure} \\
    \begin{subfigure}[b]{0.8\textwidth}
    \includegraphics[width=\textwidth,trim={0cm 0cm 0cm 0cm},clip]{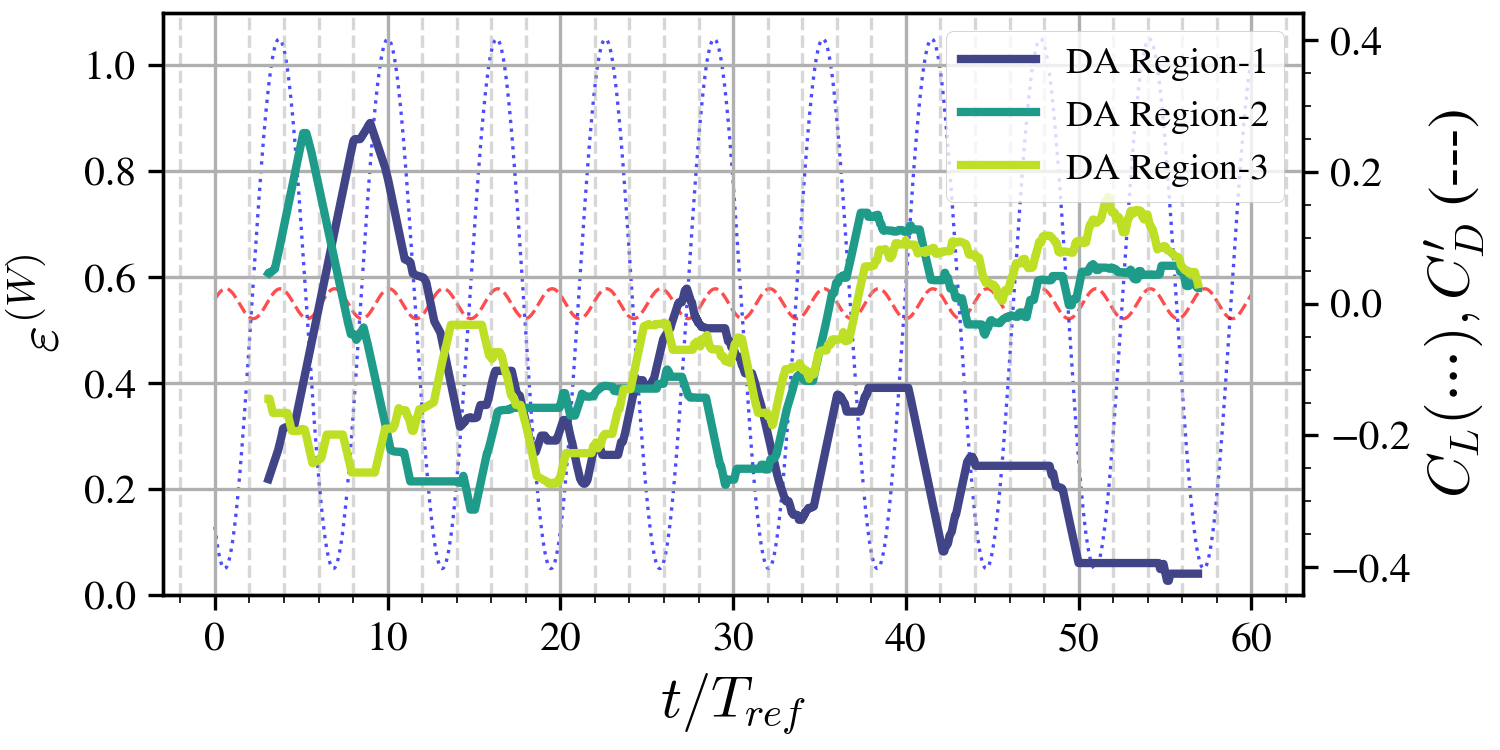}
    \caption{Correlation between $\xi_{x'}$ and $\xi_{y'}$ parameters.}
    \label{fig:sqCyl2D_CL-PBOL-ecc_xi}
    \end{subfigure}

    \caption{Time evolution of (a) $A^{(W)}/A_{max}$, which represents the size of the DA regions and (b) $\varepsilon^{(W)}$, which described the eccentricity of the ellipsoid. Results are shown over the time window of investigation for each DA region used by the PBOL method.}
    \label{fig:sqCyl2D-PBOL_params_variation}
\end{figure}

Figure \ref{fig:sqCyl2D-PBOL_params_variation} shows the variation of the normalized variable $A^{(W)}/A_{max}$ and $\varepsilon^{(W)}$ with time for each of the three DA region. $A_{max}$ is here the size of the DA region at the beginning of the optimization (i.e. sphere). Results in dark blue are for the closest DA region to the solid body while results in light green are for the furthest DA region downstream. The superscript $(W)$ indicates a moving-average operation applied to the variables as their value exhibits significant fluctuations. For a general time-dependent variable  $\phi$ the value of the average $\phi^{(W)}$ at each instant is calculated as $\phi^{(W)}(t) = \frac{1}{\Delta t_W} \int^{t+\frac{\Delta t_W}{2}}_{t-\frac{\Delta t_W}{2}}\phi(t) dt$. Here $\Delta t_W$ is the window size of the moving average and is taken as $\Delta t_W = 6.2 T_{ref}$, corresponding to the period of a full shedding cycle.

It can be observed from figure \ref{fig:sqCyl2D_CL-PBOL-DA-volume} that the size of the DA region is larger as the observation gets closer to the square cylinder at initial times. The DA regions 1 and 2 gets smaller until $t=11 T_{ref}$ with respect to their initial sizes, as more analysis phases are carried out. The DA region 1, closest to the cylinder, shows the highest variation along the time, even though it gets larger again after $t=12 T_{ref}$, the DA region shrinks significantly at $t=34 T_{ref}$, once the synchronization is mostly achieved even for $C_D$ (see figure \ref{fig:sqCyl2D_CL-CD_compare_full_dom-HL-PBOL} (e)). The other two DA regions remain limited in size compared to the first region. Due to the locations of the observation probes, assimilation of the system state over a large area in the first DA region, contributes to the synchronization in DA regions 2 and 3 as the information is advected downstream by the mean streamwise velocity.
Figure \ref{fig:sqCyl2D_CL-PBOL-ecc_xi} displays peaks for the $\varepsilon^{(W)}$ for the DA regions 1 and 2 at $t = 9 T_{ref}$ and $t = 5 T_{ref}$. Meaning that the shape of the localization functions are highly eccentric due to the relevant regions determined by PBOL lies on one axis. For DA region 1, the peak aligns with the lowest DA region area seen in figure \ref{fig:sqCyl2D_CL-PBOL-DA-volume}, meaning that the highly flat ellipse DA region is not very large in area due to its limited size in the secondary axis. On the other hand, DA region 2 has more circular shape due to the lower values of $\varepsilon^{(W)}$ when the area of the DA region is small between $t = 11 T_{ref}$ and $t=16 T_{ref}$.
After the synchronization is achieved, the DA region 1 becomes more circular while its area gets as small as possible. Meanwhile, the other two DA regions are more eccentric and their areas remains at nearly $40\%$ of $A_{max}$. It is observed that the variables oscillate significantly for the DA regions 2 and 3 after the synchronization as the $\sigma_{\enstr}$ values gets very low around these two observation points further downstream of the square cylinder.

\subsection{Turbulent flow around a 3D circular cylinder} \label{sec:sync_Cyl3D}

The developed DA methodology is now applied to a three-dimensional, turbulent test case. This increase in complexity for the physical features investigated, in particular for the multi-scale dynamic interactions, represents a test to assess the robustness and the accuracy of the methodology in more complex applications. The test case investigated is the flow around a three-dimensional circular cylinder at $Re=3900$. Similarly to flow around the square cylinder, the Reynolds number is defined as $Re = \frac{U D}{\nu}$ where the $U$ is the uniform inlet velocity, $D$ is the diameter of the cylinder and $\nu$ is the kinematic viscosity of the fluid. At this Reynolds number, a fully turbulent wake is observed downstream the cylinder, while the boundary layer on the body is laminar \cite{Breuer2000}.

The numerical simulation is performed over a parallelepiped domain, where $x$ is the streamwise direction, $y$ is the cross-stream direction and $z$ is the spanwise direction. The inlet and outlet boundaries are located $10D$ upstream and $20D$ downstream from the center of the cylinder respectively. The domain extends between $y/D=\pm10$ in the cross-stream direction and the extent in the spanwise direction is $\pi$. IBM is used to account for the immersed cylinder. The grid resolution is uniform in all three directions in region where the cylinder is the same and the size of the elements is $\Delta_x = \Delta_y = \Delta_z = 2.5 \cdot 10^{-2} D$. The resolution becomes progressively coarser moving away from the wall. More precisely, moving downstream in the wake region $\Delta _{x,y,z} = 5 \times 10^{-2} D$ for $1 < x/D < 2.5$, $\Delta_{x,y,z} = 10^{-1} D$ for $2.5 < x/D < 16$ and $\Delta_{x,y,z} = 2 \times 10^{-1} D$ in the proximity of the outflow boundary for $16 < x/D < 20$. The total number of cells in the computational grid is equal to $1.6 \times 10^6$. The grid refinement selected for this case permits to perform coarse DNS with reasonable computational resources. This point is important because the DA procedure requires an ensemble run which must be performed over a significantly long time window.  

A preliminary simulation is performed to obtain a reference solution (in terms of statistics of the flow field) and to sample data for observation. The simulation is run for $380 T_{ref}$ after the initialization, in order to dissipate the effects of the initial conditions and to grant that the flow statistics are statistically stationary. Following that , the acquisition of the data and the computation of the statistics are carried out for $400T_{ref}$. The reference time is again the characteristic advection time defined as $T_{ref} = D/U$. A time step of $dt = 4 \times 10^{-3} T_{ref}$ has been used and the Courant-Friedrichs-Lewy stability condition $max(CFL) \leq 0.42$ is verified throughout all the time range of data acquisition. The analysis of the flow statistics shows that the Strouhal number obtained from the time signal of the lift coefficient $C_L$ is $St=0.196$ and the size of the recirculation bubble is $L_r/D=1.2$ are in agreement with values reported in the literature \cite{Ma2000,Franke2002}. These quantities are associated with the main dynamics of the flow, which are captured by the grid used for calculation. Minor discrepancies of $\approx 10\%$ on average are observed for $C_L$ and $C_D$ due to lack of refinement. This point is less critical in this analysis because the numerical tool used for observation and model runs is identical and therefore the model error is identically zero.  

\subsubsection{Set-up for the DA experiment} \label{sec:config_sensors_cyl3D}

A dataset to be used as observation is generated by sampling the reference simulation. Instantaneous velocity samples are acquired for $17.7k$ probes. Among the $17.7k$ sensors, $15.7k$ are uniformly distributed in the region $x/D=[0.6, 3]$, $y/D=[-0.5, 0.5]$ and $z/D=[0.2, 3]$ with spacing equal to $\Delta_{p, x} = 0.1D$, $\Delta_{p, y} = 0.1D$ and $\Delta_{p, z} = 0.2D$. The velocity field is also sampled at $2000$ additional sensors, which are randomly distributed in the same region. The total dataset is large in terms of number of probes, and they have been distributed so that different subsets are used in the DA analyses. This point also permits to perform a sensitivity analysis to the performance of the EnKF (accuracy, cost, rate of convergence) to variations of the observation set. The velocity data at the sensors has been collected over the time range of $t/T_{ref} \in[0,\, 400]$ used to calculate statistics, at each time step. This corresponds to a time series of $100k$ samples for each of the three velocity components at each sensor.

\begin{figure}[!htb]
    \centering
    \begin{subfigure}[b]{0.48\textwidth}
    \includegraphics[width=\textwidth,trim={0cm 2cm 0cm 2cm},clip]{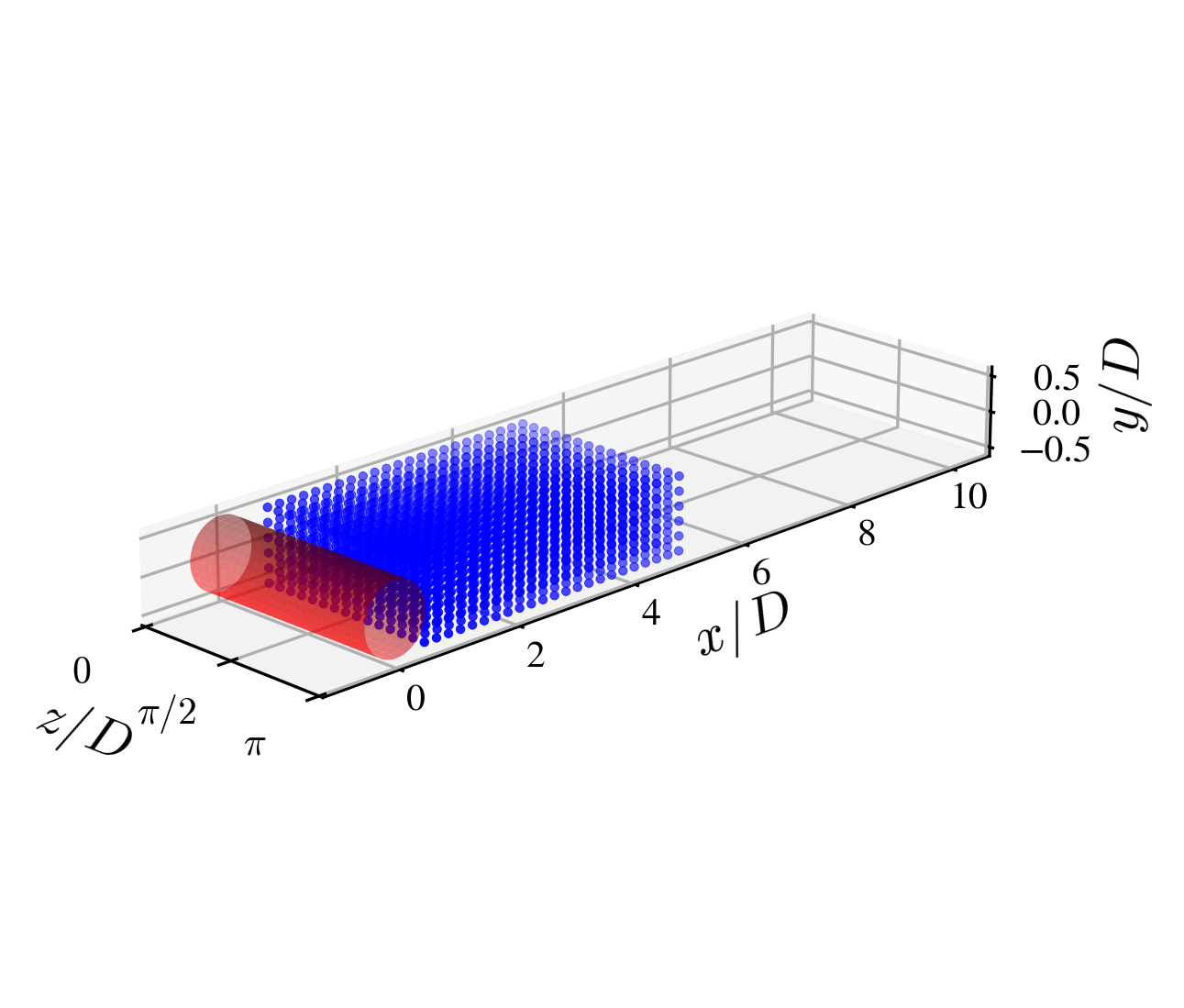}
    \caption{Vertical}
    \end{subfigure} %\\
    \begin{subfigure}[b]{0.48\textwidth}
    \includegraphics[width=\textwidth,trim={0cm 2cm 0cm 2cm},clip]{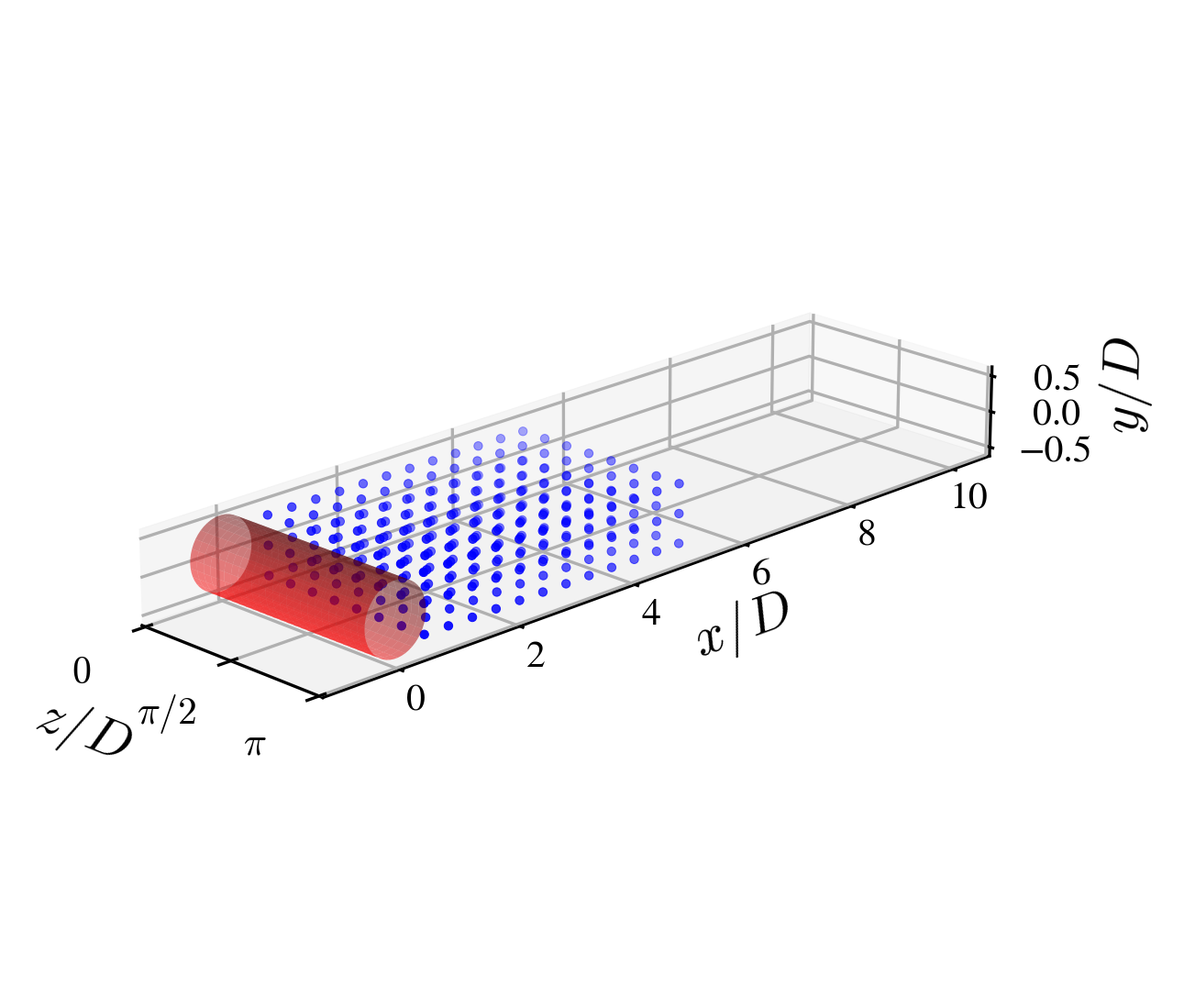}
    \caption{Vertical-sparse}
    \end{subfigure}
    \centering
    \begin{subfigure}[b]{0.48\textwidth}
    \includegraphics[width=\textwidth,trim={0cm 2cm 0cm 2cm},clip]{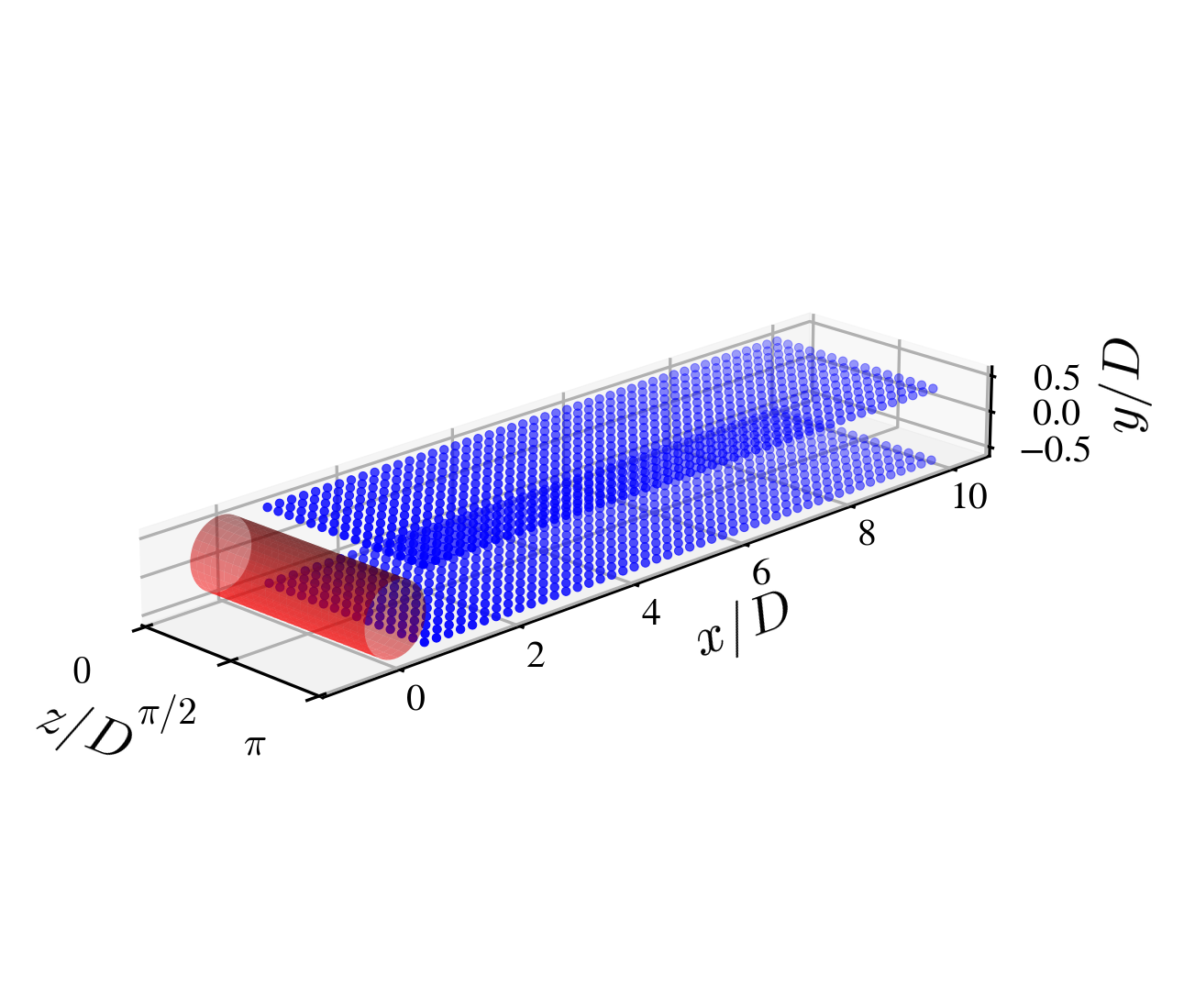}
    \caption{Horizontal}
    \end{subfigure} %\\
    \begin{subfigure}[b]{0.48\textwidth}
    \includegraphics[width=\textwidth,trim={0cm 2cm 0cm 2cm},clip]{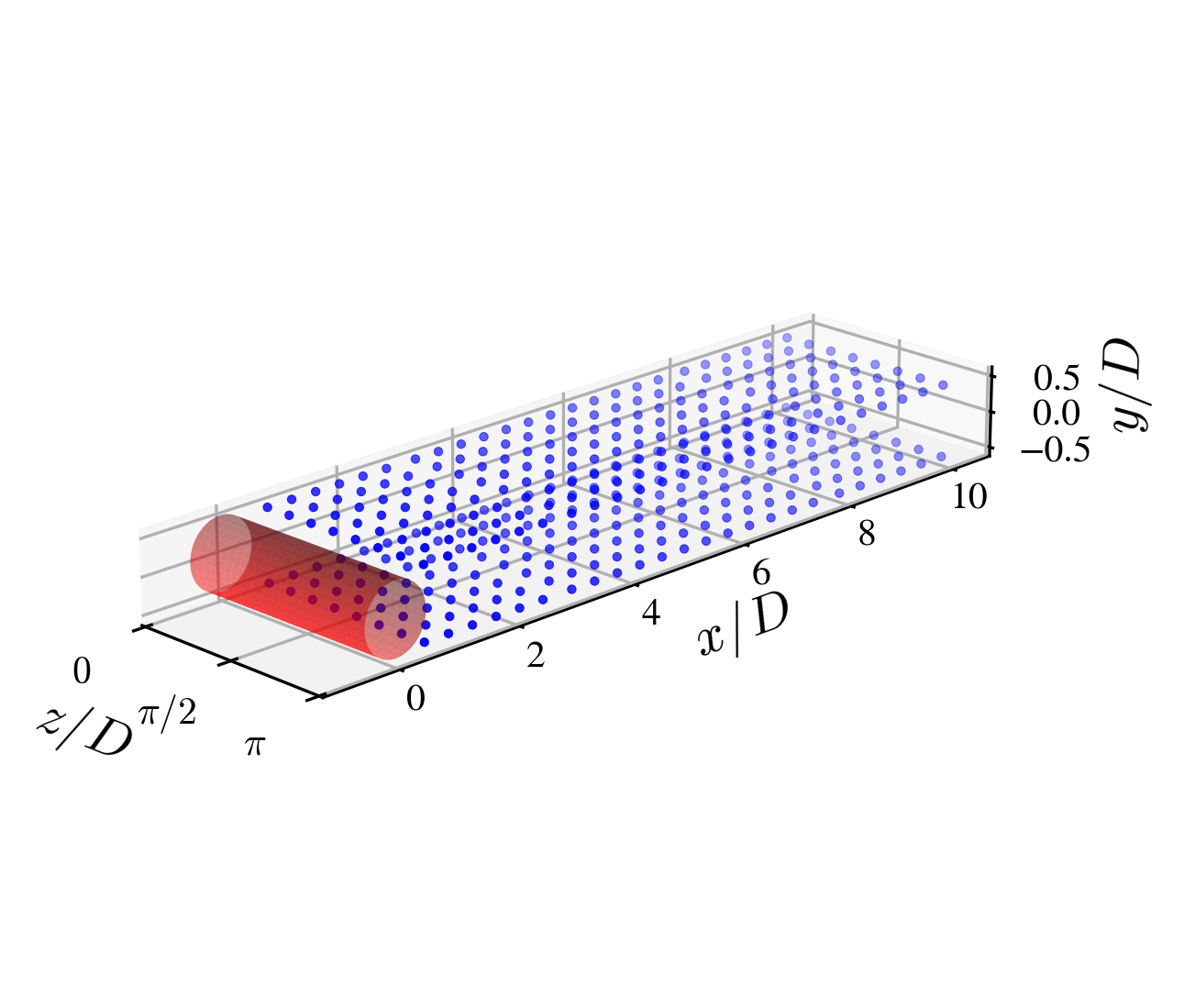}
    \caption{Horizontal-sparse}
    \end{subfigure}
    \centering
    \begin{subfigure}[b]{0.48\textwidth}
    \includegraphics[width=\textwidth,trim={0cm 2cm 0cm 2cm},clip]{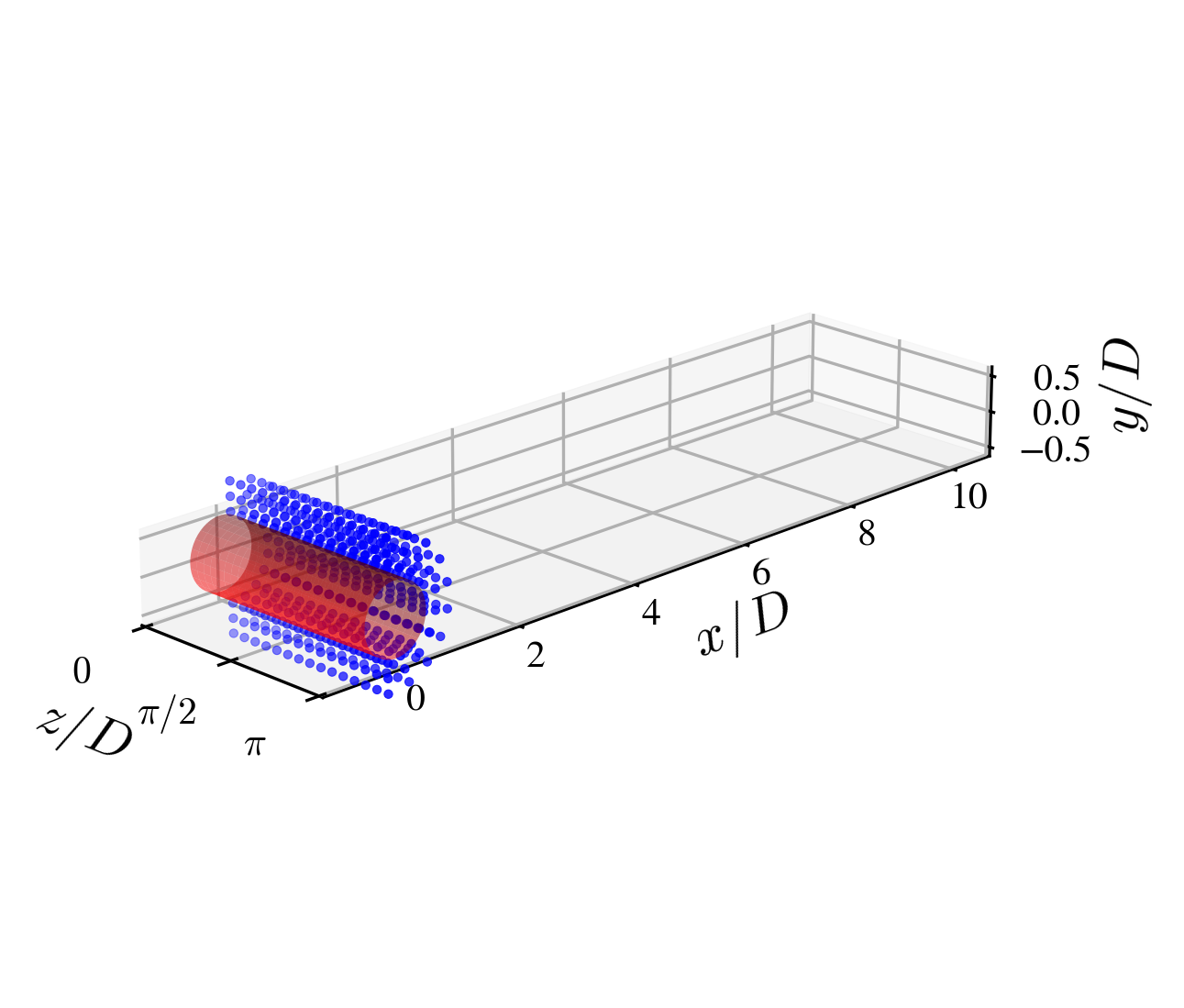}
    \caption{Azimuthal}
    \end{subfigure} %\\
    \begin{subfigure}[b]{0.48\textwidth}
    \includegraphics[width=\textwidth,trim={0cm 2cm 0cm 2cm},clip]{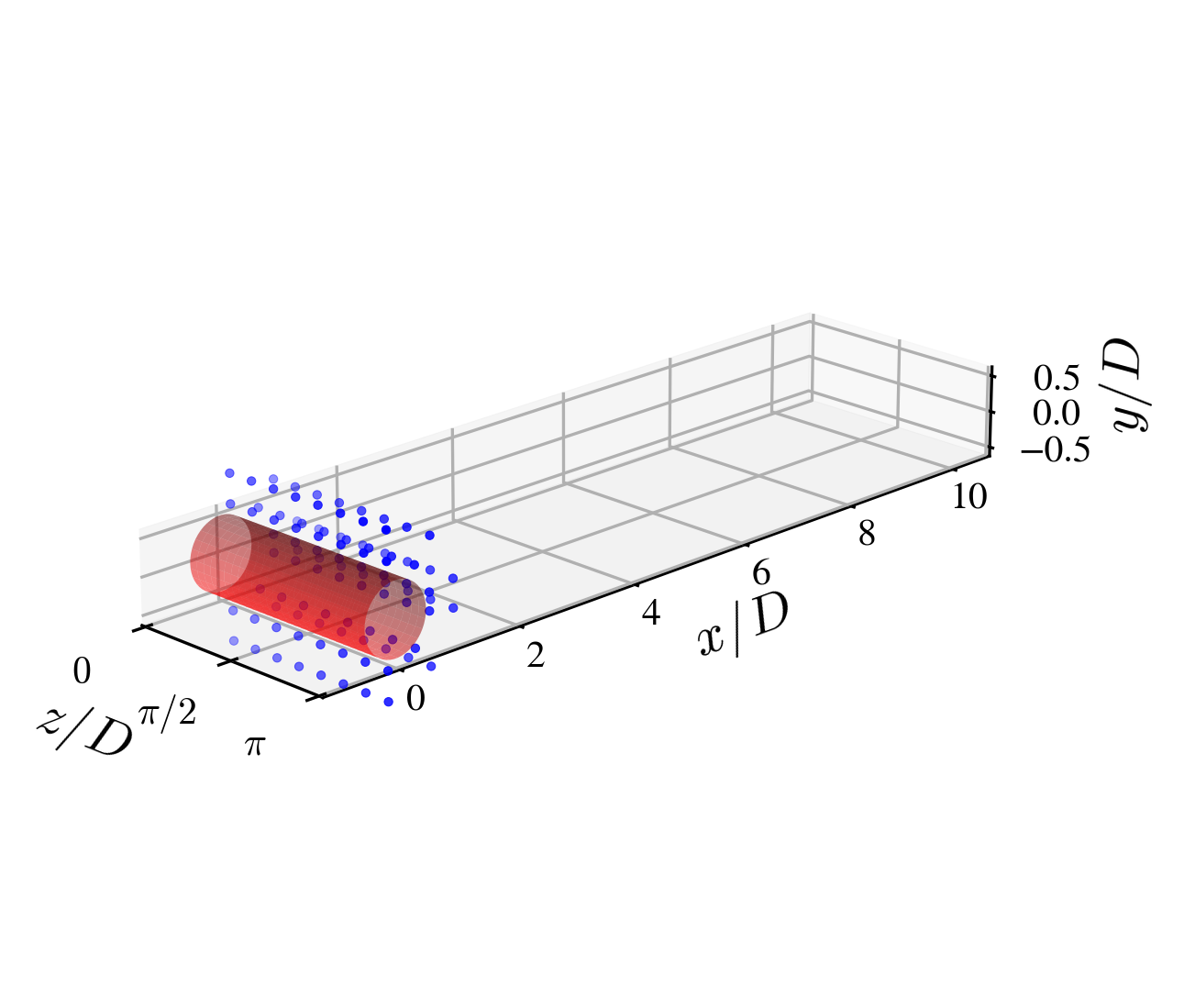}
    \caption{Azimuthal-sparse}
    \end{subfigure}
    \caption{Locations of the observations from the reference simulation on (a,b) vertical planes, (c,d) horizontal planes and (e,f) azimuthal distributed.}
    \label{fig:cyl3D_probeLoc_fig}
\end{figure}

Six different configurations for the placement of the sensors have been studied. Three distributions have been tested which are referred to as vertical, horizontal and  azimuthal. For each of these distributions, two different densities of sensors have been considered. the one corresponding to the lower density is indicated by the keyword \textit{sparse} following the name of the distribution. This selection provides a total of six sets of observation, which are going to be used for the DA runs. More details are now provided about the distribution of sensors, and the six configurations are qualitatively represented in figure \ref{fig:cyl3D_probeLoc_fig}. The set referred to as \textit{horizontal} consists of two planes for $y= \pm 0.5D$ in the wake region, $x \in [0.6D, 9.8D]$, $z \in [0.2D, 3D]$, where the distance between the neighboring observation sensors $\Delta_s$ is $\Delta_s = 0.2D$ and $\Delta_s = 0.4D$ for the \textit{horizontal} and the \textit{horizontal-sparse} cases, respectively. The \textit{vertical} set is created positioning sensors on $15$ vertical planes in the interval of $z=[0.1D, 2.9D]$ with $\Delta_s = 0.2D$ separation between consecutive planes. For the \textit{vertical-sparse} case, $8$ vertical planes are taken in the interval $z=[0.2D, 3D]$ with $\Delta_s = 0.4D$. The sensors on these planes are included in the region $x \in [0.6D, 5D]$, $y \in [-0.4D, 0.4D]$ for both cases. The \textit{azimuthal} distribution positions the sensors over circular crowns of radii $R_{min}=0.6D$ and $R_{max}=1D$. For the case \textit{azimuthal-sparse}, the distribution is the same but $R_{min}=0.7D$ and $R_{max}=1.1D$. 

A total of seven DA runs are performed. The six distributions for the sensors are used for DA runs using the HL technique, while the \textit{horizontal-sparse} case is used for DA applications with the PBOL method. For the DA cases using the HL localization, which uses the function presented in equation \ref{eq:HL_localization_fct}, the DA regions are spherical and centered around the sensors. They don't change in time. The radius $r_c$ of the DA regions for these strategies has been set to $r_c=0.1D$ and $r_c=0.2D$ for the cases \textit{vertical, horizontal, azimuthal} and for the cases \textit{vertical-sparse, horizontal-sparse, azimuthal-sparse}, respectively. For each configuration, it has been verified that neighboring DA regions does not superpose i.e. $\Delta_s >  2r_c$. Details for each of these configurations are given in table \ref{table:obsConfigs}. For the DA runs using the PBOL localization, the initial assimilation regions are set as for the HL localization. However, they are updated at each analysis phase using the dynamic algorithm, similarly to what was performed for the laminar flow around the square cylinder. The physical features used to determine the size and orientation of the ellipsoids are extensively described in Sec. \ref{sec:onlineLoc_circCyl3D}.

The numerical model used to perform the DA ensemble runs used the same solver, boundary conditions and grid used for the reference simulation. Therefore, no model error is introduced. This feature is chosen to unambiguously isolate and investigate the effect of the DA state estimation over synchronization. For both analyses using HL and PBOL localization strategies, DA experiments are performed using $20$ ensemble members which are initialized using snapshots of the flow field obtained at different times. More precisely, these initial flow fields are obtained by stored results from the reference simulation, which are sampled each $T_{ref}$ for $20 T_{ref}$. It is reminded here that $St=0.196$ for this test case and therefore a full shedding cycle takes $\Delta T_S = 5.1 T_{ref}$. Similarly with the case of the square cylinder, imposing these different initial conditions permits to assess the rate of synchronization of the ensemble towards the reference simulation, when a limited amount of sensors is used, when compared to the total number of degrees of freedom of the turbulent flow field. In order to avoid numerical issues due to discontinuities in the flow field after DA state updates, covariance localization (see eq. \ref{eq:HL_localization_fct}) is applied such that the correction is negligible approaching the boundary of the DA regions. DA analyses are performed each $0.024  T_{ref}$ i.e. $6$ time steps of the numerical solver. 

\begin{table}[h!]
\centering
\begin{tabular}{ |c|c c c c|} 
 \hline
 Configuration & \# probes & Clip Radius $r_c$ & [$x_{min}$, $x_{max}$] & [$y_{min}$, $y_{max}$]\\ 
 \hline
Horizontal & 1410 & 0.1 & $[0.6,9.8]$ & $[-0.5,0.5]$ \\
 \hline
Horizontal-sparse & 384 & 0.2 & $[0.6,9.8]$ & $[-0.5,0.5]$ \\
 \hline
 Vertical & 2070 & 0.1 & $[0.6,5.0]$ & $[-0.5,0.5]$ \\
 \hline
 Vertical-sparse & 288 & 0.2 & $[0.6,5.0]$ & $[-0.5,0.5]$ \\
 \hline
 Azimuthal & 450 & 0.1 & $[0.0,0.9848]$ & $[-1.0,1.0]$ \\
 \hline
 Azimuthal-sparse & 96 & 0.2 & $[0.0,1.083]$ & $[-1.1,1.1]$ \\
 \hline
\end{tabular}
\caption{Details of the distributions of sensors used in the DA runs.}
\label{table:obsConfigs}
\end{table}

\subsubsection{Synchronization of the flow using the HL-EnKF} \label{sec:hyperloc_circCyl3D}

\begin{figure}[!htb]
    \centering
    \begin{subfigure}[b]{0.48\textwidth}
    \includegraphics[width=\textwidth]{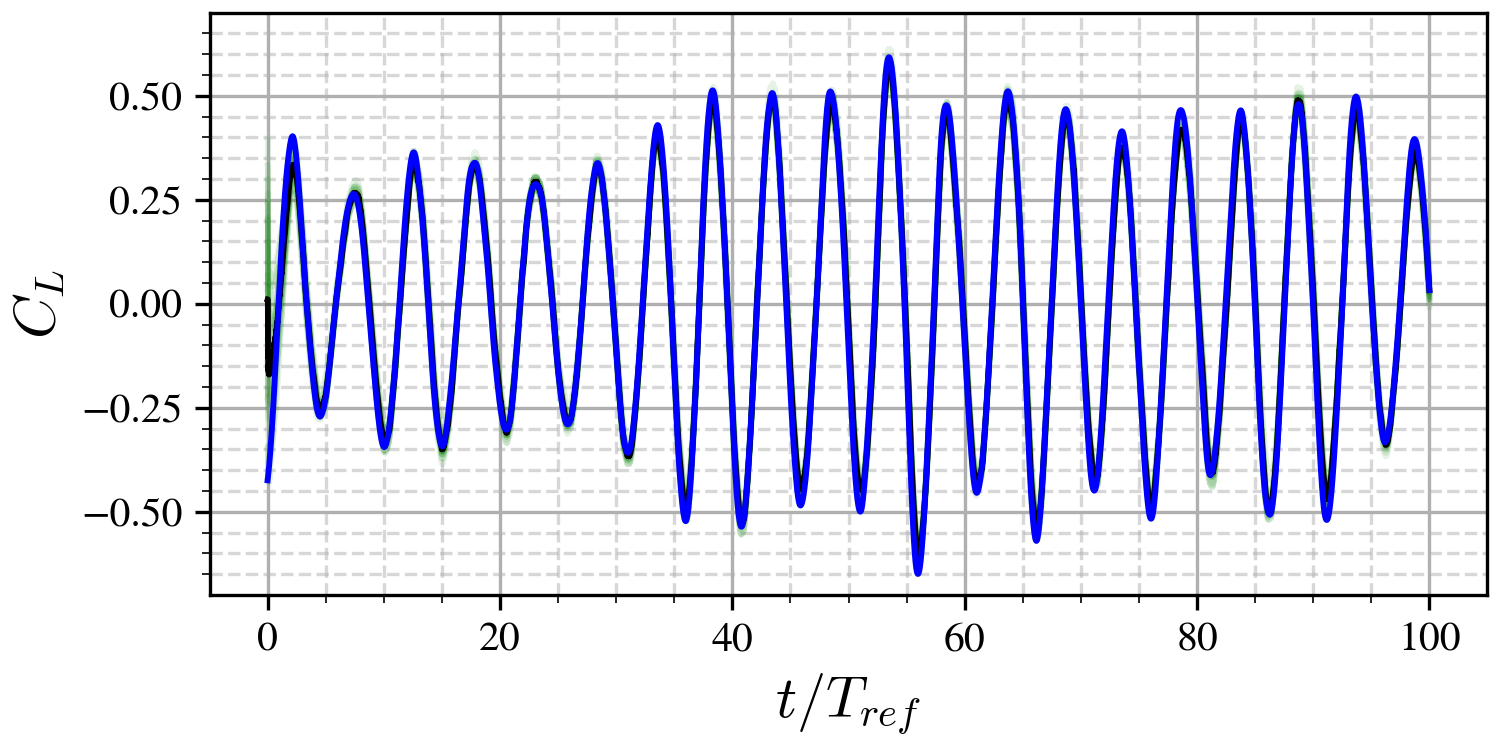}
    \caption{Vertical}
    \end{subfigure} %\\
    \begin{subfigure}[b]{0.48\textwidth}
    \includegraphics[width=\textwidth]{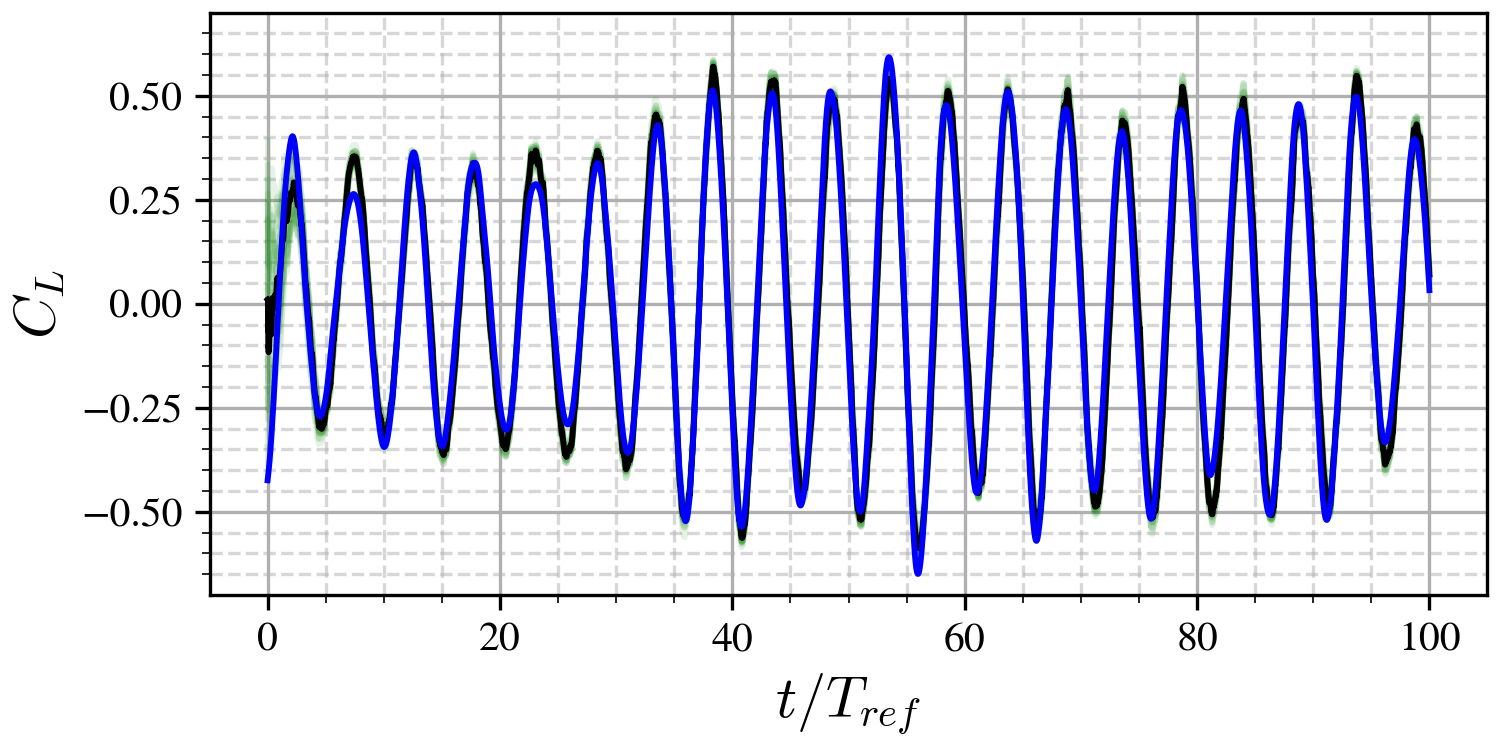}
    \caption{Vertical-sparse}
    \end{subfigure}
    \centering
    \begin{subfigure}[b]{0.48\textwidth}
    \includegraphics[width=\textwidth]{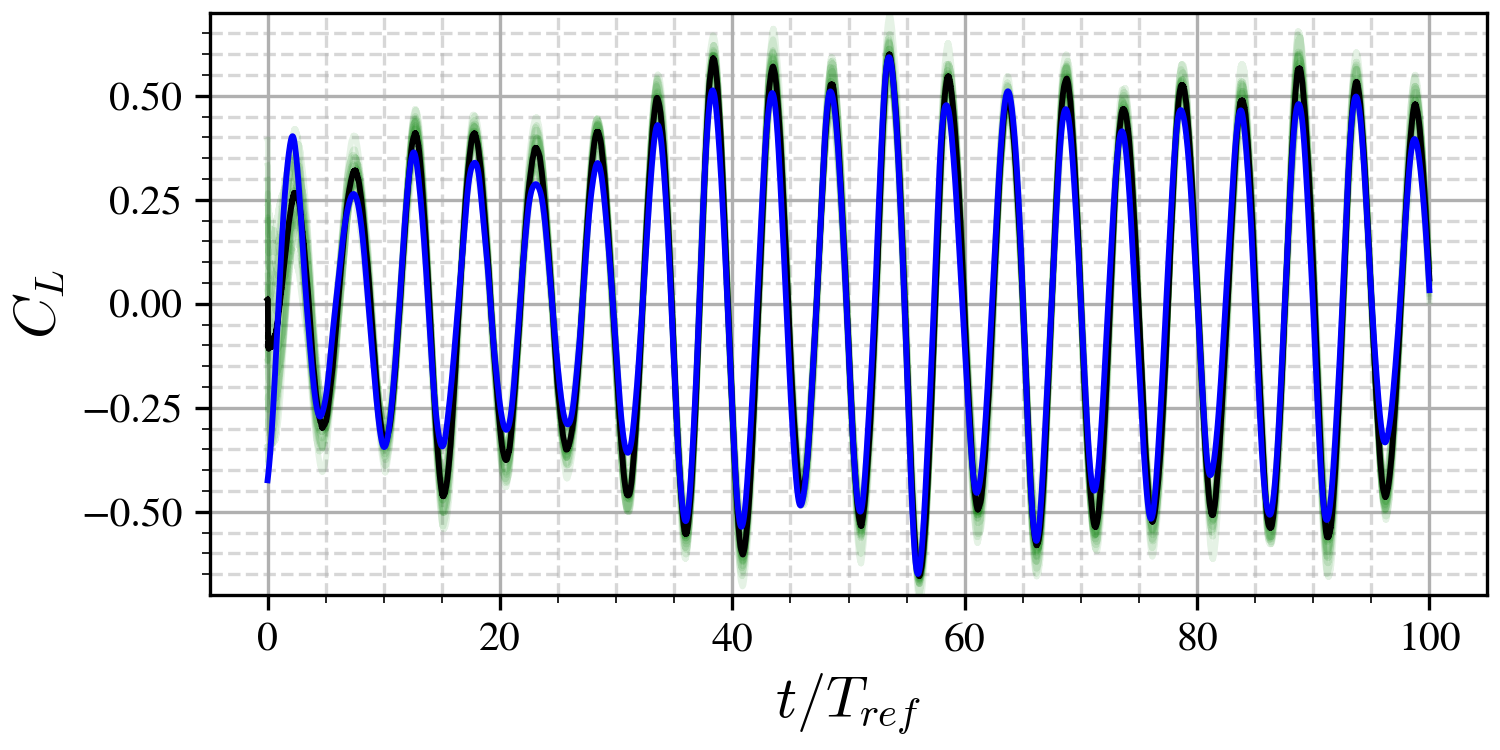}
    \caption{Horizontal}
    \end{subfigure} %\\
    \begin{subfigure}[b]{0.48\textwidth}
    \includegraphics[width=\textwidth]{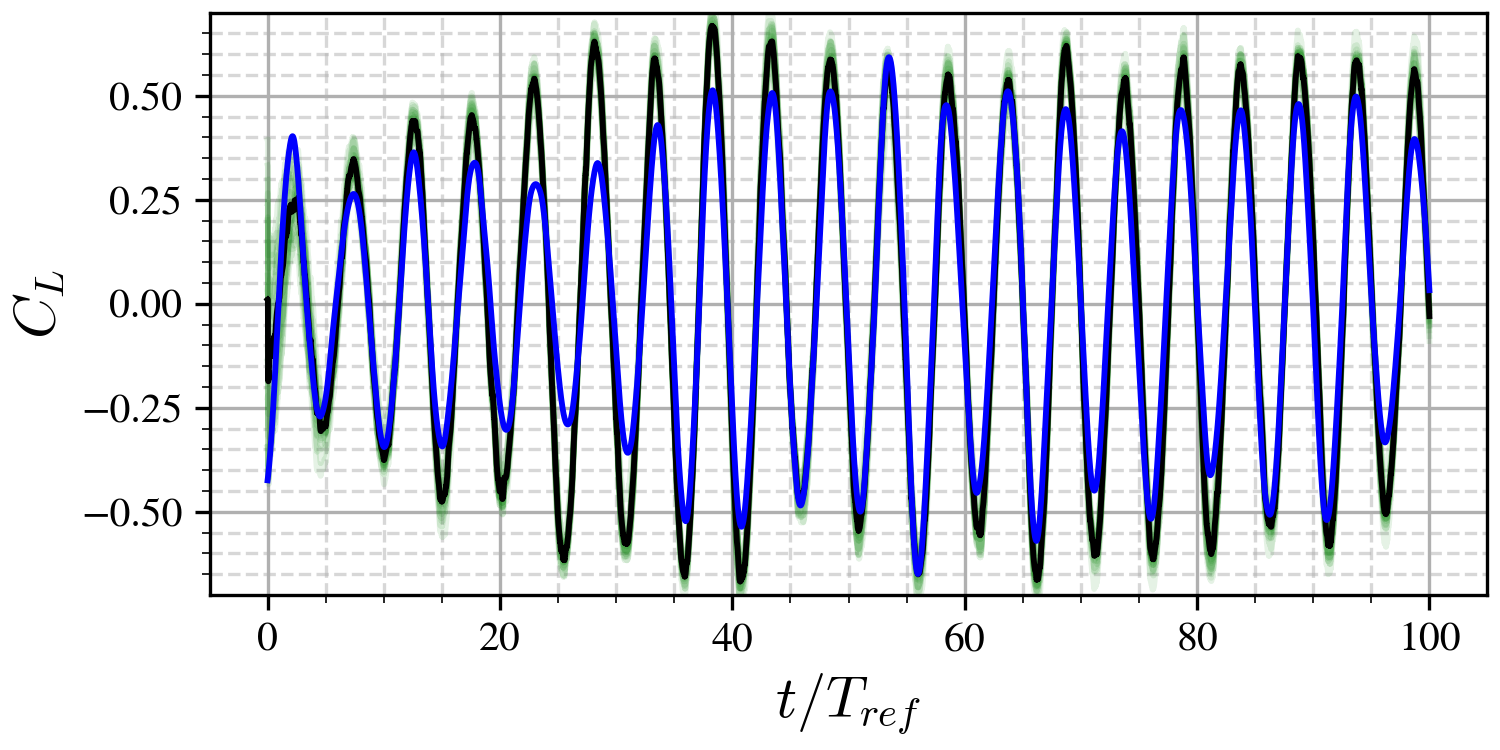}
    \caption{Horizontal-sparse}
    \end{subfigure}
    \centering
    \begin{subfigure}[b]{0.48\textwidth}
    \includegraphics[width=\textwidth]{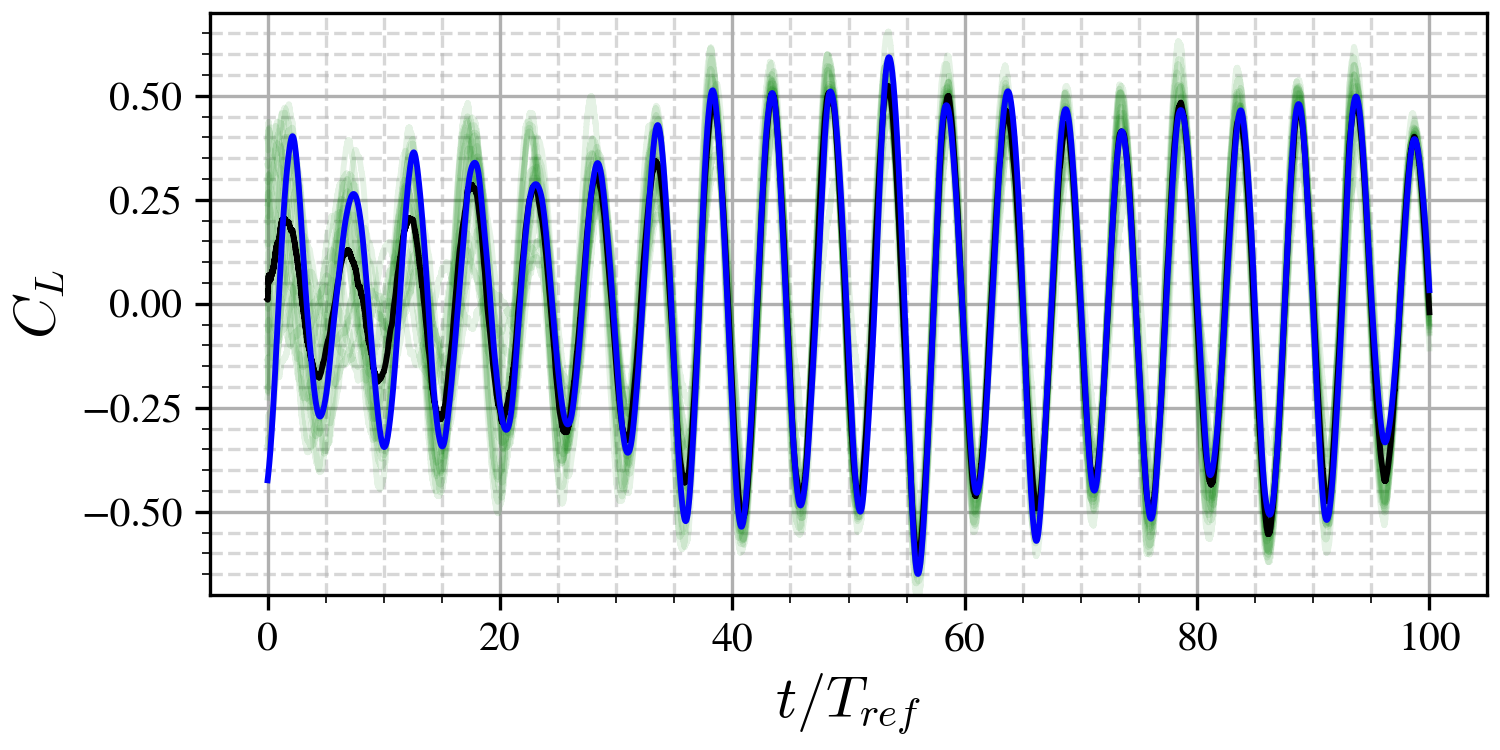}
    \caption{Azimuthal}
    \end{subfigure} %\\
    \begin{subfigure}[b]{0.48\textwidth}
    \includegraphics[width=\textwidth]{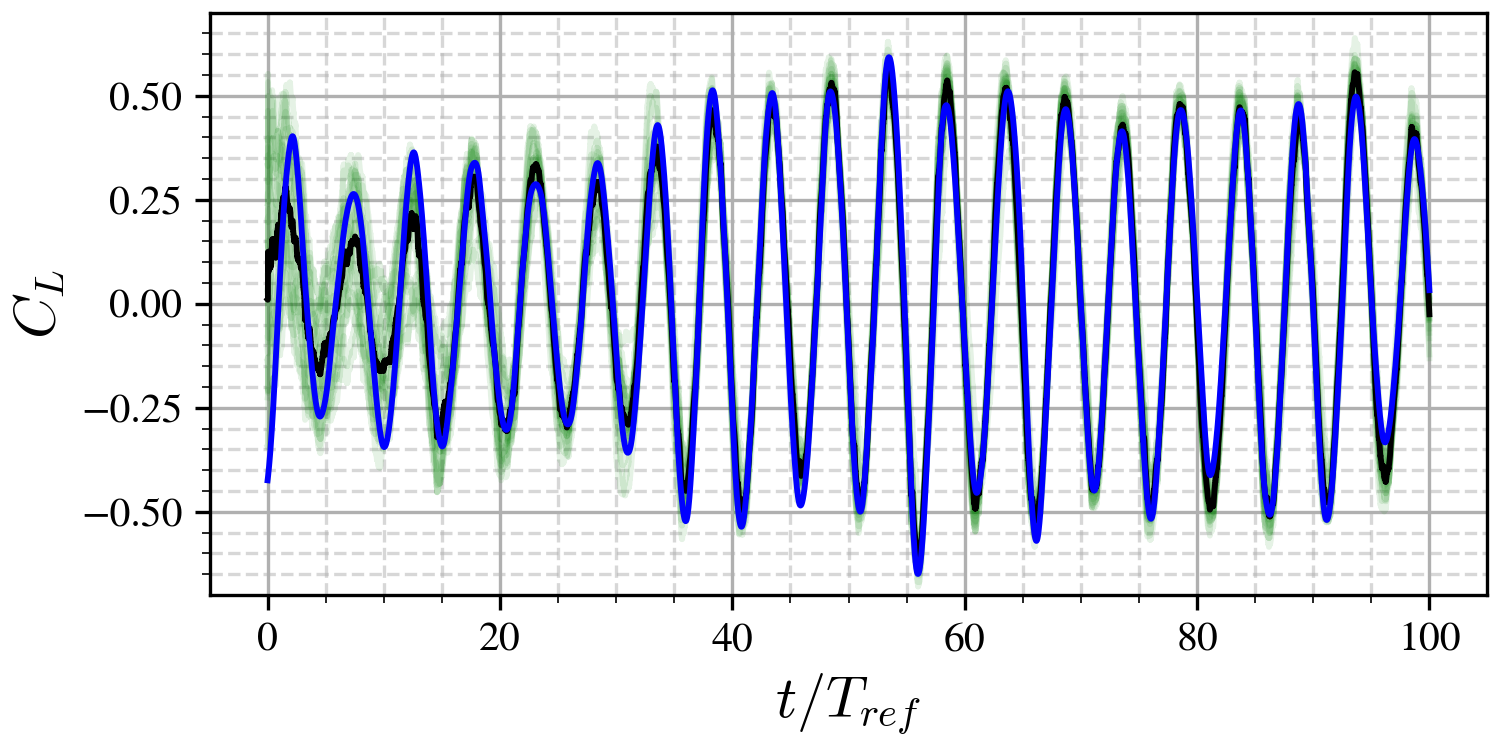}
    \caption{Azimuthal-sparse}
    \end{subfigure}
    \caption{DA synchronization of the lift coefficient $C_L$ with the reference simulation. Results are shown using the six sets of sensors for high density and sparsely distributed probes (a,b) on vertical planes, (c,d) horizontal planes and (e,f) azimuthal curves.}
    \label{fig:sync_cyl3D_CL}
\end{figure}

\begin{figure}[!htb]
    \centering
    \begin{subfigure}[b]{0.48\textwidth}
    \includegraphics[width=\textwidth]{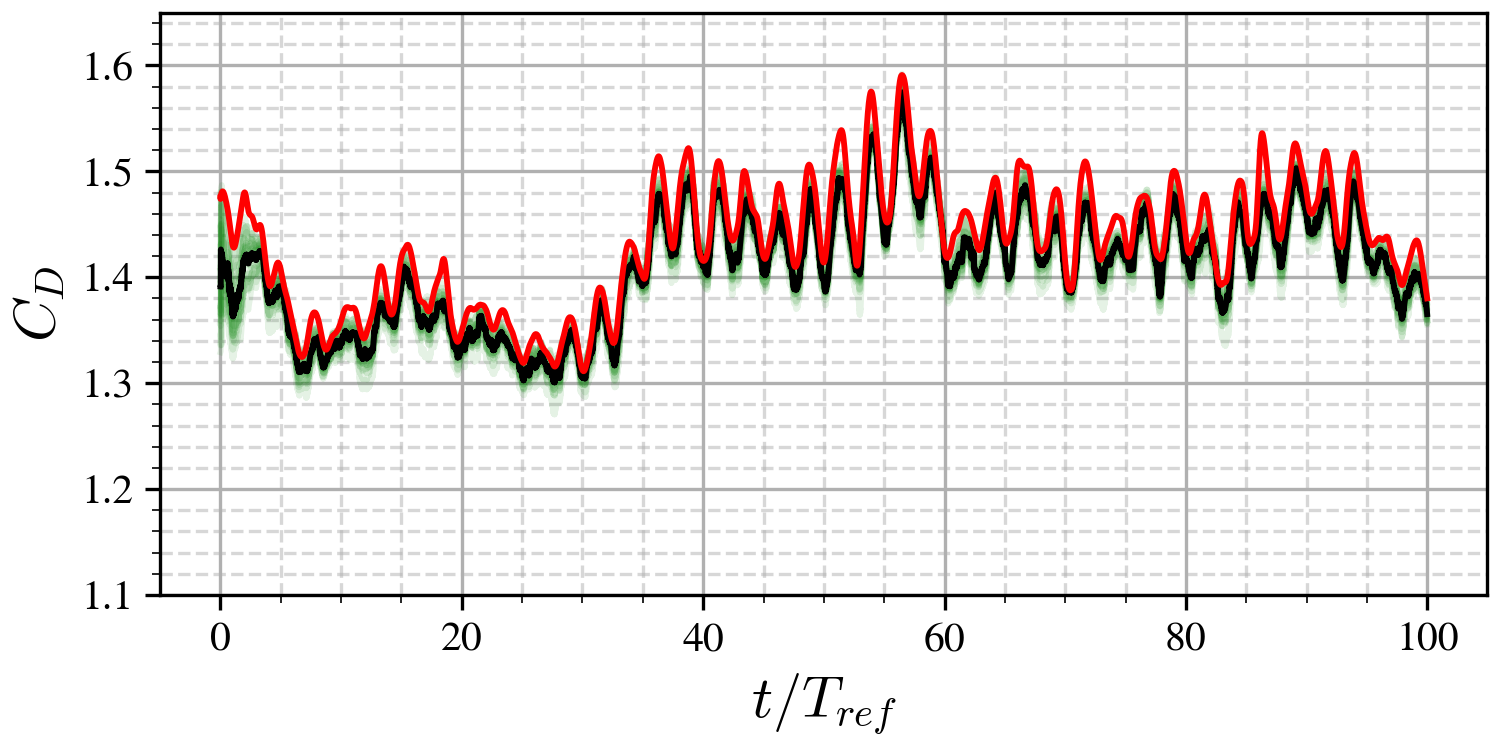}
    \caption{Vertical}
    \end{subfigure} %\\
    \begin{subfigure}[b]{0.48\textwidth}
    \includegraphics[width=\textwidth]{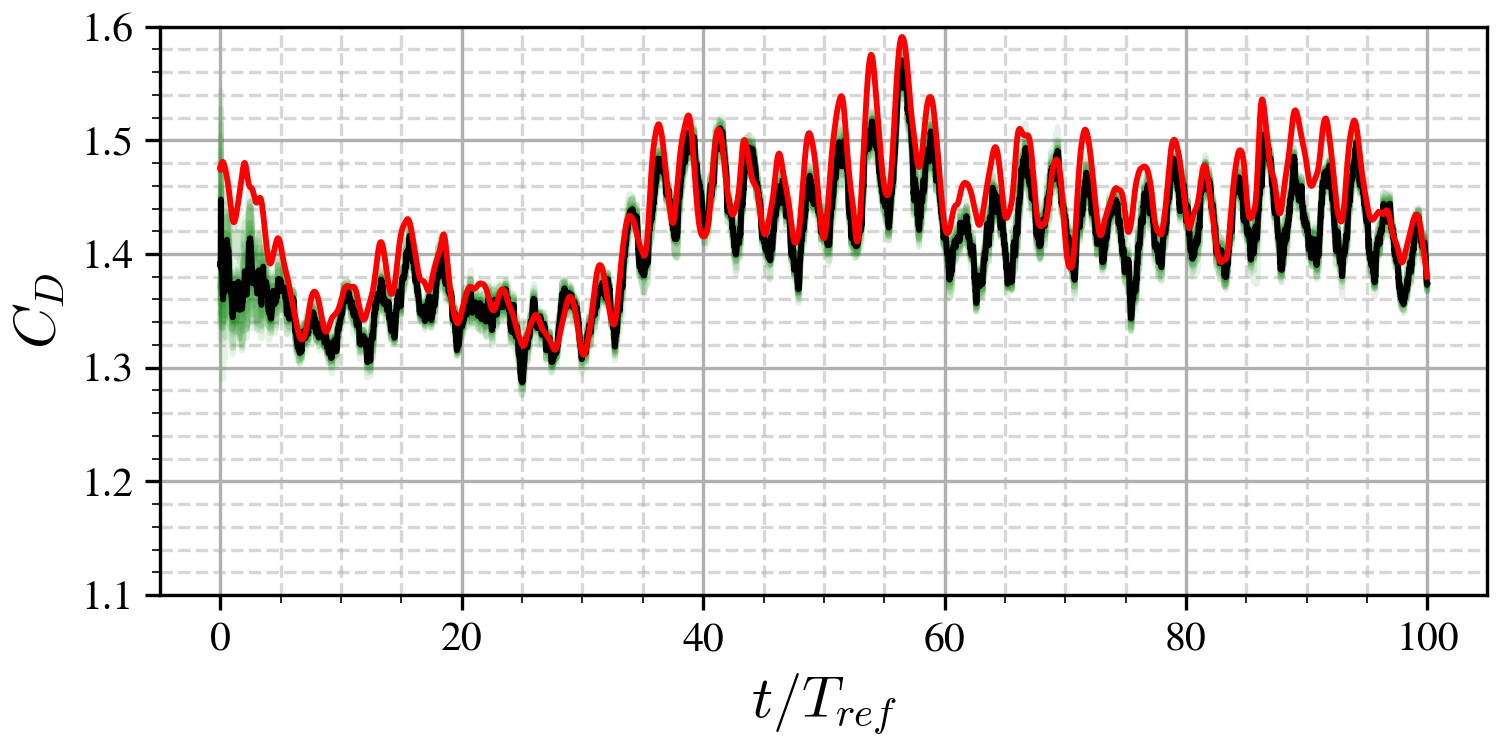}
    \caption{Vertical-sparse}
    \end{subfigure}
    \centering
    \begin{subfigure}[b]{0.48\textwidth}
    \includegraphics[width=\textwidth]{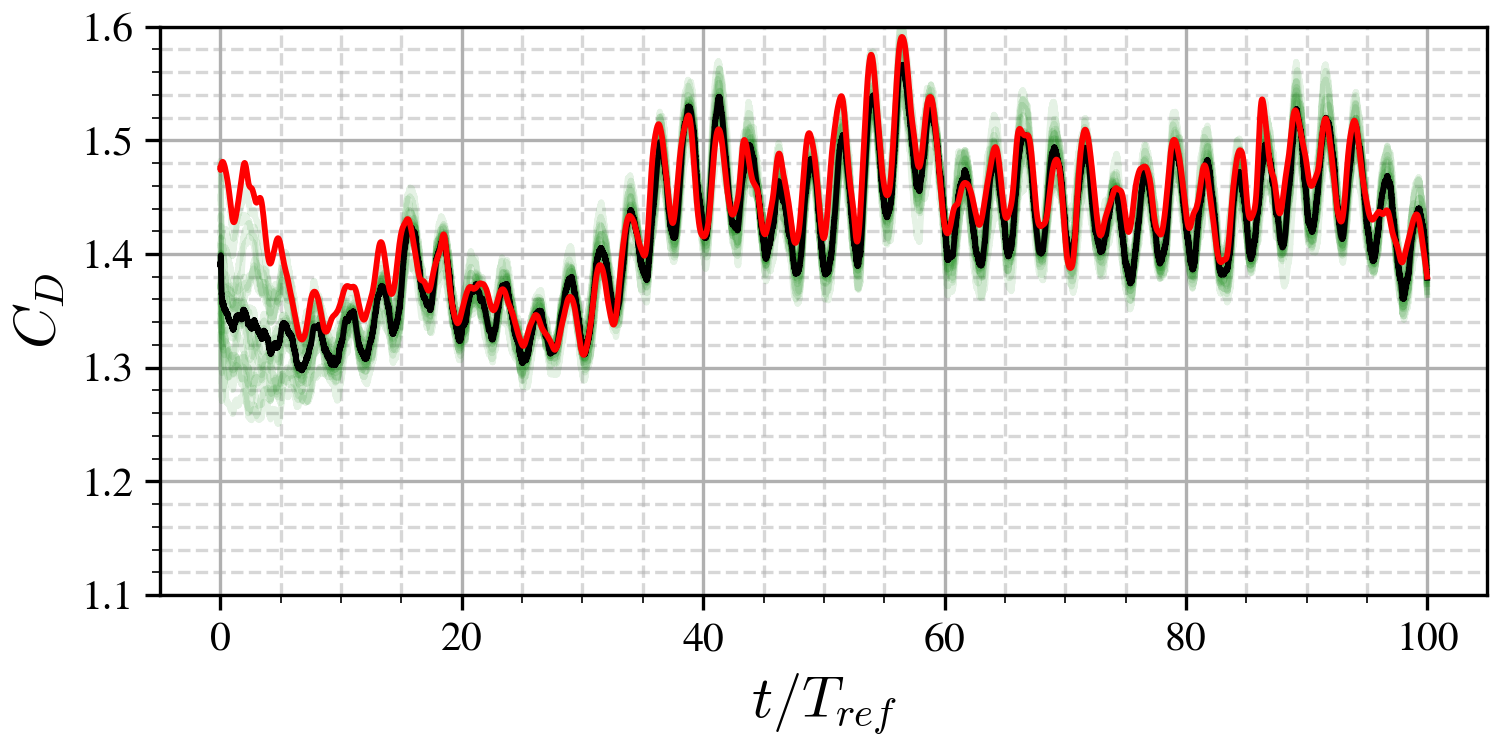}
    \caption{Horizontal}
    \end{subfigure} %\\
    \begin{subfigure}[b]{0.48\textwidth}
    \includegraphics[width=\textwidth]{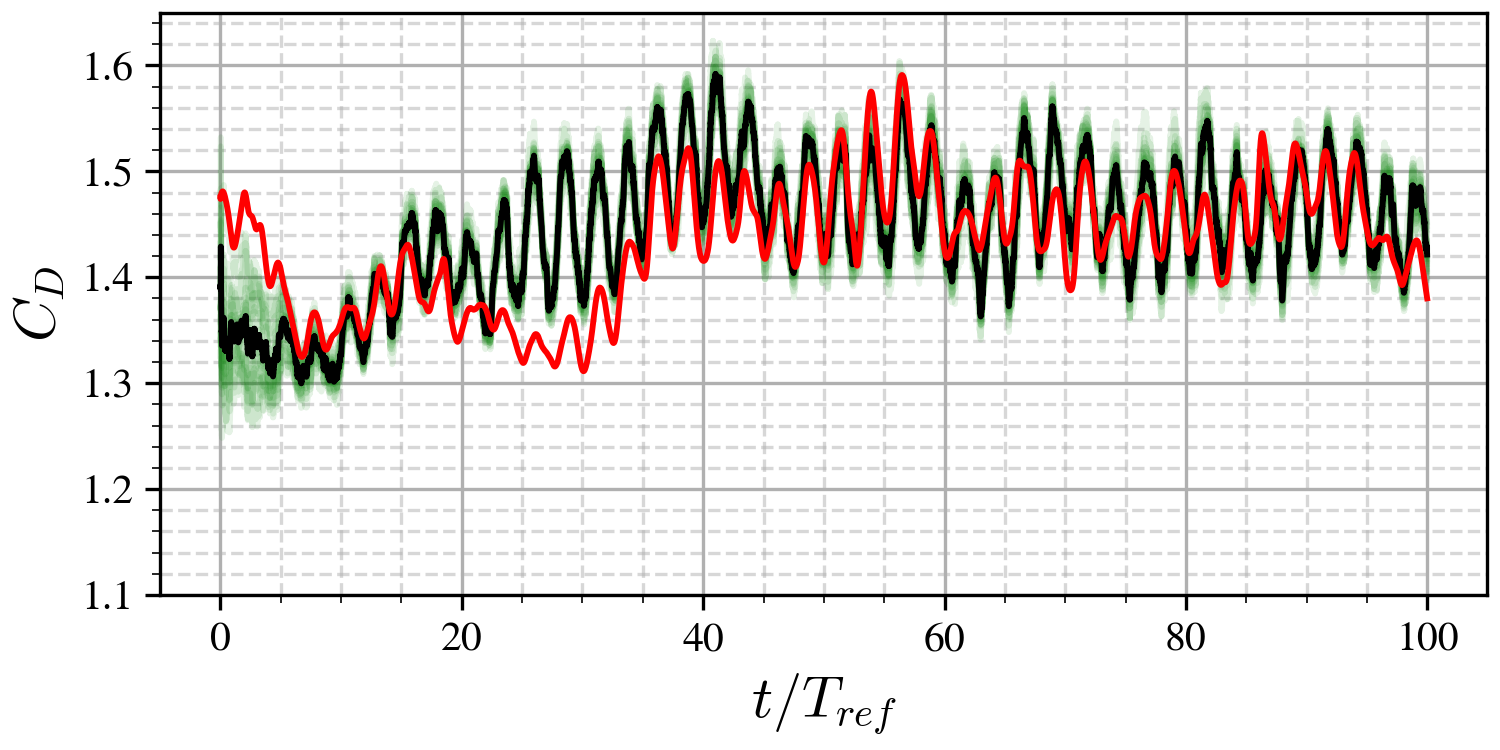}
    \caption{Horizontal-sparse}
    \end{subfigure}
    \centering
    \begin{subfigure}[b]{0.48\textwidth}
    \includegraphics[width=\textwidth]{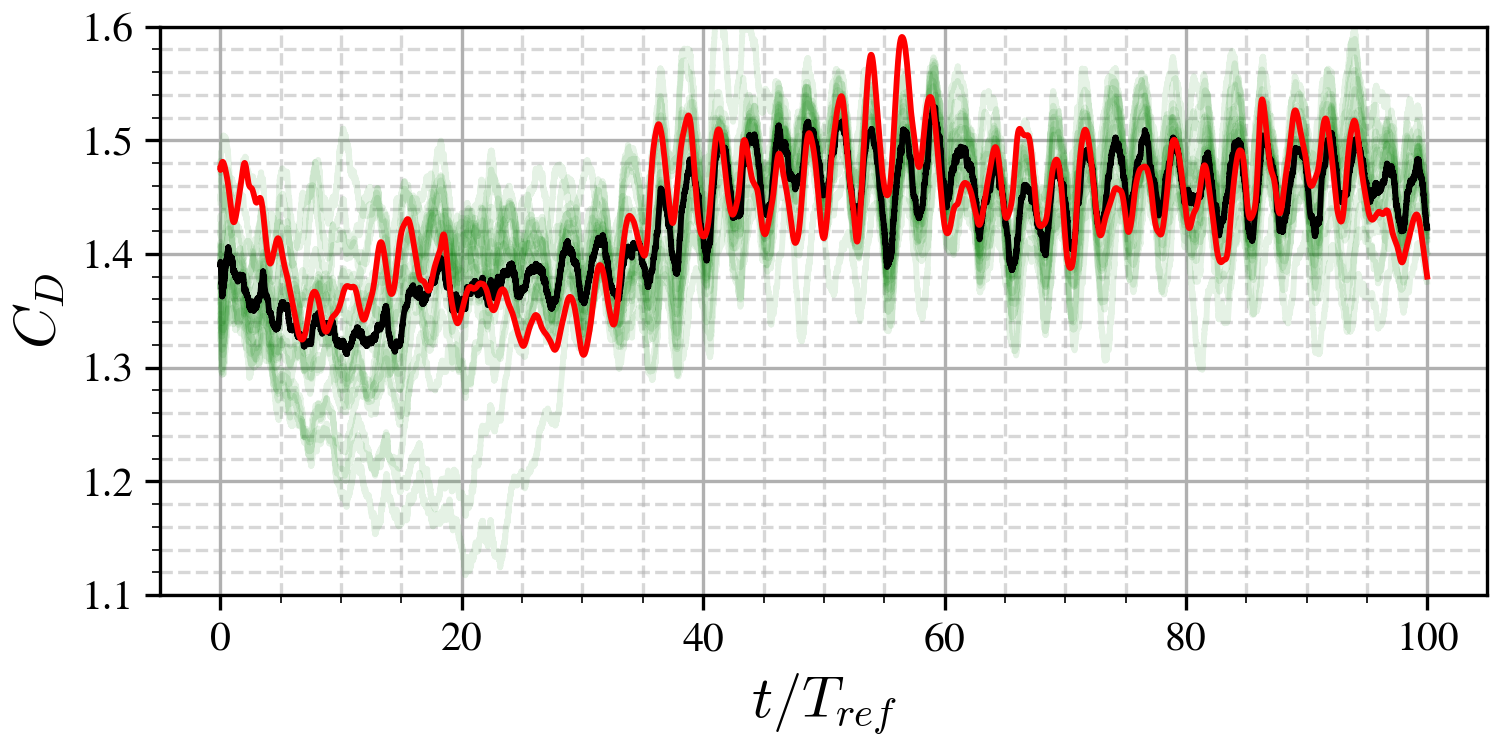}
    \caption{Azimuthal}
    \end{subfigure} %\\
    \begin{subfigure}[b]{0.48\textwidth}
    \includegraphics[width=\textwidth]{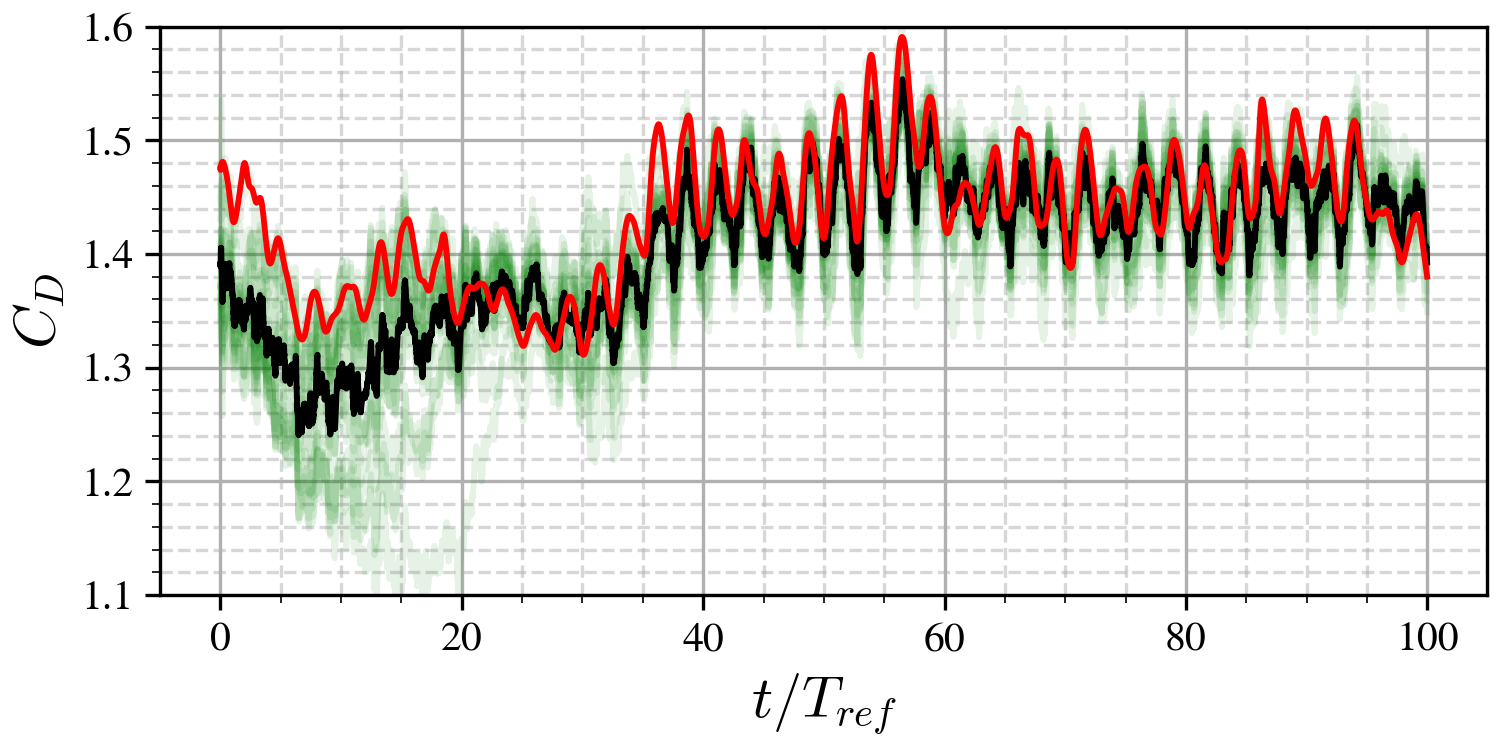}
    \caption{Azimuthal-sparse}
    \end{subfigure}
    \caption{DA synchronization of the drag coefficient $C_D$ with the reference simulation. Results are shown using the six sets of sensors for high density and sparsely distributed probes (a,b) on vertical planes, (c,d) horizontal planes and (e,f) azimuthal curves.}
    \label{fig:sync_cyl3D_CD}
\end{figure}

The first set of DA runs using the HL localization are now investigated. Six DA runs are performed using each configuration of sensors described in section \ref{sec:config_sensors_cyl3D}. Complete synchronization between model prediction and observation is obtained for every DA run. However, the rate of convergence exhibits a sensitivity to the distribution of sensors used. This sensitivity is the result of several concurring factors such as the number of grid elements in the DA regions which are affected by the state update as well as the relevance of information provided by more efficient sensor positioning. This variability can be observed by the time evolution of the lift coefficient $C_L$ predicted by DA. In figure \ref{fig:sync_cyl3D_CL}, comparisons between the DA results with the reference simulation (in blue) are presented. For all the configurations, the time to obtain suitable synchronization is shorter when a higher density of the sensors is used. A similar trend is observed also for the synchronization of the drag coefficient $C_D$ (see fig. \ref{fig:sync_cyl3D_CD}). The analysis of these features seems to indicate that the best results are obtained using vertical planes with sensors. This can be attributed to the larger number of sensors distributed in critical region for flow development in the \textit{vertical} and \textit{vertical-sparse} configurations. The \textit{azimuthal} distribution is the one requiring the largest number of analysis phases and state updates to obtain an adequate synchronization but, once it is obtained, the prediction obtained by the ensemble members almost perfectly superpose with data from the reference simulation. This result is due to the advective propagation of the state update to the wake region, which governs the evolution of the bulk flow parameters $C_L$ and $C_D$. Therefore, the high-precision state estimation obtained in the near-wall region strongly affects the accuracy of the prediction in the wake region after a sufficiently long time, which is proportional to the effect of advection. On the other hand, the usage of horizontal planes for the distribution of sensors provides the worst results. Even if initial synchronization is fast, thanks to the presence of sensors in the wake region and close to the immersed body, the lack of state update in the core region of the wake is responsible for a persistent discrepancy between the ensemble average prediction and the results from the reference simulation. This discrepancy is accentuated for the distribution with low density of sensors.

\begin{figure}[!htb]
    \centering
    \begin{subfigure}[b]{0.49\textwidth}
    \includegraphics[width=\textwidth]{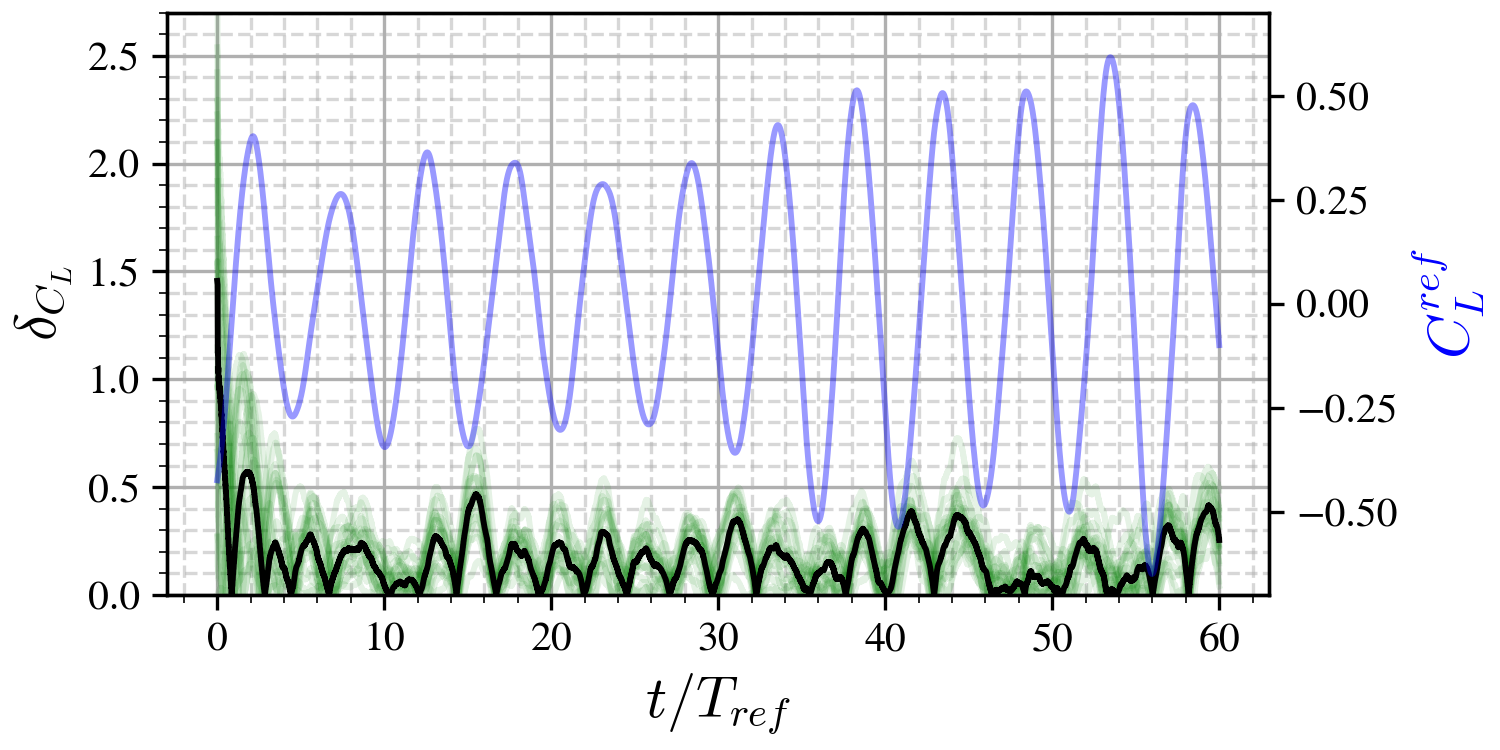}
    \caption{Horizontal}
    \end{subfigure} %\\
    \begin{subfigure}[b]{0.49\textwidth}
    \includegraphics[width=\textwidth]{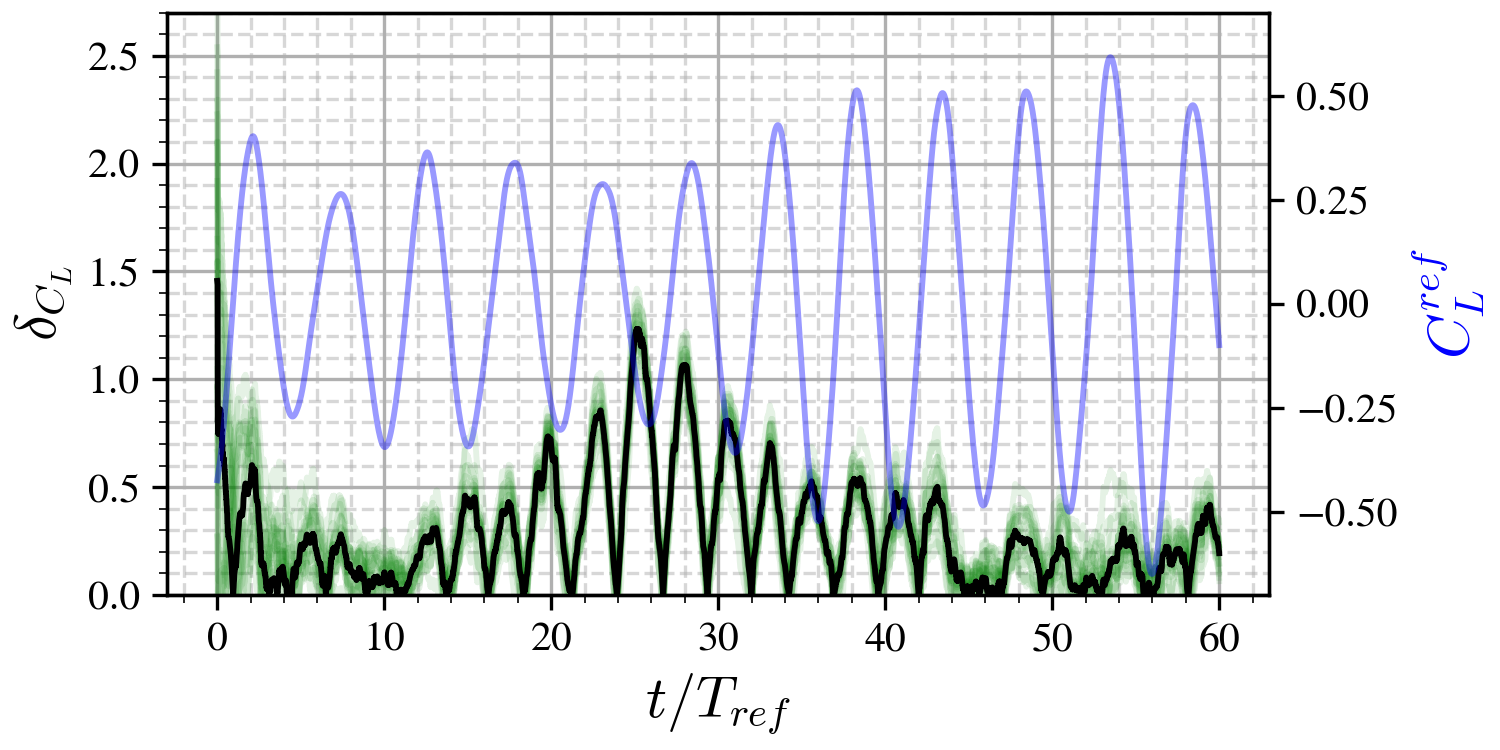}
    \caption{Horizontal-sparse}
    \end{subfigure}
    \centering
    \begin{subfigure}[b]{0.49\textwidth}
    \includegraphics[width=\textwidth]{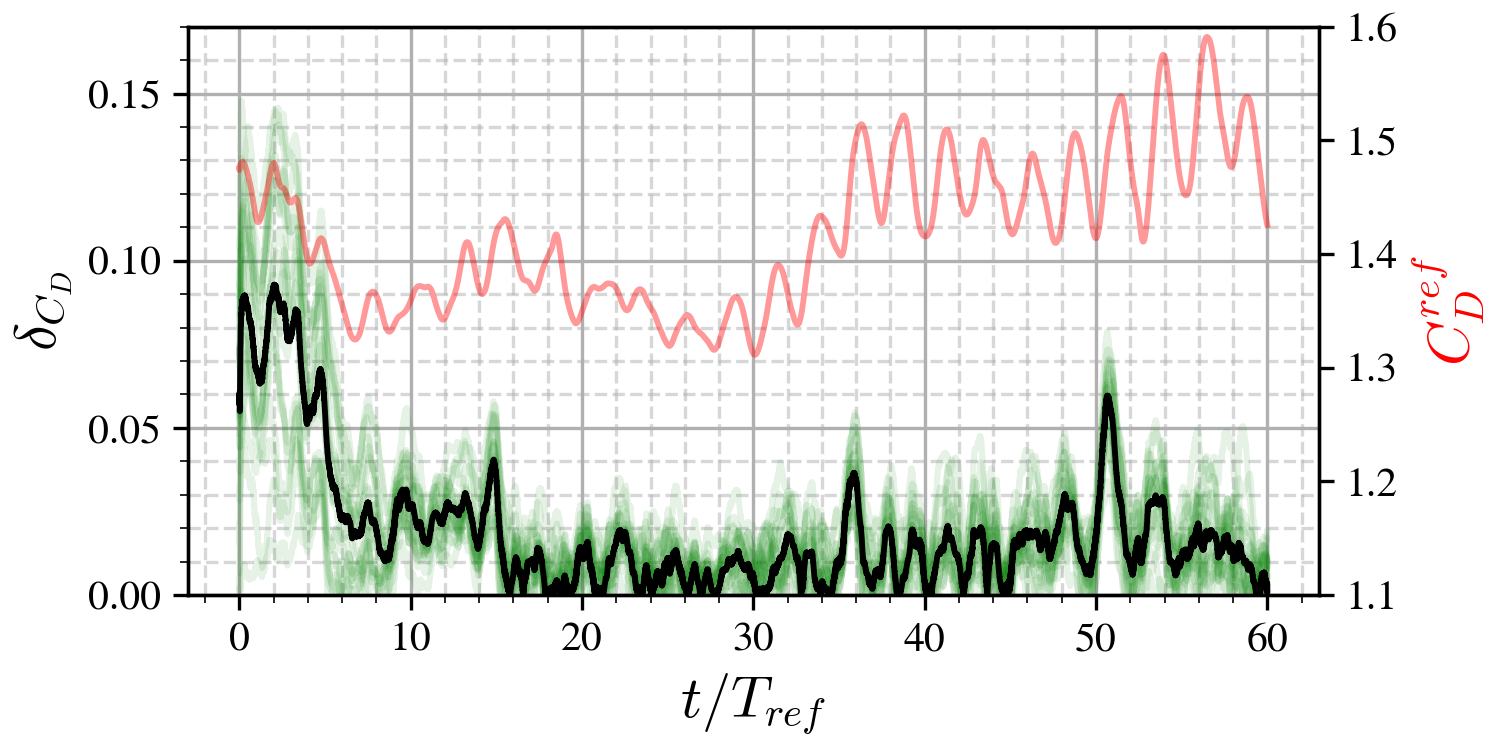}
    \caption{Horizontal}
    \end{subfigure} %\\
    \begin{subfigure}[b]{0.49\textwidth}
    \includegraphics[width=\textwidth]{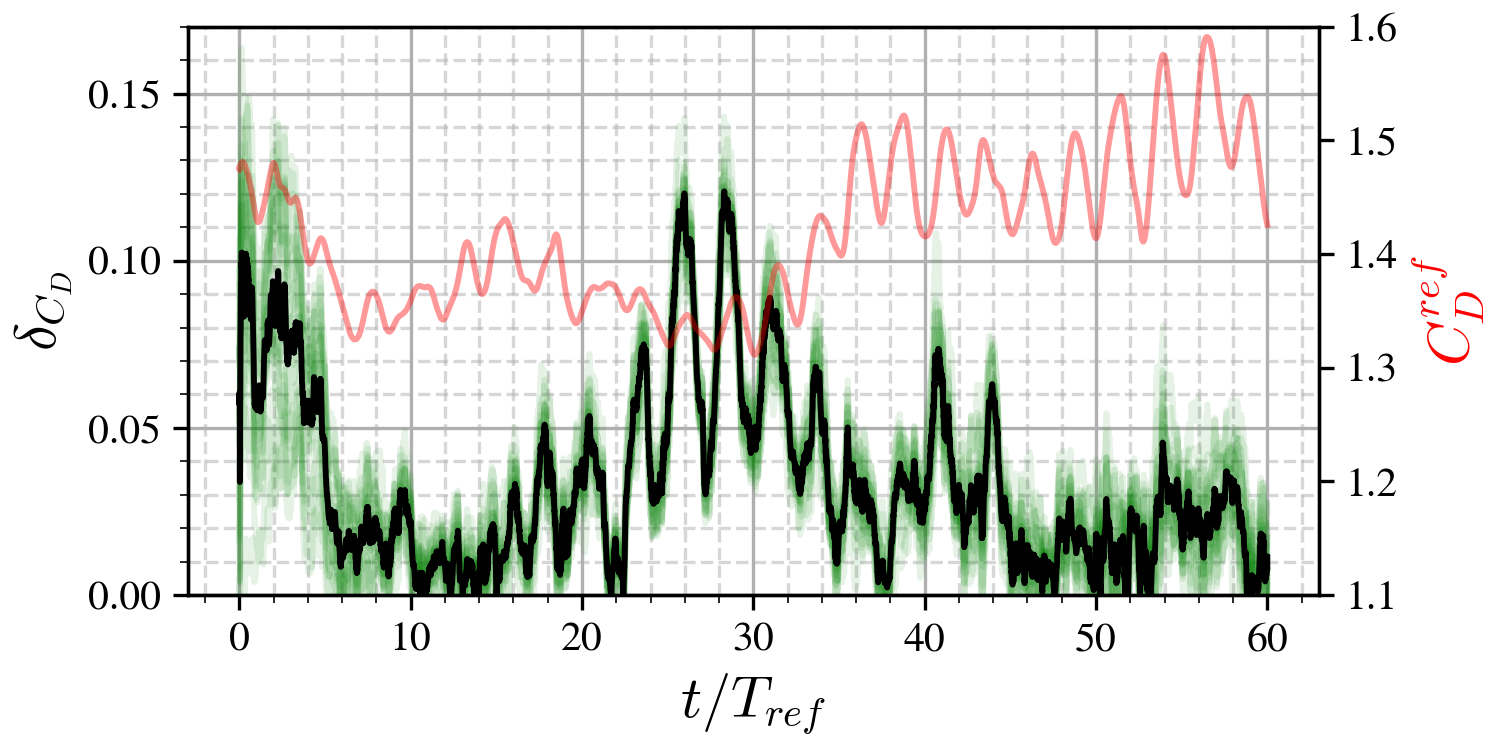}
    \caption{Horizontal-sparse}
    \end{subfigure}
    \caption{Discrepancies observed for the force coefficients between the assimilated members and the reference simulation (in green), between the ensemble mean and reference simulation (in black). The force coefficients of the reference simulation are shown in the respective right axes. DA results shown for \textit{horizontal} configuration of sensors are presented in (a) and (c), for the \textit{horizontal-sparse} configuration they are presented in (b) and (d).}
    \label{fig:delta_coeff_sync_cyl3D_hyperloc}
\end{figure}

In order to investigate the efficiency of the PBOL methodology, the \textit{horizontal-sparse} configuration is selected to perform a suitable comparison for two main reasons. First, this set is made by fewer sensors, which are clustered on the planes for $y/D = \pm 0.5$ rather than filling a volume in the wake of the cylinder. This case is common for experiments where the sensor placement can be affected by structural / safety restrictions. Finally, the \textit{horizontal-sparse} configuration is observed to be the most challenging case in terms of synchronization with the reference simulation among the cases investigated. Therefore, the larger initial discrepancies permit to observe the variations in the accuracy of the DA methodology more clearly. An additional analysis of the results is performed for the DA run using \textit{horizontal} configuration. The normalized discrepancy between the force coefficients of the assimilated members and the reference simulation is shown in figure \ref{fig:delta_coeff_sync_cyl3D_hyperloc}, which are computed by using the formulas $\delta_{C_L}(t) = |C_L^{ref} - C_L|/\sigma_{C_L^{ref}}$ and $\delta_{C_L}(t) = |C_D^{ref} - C_D|/\overline{C_D^{ref}}$. The discrepancy for the lift coefficient $C_L$ is  normalized over the standard deviation of $C_L$, referred to as $\sigma_{C_L}$, and discrepancy for the drag coefficient $C_D$ is normalized over its mean value $\overline{C_D}$.

One can see that the magnitude of these errors is within $50\%$ for the $C_L$ and $5\%$ for the $C_D$ for the DA run with \textit{horizontal} configuration. On the other hand, a significantly larger discrepancy is observed for the \textit{horizontal-sparse} configuration, particularly visible around $t/T_{ref} = 27$. Also, a significantly higher variability is observed for the latter case, highlighting the importance of additional sensors in the DA analysis for this case.

\subsubsection{Results using the PBOL-EnKF strategy} \label{sec:onlineLoc_circCyl3D}

In this section the synchronization capabilities of the EnKF with PBOL localization are investigated. A DA run, referred to as DA-PBOL, has been run with the \textit{horizontal-sparse} sensor distribution. It is important to stress that the initial volume of each of the DA regions is a sphere with radius $r_c=0.2D$, and that the PBOL procedure determines the localization function inside this volume at each analysis phase to preserve the part of the region determined by the physical criteria employed.

% ONLINE-LOCALIZATION CYL3D METHODOLOGY ---------------------

In terms of DA set-up of the problem, the observation, model and initial conditions used for this analysis are the ones presented in sections \ref{sec:sync_Cyl3D}, \ref{sec:config_sensors_cyl3D} and \ref{sec:hyperloc_circCyl3D}. The determination of the sets of parameters $[\bm{e}_{x'}, \, \bm{e}_{y'}, \, \bm{e}_{z'}, \, \xi_{x^\prime}, \, \xi_{y^\prime}, \, \xi_{z^\prime}]$ used in the PBOL localization for each DA region is carried out with the methodology described in section \ref{sec:onlineLoc_methodology}. The rationale between the strategy presented here 
is to exploit the correlations observed in turbulent flows, for which theoretical frameworks available in the literature can be infused in the mathematical formulation. The orientation of the vector $\bm{e}_{x'}$ is chosen to be the direction of the local ensemble mean velocity vector $\ensavg{\bm{u}}$ at each sensor probe location. \cite{AlvesPortela2017} presents the autocorrelation functions of streamwise velocity component computed by both spatial and temporal displacements, $\Delta x$ and $\Delta t$, using the relation
\begin{equation}
    R_u(x, \Delta x, \Delta t) = \frac{\langle u(x,t) u(x+ \Delta x, t + \Delta t) \rangle}{\sqrt{\langle u(x,t)^2 u(x + \Delta x, t + \Delta t)^2}},
\end{equation}
in the downstream wake of a square prism. It is shown that there exists a linear relationship between the parameter $\Delta t$ corresponding the peak of $R_u$ and the parameter $\Delta x$, suggesting a clearly defined convection velocity in the wake flow. In \cite{AlvesPortela2017}, the analysis of $R_u$ is performed considering spatial displacements only in the streamwise direction. On the other hand, the aim of the present methodology is to obtain generalized and robust criteria to be applied in various flow fields. Therefore, in the present case, it is chosen to utilize $\ensavg{\bm{u}}$ here to determine the direction of $\bm{e}_{x'}$, which can be in any spatial direction, without any constraints. 

Taking into account the relatively high frequency of DA analysis phase, given by the relation $\Delta t_{DA}/T_{ref} << 1$, the quantity $\ensavg{\bm{u}}$ is considered as the local convection velocity at the location of the sensor at a given moment. The procedure then determines $\bm{e}_{y'}$, which is orthogonal to $\bm{e}_{x'}$. As $\bm{e}_{x'}$ defines a plane containing infinite amount of vectors orthogonal to itself, another condition must be applied for the determination of $\bm{e}_{y'}$. It is proposed to select the direction for which the projection of $\bm{e}_{y'}$ on global cross-stream direction, $ \bm{e}_{y'} \cdot \bm{j}$, is highest. This condition corresponds to the two vectors, which are located at the intersection of the planes defined by the surface normals $\bm{e}_{x'}$ and $\bm{k}$, where $\bm{k}$ is the unit vector in the global z-direction i.e. spanwise direction. This choice is motivated by the fact that the directions of $\bm{e}_{x'}$, $\bm{e}_{y'}$ and $\bm{e}_{z'}$ will be used to determine the extent of the localization function, and choosing $\bm{e}_{y'}$ as the vector with most significant cross-stream component, relates the large scale dynamics of the flow and PBOL methodology, when compared to fully random choice of $\bm{e}_{y'}$ vector on the plane defined by $\bm{e}_{x'}$. Finally the $\bm{e}_{z'}$ is determined by the cross-product $\bm{e}_{x'} \times \bm{e}_{y'}$. Thus the conditions for the determination of the unit vectors of the local coordinate system can be summarized as
\begin{align}
    \bm{e}_{x'} &= \ensavg{\bm{u}} / |\ensavg{\bm{u}}|, \nonumber \\ 
    \bm{e}_{y'} \cdot \bm{e}_{x'} &= 0, \nonumber \\
    \bm{e}_{y'} \cdot \bm{k} &= 0, \\ 
    |\bm{e}_{y'}| &= 1, \nonumber \\
    \bm{e}_{z'} &= \bm{e}_{x'} \times \bm{e}_{y'}. \nonumber
\end{align}

As it was discussed in section \ref{sec:onlineLoc_methodology}, the shape of the ellipsoid determining the DA region in the local coordinate system is controlled by the parameters $\xi_{x^\prime}$, $\xi_{y^\prime}$ and $\xi_{z^\prime}$. In order to determine these parameters, the local velocity vector is interpolated on 1D grids created along the directions of $\bm{e}_{x'}$, $\bm{e}_{y'}$ and $\bm{e}_{z'}$ in the DA region. A projection is applied to this interpolated velocity field $\bm{u}$ to obtain its components along the local coordinate system. This new vector field will be referred to as $\bm{u}'$. By using the scalar components of $\bm{u}'$ along the $x'$, $y'$ and $z'$ directions, the auto-correlation functions for each of the velocity components are computed along their respective axes i.e. $R_{u'}(\bm{x}_{p}, \Delta x')$, $R_{v'}(\bm{x}_{p}, \Delta y')$, $R_{w'}(\bm{x}_{p}, \Delta z')$ with
\begin{equation}
    R_{u_i} (\bm{x}_o, \Delta x_i') = \ensavg{ u_i'(\bm{x}_o+\Delta x_i') u_i'(\bm{x}_o) },
    \label{eq:1D_corr_spatial}
\end{equation}
where $\bm{x}_o$ stands for the location of the observation sensor for each DA region, similar to the section \ref{sec:squareCyl2D}. The integral length scales are then computed by using the relation,
\begin{equation}
    \mathscr{L}_{i} = \int_{-\infty}^{\infty} R_{u'_i}(\Delta x'_i) d \Delta x^{\prime}_i.
    \label{eq:integral_len_sc}
\end{equation}

Finally, the values of $\xi_{x^\prime}$, $\xi_{y^\prime}$ and $\xi_{z^\prime}$ are determined by the conditions $L(\mathscr{L}_{x}) = L(\mathscr{L}_{y}) = L(\mathscr{L}_{z}) = \alpha$ along the axes $x'$, $y'$ and $z'$ using the relation
\begin{equation}
    \xi_i = \left(\frac{\mathscr{L}_i^2}{-2 \ln(\alpha)} \right)^{1/2},
    \label{eq:xi_formula}
\end{equation}
where $\alpha$ is the value of the localization function at a distance of $\mathscr{L}$ from the location of the observation sensor and it is taken $\alpha=0.85$ in the present study. This ensures that the $85\%$ of the DA correction is retained at a distance of local integral length scale from the observation probe location. Once the $\xi_{i}$ coefficients are determined, the DA procedure can be applied in the updated DA volume. The process is repeated for each sensor and the associated DA region, calculating therefore localization functions using the local features of the flow field.

%%% -----------------------------------------------------

\begin{figure}[!htb]
    \centering
    \includegraphics[width=0.9\textwidth]{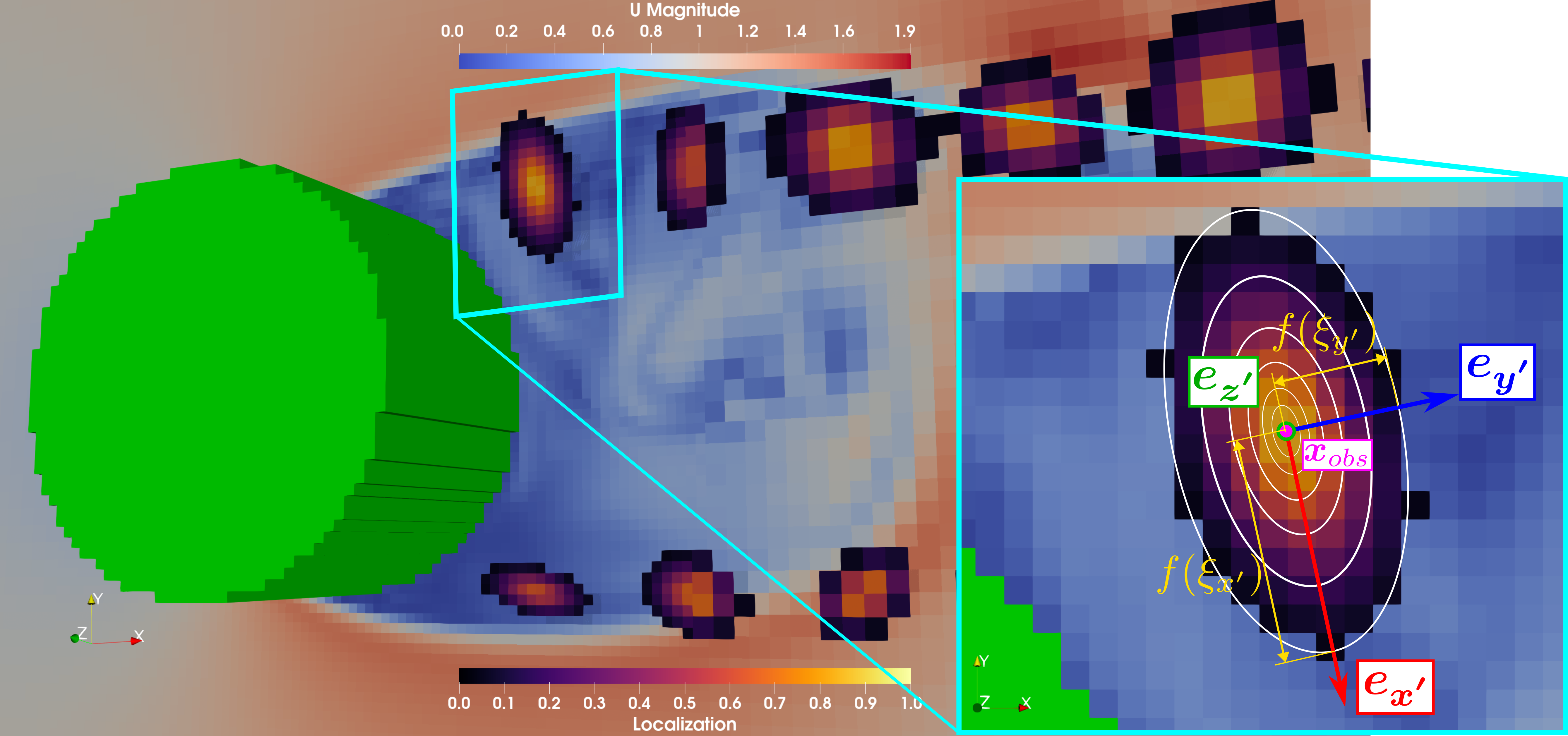}
    \caption{A cut-section normal to spanwise direction, demonstrating the contour fields of the localization functions and the velocity magnitude. Figure inset shows one of the 3D localization functions determined at a DA region.}
    \label{fig:example_field_onlineLoc}
\end{figure}

The resulting localization values in the DA regions are exemplified in figure \ref{fig:example_field_onlineLoc} at a given analysis phase. The regions plotted by the dark shade are characterized values of the localization function larger than $0.05$ i.e. regions where the DA state update is applied for more than $5\%$ of its initial magnitude.
%=========================================================
%   [[[  Example of 3D localization functions ]]]
%=========================================================
It can also be seen in figure \ref{fig:example_field_onlineLoc} that the localization function determined by PBOL varies for each observation probe in relation with the instantaneous, local features of the flow field. Similarly, each localization function evolves throughout the time with the variations of the unsteady flow field. It follows that the total number of cells used in the EnKF procedure and the volumes of DA regions with significant DA correction vary in time.

\begin{figure}[!htb]
    \centering
    \begin{subfigure}[b]{0.49\textwidth}
    \includegraphics[width=\textwidth]{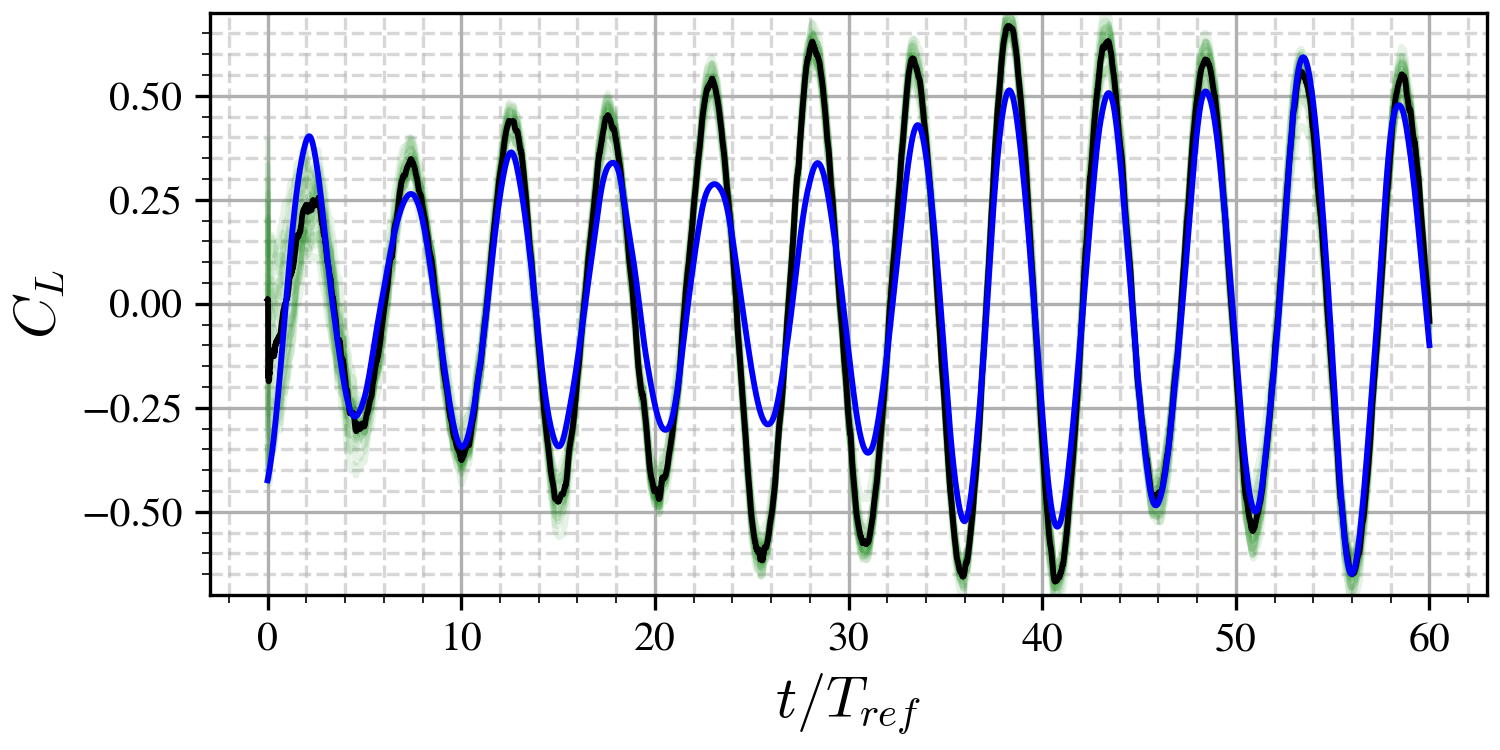}
    \caption{DA-HL}
    \end{subfigure} %\\
    \begin{subfigure}[b]{0.49\textwidth}
    \includegraphics[width=\textwidth]{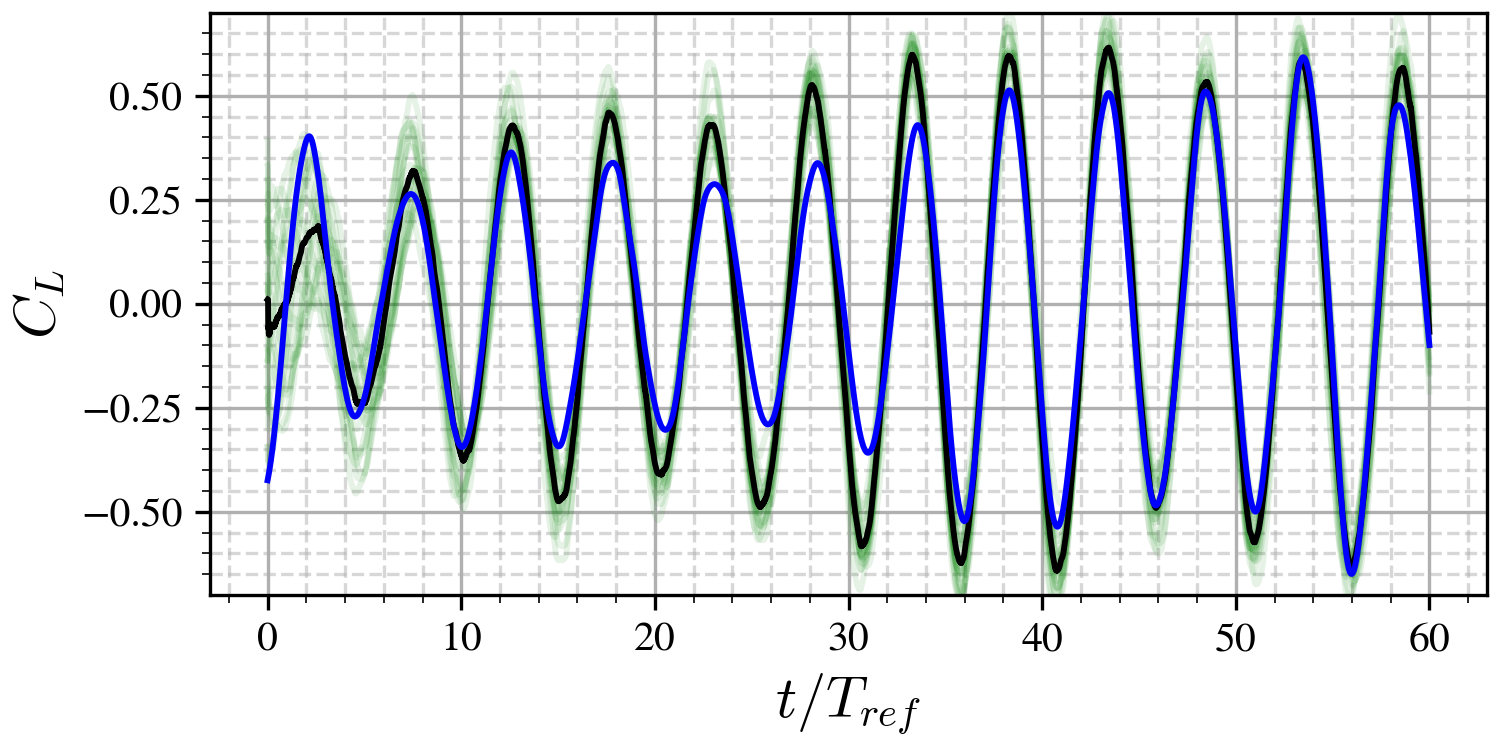}
    \caption{DA-PBOL}
    \end{subfigure}
    \centering
    \begin{subfigure}[b]{0.49\textwidth}
    \includegraphics[width=\textwidth]{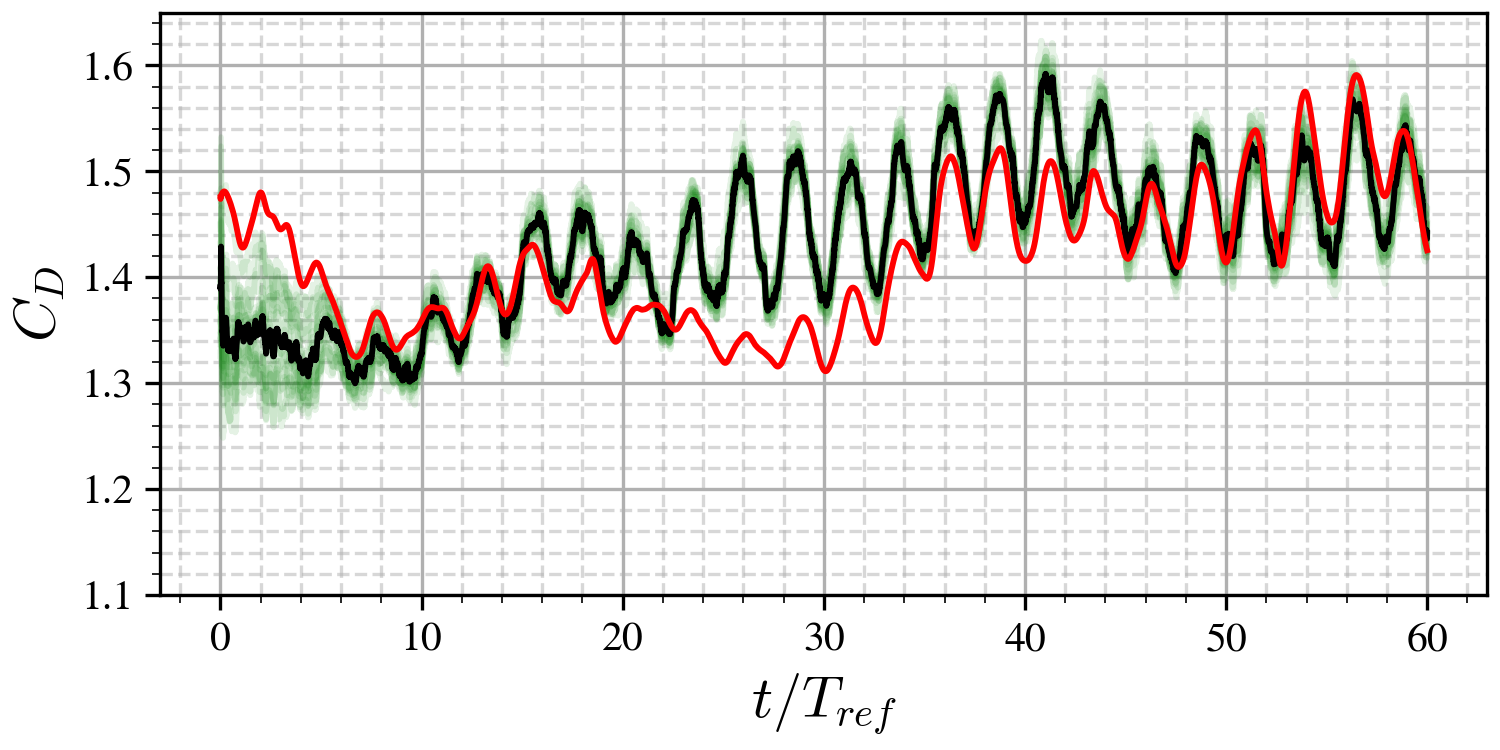}
    \caption{DA-HL}
    \end{subfigure} %\\
    \begin{subfigure}[b]{0.49\textwidth}
    \includegraphics[width=\textwidth]{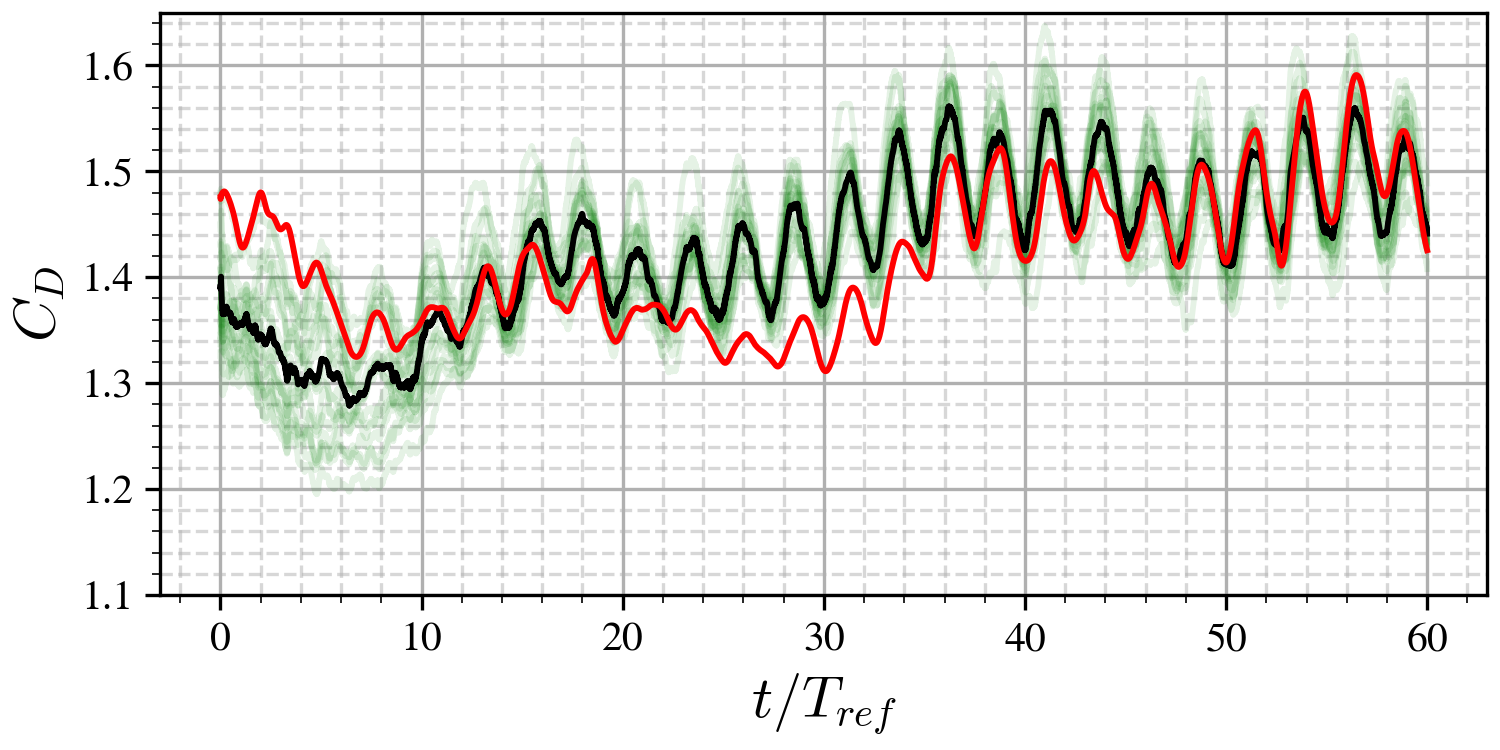}
    \caption{DA-PBOL}
    \end{subfigure}
    \caption{Time variation of the $C_L$ and $C_D$ of the reference simulation, assimilated members and their mean. Both DA-HL and DA-PBOL cases are run with the \textit{horizontal-sparse} configuration of the observation probes.}
    \label{fig:forceCoeffs_HL_OL_compare}
\end{figure}

Results of the EnKF procedures using HL and PBOL localizations are now compared with the results from two DA runs using the HL and PBOL methodologies, which will be referred as DA-HL and DA-PBOL. Figure \ref{fig:forceCoeffs_HL_OL_compare} shows the synchronization of the force coefficients for DA-HL and DA-PBOL, in time. In both cases, the ensemble mean approaches to the reference signal with increasing DA stages. The qualitative behavior of the synchronization are similar for DA-HL and DA-PBOL cases, even though the regions where the state vector is assimilated are smaller in DA-PBOL case, after choosing the optimal localization function. This shows the PBOL methodology is able to exclude the regions, which are not correlated with the observation data, from the assimilation process.

%=========================================================
% [[[ Online-localization Delta CL & Delta CD figures ]]]
%=========================================================

\begin{figure}[!htb]
    \centering
    \begin{subfigure}[b]{0.49\textwidth}
    \includegraphics[width=\textwidth]{figs/sync_cyl3D/horizontal_sparse/disparity_lift_coeff-time-members.png}
    \caption{DA-HL}
    \end{subfigure} %\\
    \begin{subfigure}[b]{0.49\textwidth}
    \includegraphics[width=\textwidth]{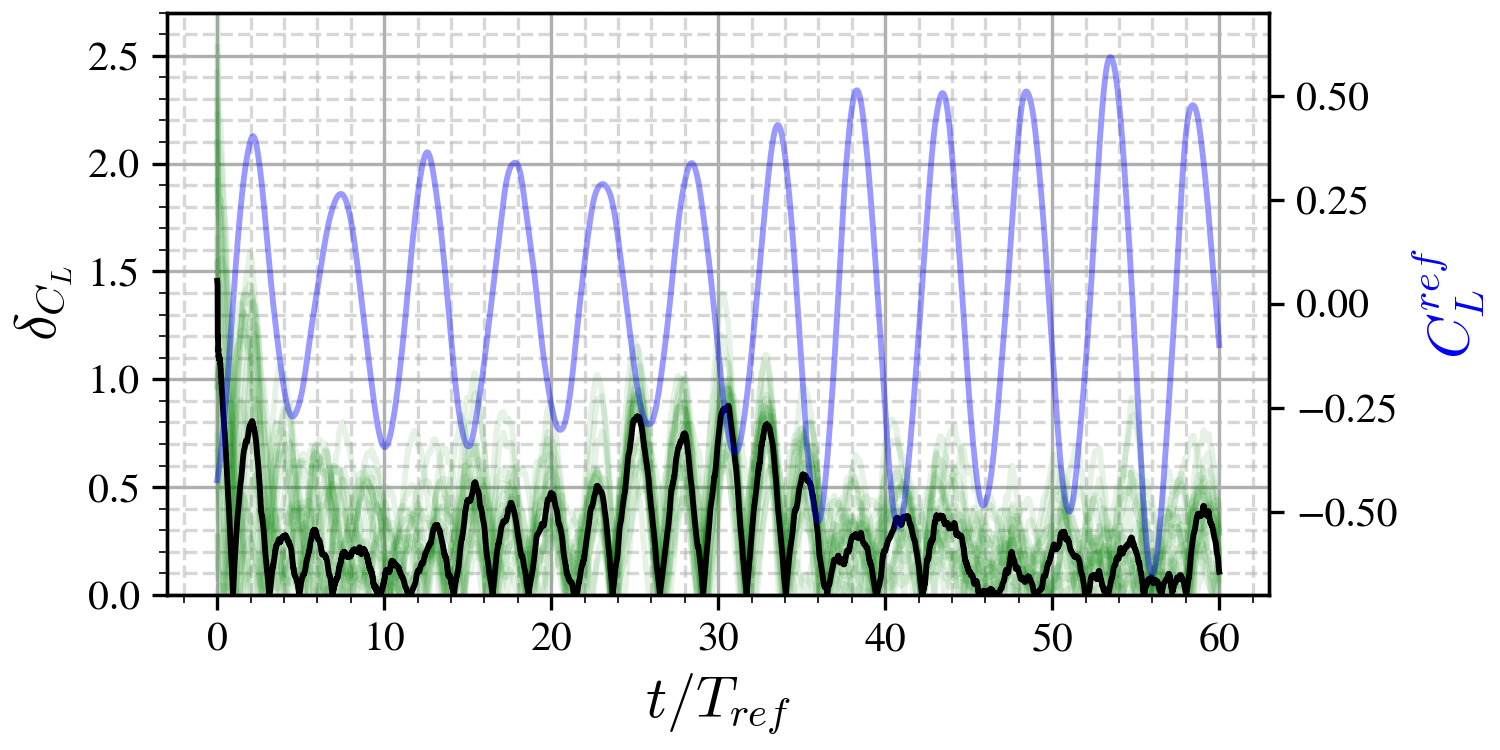}
    \caption{DA-PBOL}
    \end{subfigure}
    \centering
    \begin{subfigure}[b]{0.49\textwidth}
    \includegraphics[width=\textwidth]{figs/sync_cyl3D/horizontal_sparse/disparity_drag_coeff-time-members.png}
    \caption{DA-HL}
    \end{subfigure} %\\
    \begin{subfigure}[b]{0.49\textwidth}
    \includegraphics[width=\textwidth]{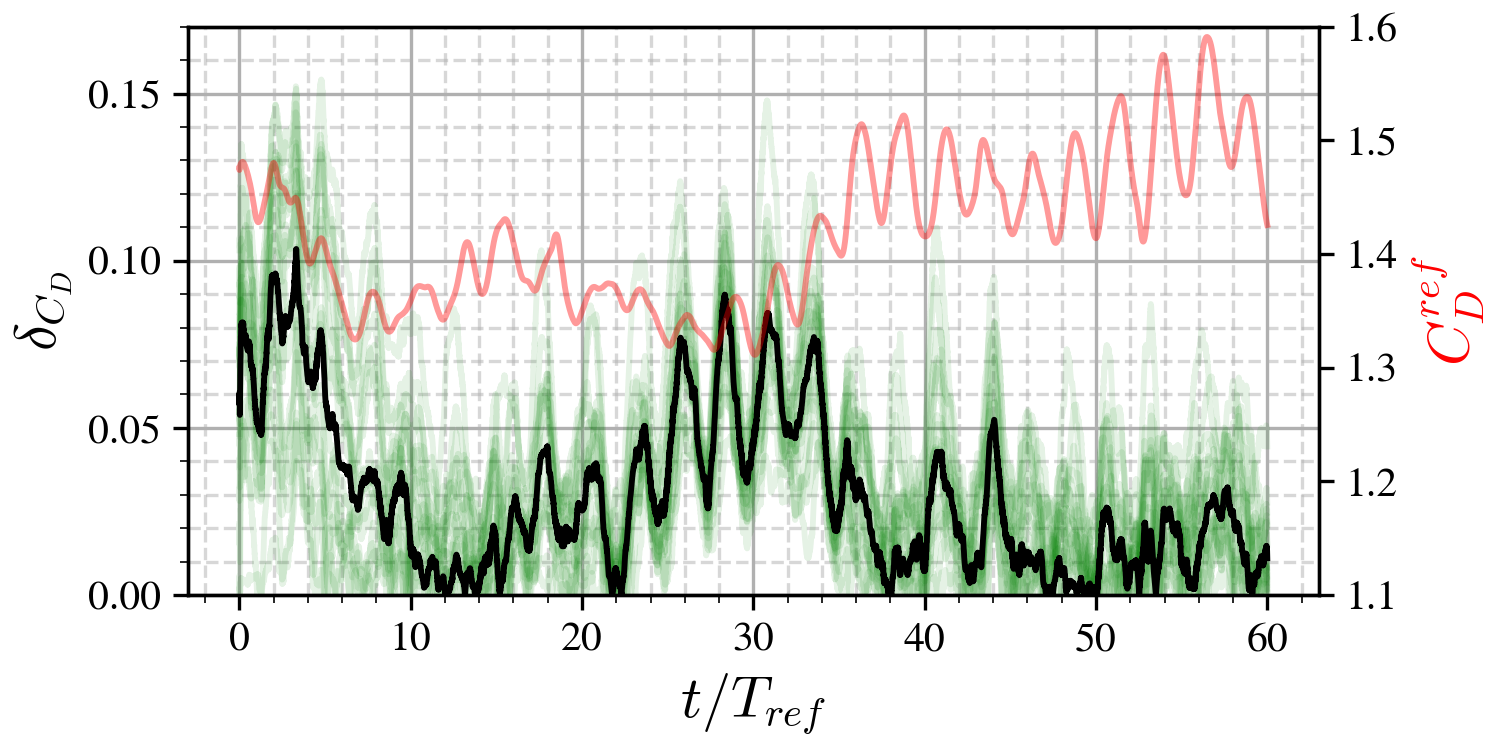}
    \caption{DA-PBOL}
    \end{subfigure}
    \caption{Time variation of the discrepancy between the $C_L$ and $C_D$ of the reference simulation, assimilated members and their mean. Both DA-HL and DA-PBOL cases are run with the \textit{horizontal-sparse} configuration of the observation probes.}
    \label{fig:delta_coeff_HL_OL_compare}
\end{figure}

Figure \ref{fig:delta_coeff_HL_OL_compare} shows the discrepancy between the force coefficients $C_L$, $C_D$ of the reference simulation, the ensemble members and their mean for the two DA runs considered. It can be seen that the discrepancy between the ensemble mean curves for $C_L$ and $C_D$ using PBOL localization are qualitatively similar to the ones observed for the HL case. However, one can see that the variance between the ensemble members appears to be larger for the case of PBOL. This observation can be justified by the fact that the PBOL localization limits the state update to the regions around the that are relevant for the physical behavior of the flow. Therefore, smaller volumes are affected by the DA procedure and a globally higher variance is conserved, which is a positive point indicating that the system should be less prone to collapse of the variance, which is one on the main problems of the classical EnKF approaches \cite{Asch2016} and for which inflation must be systematically used to palliate such issue and  it precludes efficient optimization and synchronization in the case of time evolving physical phenomena, such as turbulence.

%=========================================================
% [[[ Online-localization Delta CL & Delta CD figures ]]]
%=========================================================

\begin{figure}[!htb]
    \centering
    \begin{subfigure}[t]{0.49\textwidth}
    \includegraphics[width=\textwidth]{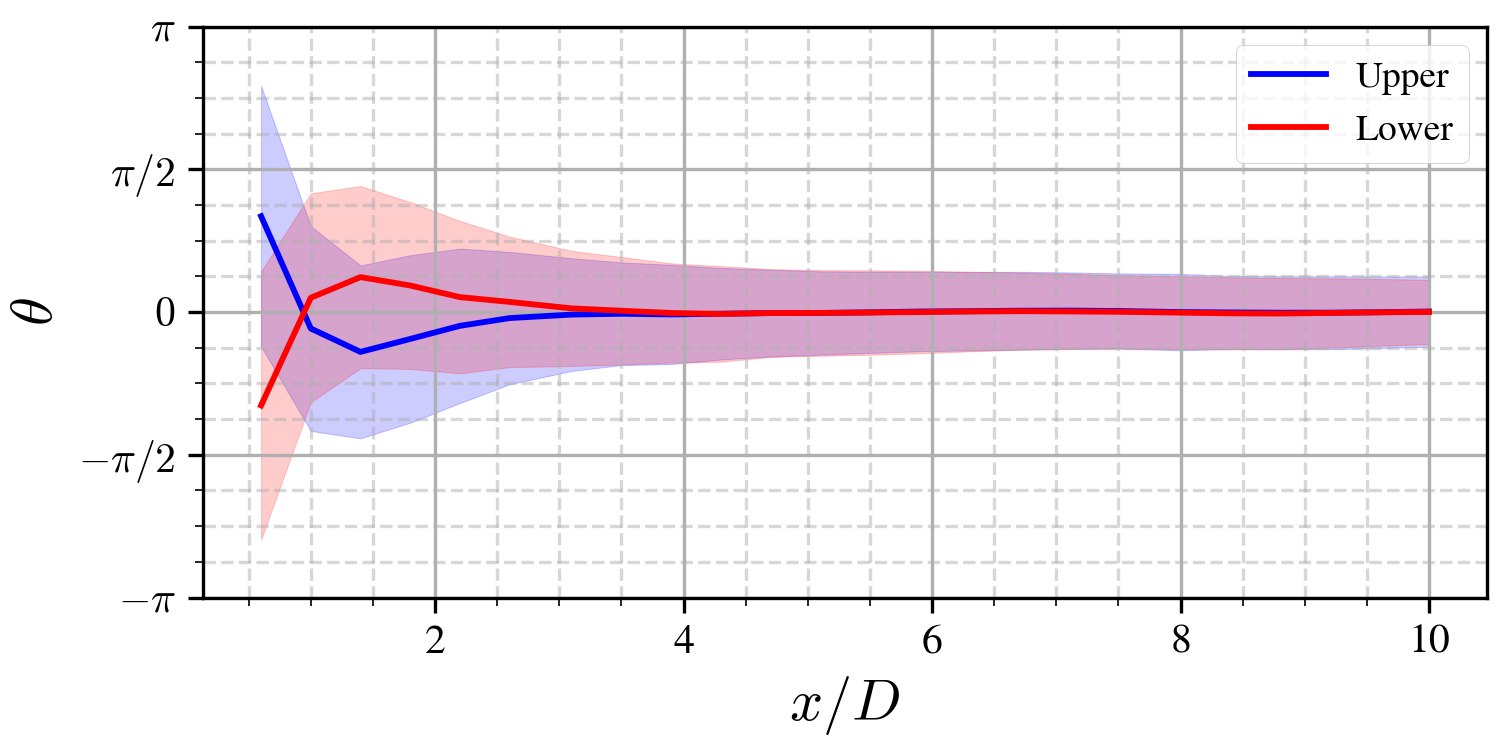}
    \caption{Circular statistics of the orientation of $\bm{e}_{x'}$.}
    \label{fig:stats_theta_streamwise}
    \end{subfigure} \hfill %\\
    \begin{subfigure}[t]{0.49\textwidth}
    \includegraphics[width=\textwidth]{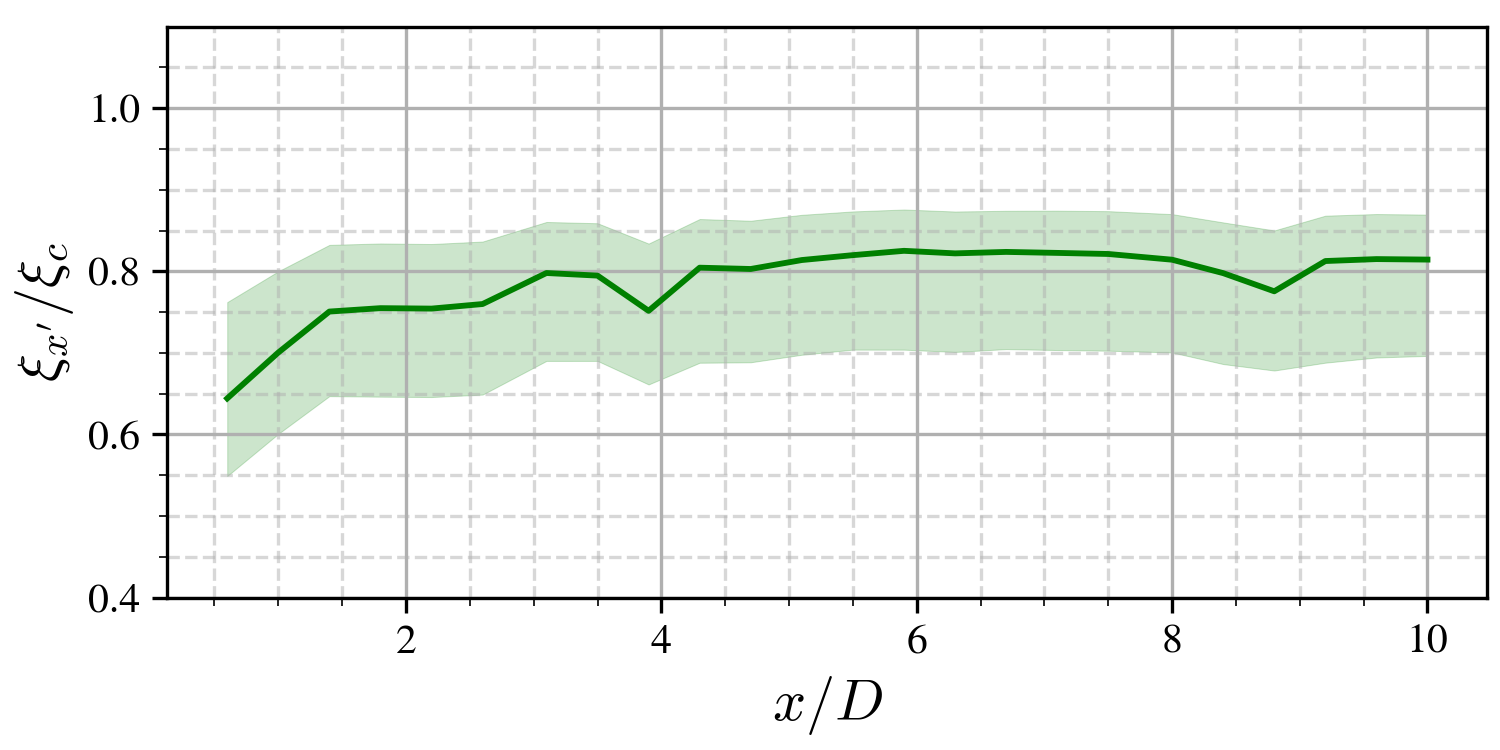}
    \caption{Mean and standard deviation of $\xi_{x^\prime}/\xi_c$.}
    \label{fig:stats_sigmaX_streamwise}
    \end{subfigure}
    \begin{subfigure}[t]{0.49\textwidth}
    \includegraphics[width=\textwidth]{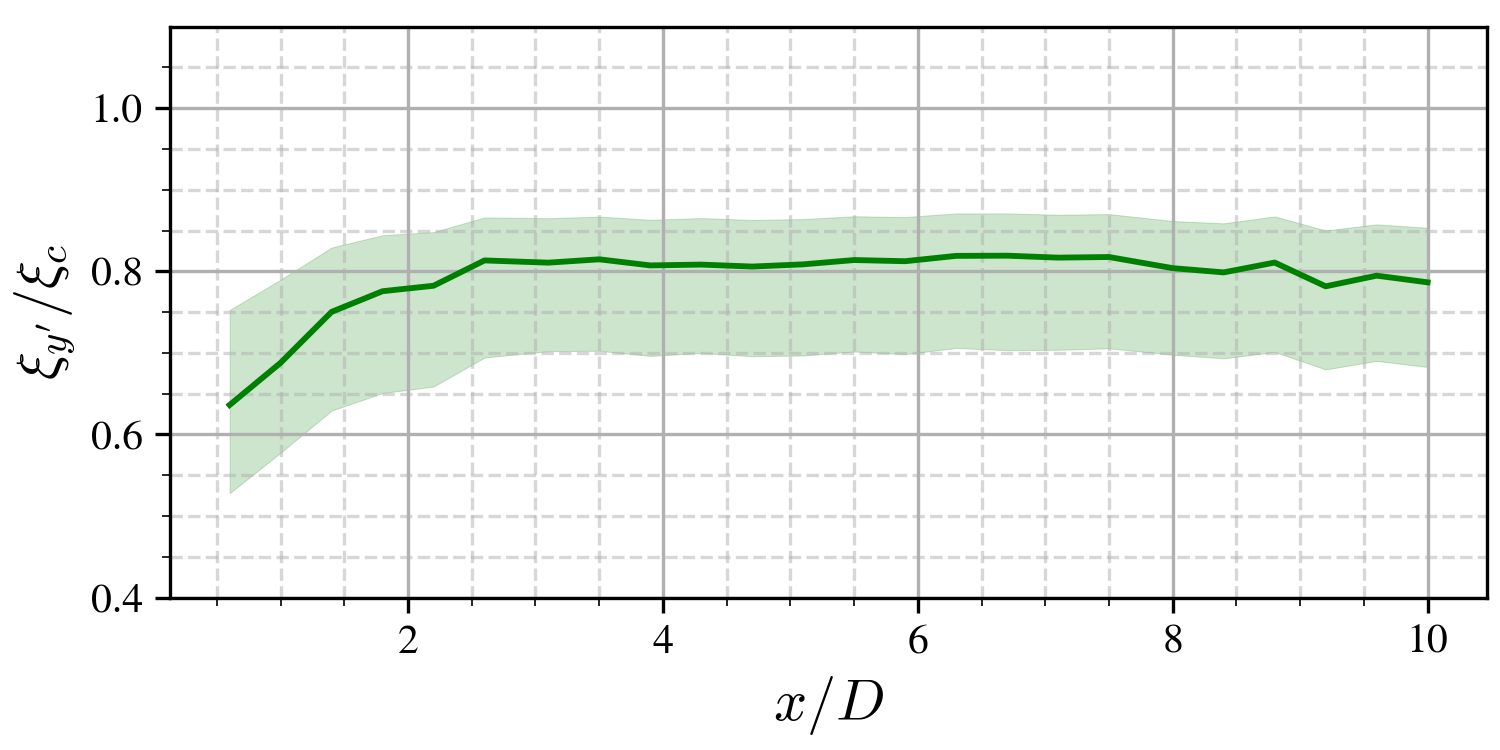}
    \caption{Mean and standard deviation of $\xi_{y^\prime}/\xi_c$.}
    \label{fig:stats_sigmaY_streamwise}
    \end{subfigure}
    \begin{subfigure}[t]{0.49\textwidth}
    \includegraphics[width=\textwidth]{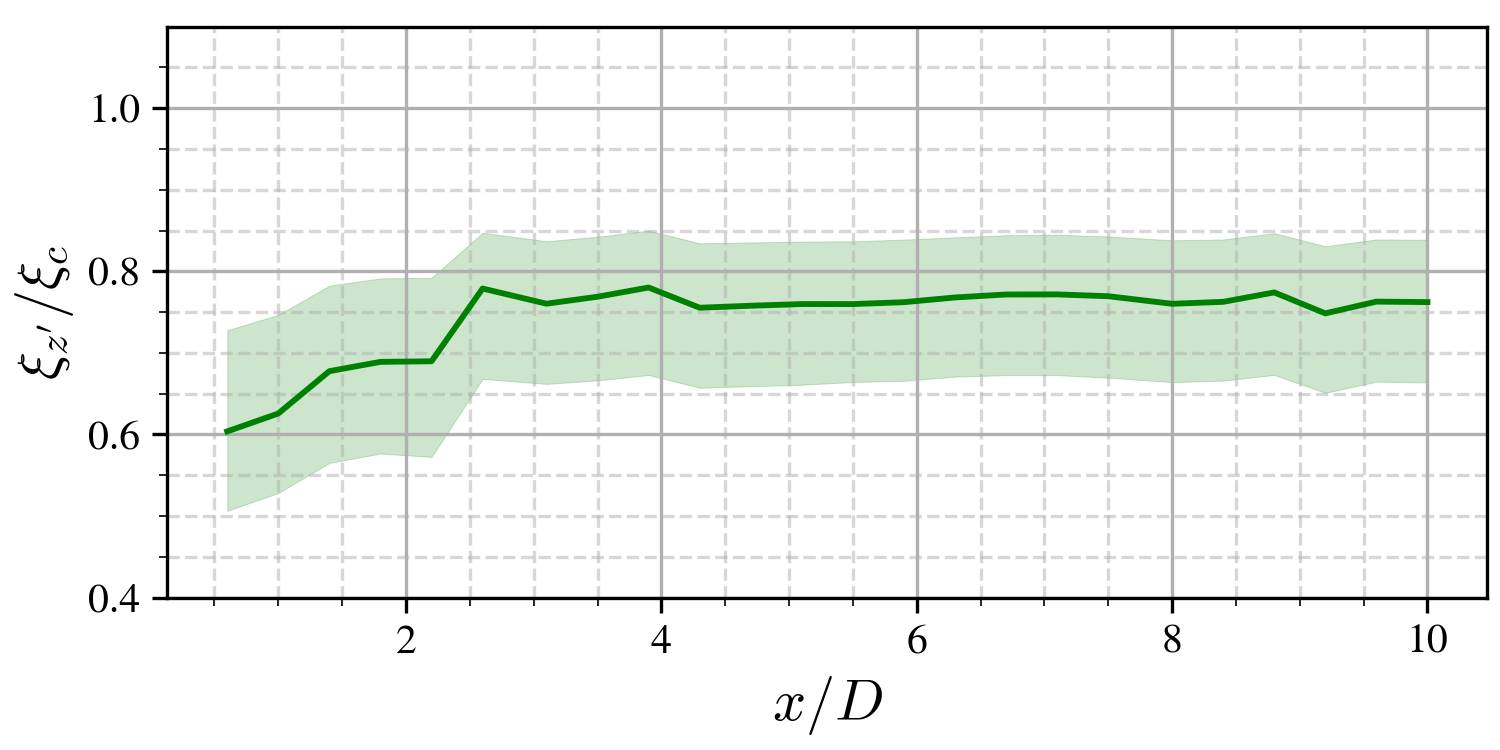}
    \caption{Mean and standard deviation of $\xi_{z^\prime}/\xi_c$.}
    \label{fig:stats_sigmaZ_streamwise}
    \end{subfigure}
    \caption{Statistics of the localization parameters for the PBOL case. (a) Mean value and the standard deviation of $\theta$, the angle between mean $\bm{e}_{x'}$ and the streamwise vector $\bm{i}$, and of of (b) $\xi_{x^\prime}/\xi_c$, (c) $\xi_{y^\prime}/\xi_c$, (d) $\xi_{z^\prime}/\xi_c$ along the streamwise direction.}
    \label{fig:stats_params_PBOL}
\end{figure}

The sensitivity of the parameters characterizing the PBOL to the positioning of the sensor is now investigated. To this purpose, averages are performed in the spanwise direction $z$, so that results are obtained for the two plane for $y= \pm 0.5D$, $x \in [0.55D, \, 10D]$. The investigation is focused on how local physical features of the flow affect the determination of such model constants. Figure \ref{fig:stats_theta_streamwise} shows the spanwise and time averaged angle $\theta$ between the projection of $\bm{e}_{x'}$ vector on $x-y$ plane and the streamwise unit vector $\bm{i}$ along x-axis.

Circular averaging operation \cite{Mardia1999} is applied to find the time-averaged $\theta$ values, $\overline{\theta}$, for a given DA region. Furthermore, by taking advantage of the the spanwise homogeneity of the case, the $\overline{\theta}$ values of DA regions are averaged by circular mean for the DA regions sharing the same streamwise location. The statistics are observed to be symmetric with respect to the $y=0$ plane for the upper and lower observation planes, as shown in figure \ref{fig:stats_theta_streamwise}. In the zone closer to the cylinder, for $x < 3.5 D$, the parameter $\theta$ is affected by the recirculation bubble forming behind the solid body. Moving downstream in the streamwise direction, it can be seen that the orientation of $\bm{e}_{x'}$ and therefore $\theta$ is mainly determined by the mean streamwise flow direction. It can also be observed that the standard deviation of the parameter $\theta$ (shaded region) is significantly higher in the regions affected by the separation in the proximity of the cylinder, exhibiting a maximum value of $\sigma_{\theta} = 83^{\circ}$ for $x/D = 0.6$. The variation of $\theta$ reduces and becomes constant moving downstream in the wake region. This shows that the instantaneous local convection velocity, which dominates the local correlation field in the DA regions becomes more similar to the mean streamwise velocity.

%=========================================================
% [[[ Stats of the PBOL parameters ]]]
%=========================================================

Figure \ref{fig:stats_sigmaX_streamwise} shows the mean and standard deviation of the $\xi_{x^\prime}/\xi_c$ where $\xi_c$ is the maximum value for $\xi_{x^\prime}$, $\xi_{y^\prime}$ and $\xi_{z^\prime}$ which leads to $L=0.05$ at $r = r_c$. The average $\xi_{x^\prime}/\xi_c$ value determined by the PBOL methodology is $0.65$ at the streamwise location of $x/D = 0.6$, where the effects of the separation just after the cylinder is still very effective. The mean value of $\xi_{x^\prime}/\xi_c$ increases when going in the streamwise direction and it approaches to $\xi_{x^\prime}/\xi_c = 0.8$ and remains at that value after $x/D \approx 3$. This means that the DA volumes determined by the PBOL are smaller on average at the locations close to the cylinder, when compared to downstream locations. Even though the evolution of $\xi_{y^\prime}/\xi_c$ and $\xi_{z^\prime}/\xi_c$ remains qualitatively similar (see figures \ref{fig:stats_sigmaY_streamwise} and \ref{fig:stats_sigmaZ_streamwise}), it can also be observed that the mean values of $\xi_{z^\prime}/\xi_c$ remains lower than the $\xi_{x^\prime}/\xi_c$ and $\xi_{y^\prime}/\xi_c$ at all the streamwise locations. This shows that the local velocity correlation lengths in $z'$ direction remains lower than the other two directions in the downstream direction.

The standard deviations shown by the shaded regions in figures \ref{fig:stats_sigmaX_streamwise}, \ref{fig:stats_sigmaY_streamwise} and \ref{fig:stats_sigmaZ_streamwise}, start with equal values for positive and negative sides of the mean curve in the range $x/D \lessapprox 1$ for $\xi_{x^\prime}/\xi_c$, $\xi_{y^\prime}/\xi_c$ and $x/D \lessapprox 2$ for $\xi_{z^\prime}/\xi_c$. For these streamwise locations, where the DA regions are located near the recirculation zone, strongly unsteady motions are driven by the periodic flow separation. The optimized DA regions are here relatively small on average, with relatively large, positive and negative variations of the DA volume. Similarly, a high variance is observed for the orientation of the primary axis (see figure \ref{fig:stats_theta_streamwise}). Going downstream, the tendency of having DA regions larger than the average size gets lower after $x/D \gtrapprox 2$, while the tendency of having smaller DA regions than the average size remains similar, manifested by the unequal standard deviations at the upper/lower sides of the mean curve.

\begin{figure}[!htb]
    \centering
    \includegraphics[width=0.7\textwidth]{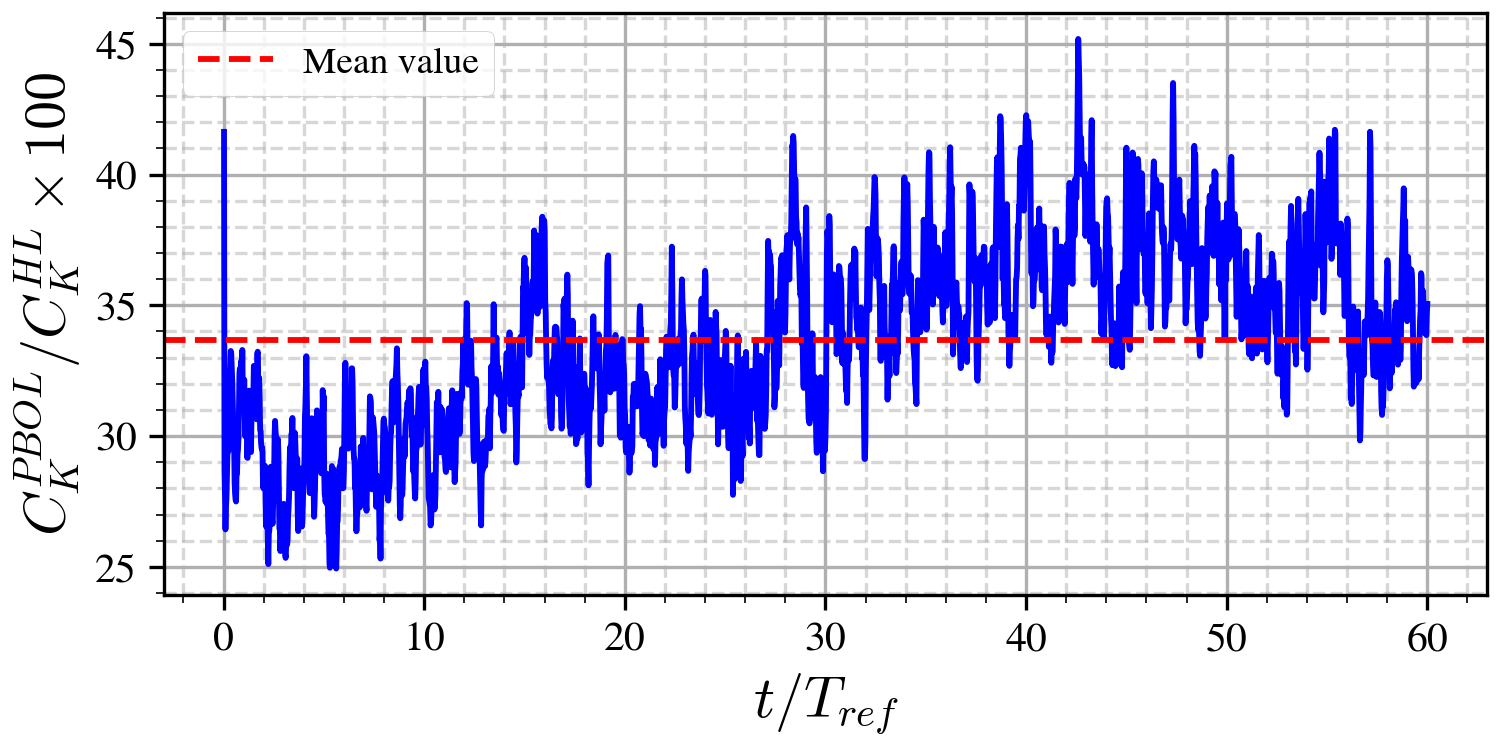}
    \caption{Computational cost of EnKF with the application of the PBOL as a percentage of the cost of HL.}
    \label{fig:cost_comparison_HL_OL}
\end{figure}

Finally, a comparison is performed between the computational costs associated with DA-HL and DA-PBOL methodologies, in order to highlight the computational gains of the latter. The computational tasks to be performed for the calculation of the Kalman gain for both cases are evaluated using the relation \cite{Villanueva2024_PhDThesis},
\begin{equation}
    C_K = \mathcal{O}(K) = \mathcal{O} \left[ N_r N_o^3 + \left( \sum^{N_R}_{i=1} N_{DOF,i} + N_e + N_r \right) N_o^2 + N_{DOF,i} N_e N_o \right],
    \label{eq:complexity_DA_cost}
\end{equation}
where $N_r$ is the number of DA regions, $N_{DOF}$ is the number of degrees of freedom of each of those regions, $N_e$ is the number of ensemble members and $N_o$ is the number of observations at each sensor e.g. $N_o = 3$ if the three components of the velocity are observed. Figure \ref{fig:cost_comparison_HL_OL} shows the ratios of $C_K^{PBOL}$ and $C_K^{HL}$ associated with the computational cost of PBOL, as a percentage of the cost of HL. The cost is not evaluated here as a real computational cost, but it is estimated by the eq. \ref{eq:complexity_DA_cost} and presented as a ratio of the both cases. It can be emphasized here that the difference in the performance is due to the variation of the $N_{DOF}$ values for the PBOL and HL procedures. The computational cost of the HL remains constant during the investigation, as the localization functions applied at each sensor do not change, thus the volumes of DA regions remain same throughout the time. For the time interval  $t/T_{ref} \in [0, \, 60]$, one can see that the computational cost of the DA-PBOL corresponds to around $33.7\%$ of the cost of DA-HL. Therefore, while the calculation of physical quantities to select DA physical domains increases complexity and calculations to be performed, the reduction of the global DA space associated with PBOL localization more than balances that increase, providing an efficient localization strategy tailored to the physical features of the flow investigated.  

\section{Conclusion} \label{sec:conclusion}

This work presents a physics-based online-localization methodology, referred to as PBOL, for data assimilation by EnKF. The algorithm is developed for the applications of simulations of turbulent flows with complex boundaries. Contrary to the more classical, distance-based covariance localization scheme, a methodology where the localization function varies as a function of the flow conditions in the vicinity of each observation probe is proposed with the aim of conducting the DA only in the relevant regions i.e. where the expected correlation with the observation data is significant. The results from the synchronization test cases has been presented for both 2D and 3D simulations of fluid flows, where the former being laminar, unsteady and the latter being in fully turbulent regime.

The results from the 2D test case, the flow around a square cylinder, shows a gain in the computational costs. The synchronization of the model achieved rapidly, while the speed of convergence is observed to be affected highly by the number of cells being assimilated. In the 3D test case, the PBOL is applied for the synchronization of the flow around a circular cylinder. First, the performance of the EnKF for the synchronization of the turbulent flow simulations with various observation probe configurations are presented with the HL methodology. Following that, the synchronization case is re-run, using the PBOL methodology and the results are compared with the HL case. The variation of the localization function in space and in time for different DA regions are presented. It is shown that the statistics of the localization parameters display their connection with the physics of the flow e.g. the symmetry of the mean flow field with respect to the center-plane.

The synchronization performance remains similar when using PBOL and HL methodologies, even though the DA regions are much more restricted and the number of cells being assimilated are much lower with the former with the choice of the optimal localization extent. Meanwhile, the collapse of the ensemble member states are less dramatic when PBOL is utilized, which grants an advantage for the convergence towards the reference state. In addition to that, the computational cost for the DA stage is shown to reduce significantly, down to $34\%$ of the DA using HL strategy, thanks to the dynamic choice of the optimal DA regions based on the physics of the problem.

Future development will deal with more general formulation of the PBOL method, in order to prescribe arbitrary shapes of the DA regions containing potentially multiple sensors. In addition, the possibility to automatize the choice of the physical constraints by machine learning will be investigated. This last point will permit to exploit the potential of artificial intelligence to reduce human supervision in the process and to recognize the best physical criteria to be selected according to the case of investigation.

\section{Acknowledgments} \label{sec:acknowledgements}

This research activity was performed in the framework of the project ANR
JCJC-2021 IWP-IBM-DA. Numerical computations were performed using resources by GENCI at the TGCC supercomputer obtained via the grant 2023-A0142A01741 on the supercomputer Joliot Curie’s SKL partition.

%% The Appendices part is started with the command \appendix;
%% appendix sections are then done as normal sections
%\appendix
%\section{Example Appendix Section}
%\label{app1}

%% For citations use: 
%%       \cite{<label>} ==> [1]

%% If you have bib database file and want bibtex to generate the
%% bibitems, please use
%%
%\bibliographystyle{elsarticle-num} 
\bibliographystyle{elsarticle-num-names} 
\bibliography{references}

\end{document}